\documentclass[11pt,a4paper]{article}
\usepackage{amssymb}
\usepackage{amscd}
\usepackage{amsmath}
\usepackage{graphicx}

\newcounter{mo}

\newcounter{bk}

\newcounter{sa}

\newcounter{aag}

\newcommand{\Si}{\Sigma}
\newcommand{\tr}{{\rm tr}}

\newcommand{\ti}[1]{\tilde{#1}}
\newcommand{\om}{\omega}
\newcommand{\Om}{\Omega}
\newcommand{\de}{\delta}
\newcommand{\al}{\alpha}
\newcommand{\te}{\theta}
\newcommand{\vth}{\vartheta}
\newcommand{\be}{\beta}
\newcommand{\la}{\lambda}
\newcommand{\La}{\Lambda}

\newcommand{\ve}{\varepsilon}
\newcommand{\ep}{\epsilon}
\newcommand{\vf}{\varphi}
\newcommand{\G}{\Gamma}
\newcommand{\ka}{\kappa}

\newcommand{\ga}{\gamma}

\newcommand{\si}{\sigma}
\newcommand{\wt}[1]{\widetilde{#1}}
\newcommand{\brc}[1]{\left(#1\right)}
\newcommand{\rme}{\textrm{e}}
\newcommand{\rmd}{\textrm{d}}
\newcommand{\bfi}[1]{\left\{ #1\right\}}
\newcommand{\bsq}[1]{\left[#1\right]}
\newcommand{\matr}[2]{\begin{array}{#1}#2\end{array}}
\newcommand{\svs}{\\[7pt]}

\def\bfe{{\bf e}}

\def\bft{{\bf t}}
\def\bfx{{\bf x}}

\def\bfS{{\bf S}}
\def\bfJ{{\bf J}}
\def\bfC{{\bf C}}
\def\bfF{{\bf F}}

\def\bfW{{\bf W}}

\def\cC{{\cal C}}
\def\cD{{\cal D}}
\def\cF{{\cal F}}

\def\cO{{\cal O}}
\def\cS{{\cal S}}
\def\cA{{\cal A}}
\def\cG{{\cal G}}
\def\cL{{\cal L}}
\def\cM{{\cal M}}
\def\cP{{\cal P}}
\def\cR{{\cal R}}
\def\cT{{\cal T}}
\def\cU{{\cal U}}

\def\cW{{\cal W}}
\def\cX{{\cal X}}
\def\cY{{\cal Y}}

\def\mC{{\mathbb C}}
\def\mZ{{\mathbb Z}}
\def\mR{{\mathbb R}}
\def\mN{{\mathbb N}}

\newcommand{\mat}[4]{\left(\begin{array}{cc}{#1}&{#2}\\{#3}&{#4}
\end{array}\right)}

\newcommand{\beq}[1]{\begin{equation}\label{#1}}
\newcommand{\beqnl}{\begin{equation}}
\newcommand{\eq}{\end{equation}}
\newcommand{\beqn}[1]{\begin{eqnarray}\label{#1}}
\newcommand{\eqn}{\end{eqnarray}}
\newcommand{\p}{\partial}
\newcommand{\di}{{\rm diag}}
\newcommand{\oh}{\frac{1}{2}}
\def\1m{^{-1}}

\newcommand{\SL}{{\rm SL}(2,{\mathbb C})}
\newcommand{\slt}{{\rm sl}(2,{\mathbb C})}
\newcommand{\GLN}{{\rm GL}(N,{\mathbb C})}
\newcommand{\GLT}{{\rm GL}(2,{\mathbb C})}
\def\sln{{\rm sl}(N, {\mathbb C})}
\def\SLN{{\rm SL}(N, {\mathbb C})}
\newcommand{\gln}{{\rm gl}(N, {\mathbb C})}

\def\f1#1{\frac{1}{#1}}

\newcommand{\flr}[1]{\left\lfloor #1\right\rfloor}

\newcommand{\rar}{\rightarrow}

\newcommand{\bp}{\bar{\partial}}
\newcommand{\bz}{\bar{z}}
\newcommand{\bA}{\bar{A}}
\newcommand{\bL}{\bar{L}}

\def\frak{\mathfrak}
\def\gg{{\frak g}}

\def\gS{{\frak S}}

\def\gga{{\frak a}}
\def\gb{{\frak b}}
\def\gh{{\frak h}}
\def\gn{{\frak n}}
\def\gM{{\frak M}}

\newcommand{\ran}{\rangle}
\newcommand{\lan}{\langle}

\def\bw{\bar{w}}

\renewcommand{\theequation}{\thesection.\arabic{equation}}
\newtheorem{predl}{Proposition}[section]

\newtheorem{rem}{Remark}[section]

\newtheorem{lem}{Lemma}[section]

\def\teal{\te\ep\cdot\al}

\def\eal{\ep\cdot\al}
\def\ega{\ep\cdot\ga}

\begin{document}

\vspace{0.3in}
\begin{flushright}
 ITEP-TH-53/12\\
%{\large\today}\\
%\today
\end{flushright}
\vspace{15mm}
\begin{center}
{\Large{\bf
 Painlev{\'e} Field Theory}
}\\
\vspace{12mm} {\bf  G. Aminov},$^{\,\flat\,\sharp}$ {\bf S. Arthamonov},$^{\,\flat\,\dag}$
 {\bf A. Levin},$^{\,\natural\,\flat}$ {\bf M. Olshanetsky},$^{\,\flat\,\sharp}$
 {\bf A. Zotov}$^{\,\flat\,\sharp\,\diamondsuit}$\\ \vspace{7mm}
 \vspace{3mm} $^\flat$ - {\sf Institute of Theoretical and Experimental Physics,
  Moscow, 117218 Russia}\\
 \vspace{2mm}$^\natural$ - {\sf Laboratory of Algebraic Geometry, GU-HSE,
 7 Vavilova Str., Moscow, 117312 Russia}\\
 %\vspace{2mm} $^\sharp$ - {\sf Steklov Mathematical Institute, 8 Gubkina Str.,
 %Moscow, 119991, Russia}\\
 \vspace{2mm} $^\sharp$ - {\sf Moscow Institute of Physics and Technology,
 %Inststitutskii per.  9,
  Dolgoprudny, 141700
 %Moscow Region,
 Russia}\\
 \vspace{2mm} $^\diamondsuit$ - {\sf Steklov Mathematical Institute, RAS, 8 Gubkina Str., Moscow, 119991 Russia
 }\\
 \vspace{2mm} $^\dag$ - {\sf Rutgers University, New Brunswick, NJ, USA
 }
 %\vspace{4mm} E-mails:
 %{\em  aminov@itep.ru}; {\em  artamonov@itep.ru}; {\em alevin57@gmail.com}; {\em olshanet@itep.ru};  {\em zotov@itep.ru}\\
 \vspace{5mm}
 \let\thefootnote\relax\footnote{}
 \let\thefootnote\relax\footnote{E-mails:
 {\em  aminov@itep.ru}; {\em  artamonov@itep.ru}; {\em alevin57@gmail.com}; {\em olshanet@itep.ru};  {\em zotov@itep.ru}.}
\end{center}

\vspace{0mm}

\begin{abstract}
\vskip2mm

We propose multidimensional versions of the  Painlev{\'e} VI equation and its degenerations.
These field theories are related to the isomonodromy problems of flat holomorphic infinite
 rank bundles over elliptic curves and take the form of non-autonomous Hamiltonian equations.
 The modular parameter of curves plays the role of "time".
 Reduction of the field equations to the zero modes leads to $\SLN$ monodromy preserving equations.
The latter coincide with the  Painlev{\'e} VI equation for $N\!=\!2$. We consider two types of the bundles. In the first one
the group of automorphisms is the centrally and cocentrally extended loop group $L(\SLN)$ or some multiloop group. In the
case of the Painlev{\'e} VI field theory in D=1+1 four constants of the  Painlev{\'e} VI equation become dynamical fields.
%\mo{bf 1}
The second type of bundles are defined by the group of  automorphisms
of the noncommutative torus. They
 lead  to the equations in dimension 2+1. In
both cases we consider trigonometric, rational and scaling limits of the theories. Generically (except some degenerate cases)
the derived equations are nonlocal.
 We consider Whitham  quasiclassical limit to integrable systems.
 In this way we derive two and three dimensional integrable nonlocal versions of the
integrable Euler-Arnold tops.
\end{abstract}

\vspace{10mm}

{\small{ \tableofcontents}}

\section{Introduction}
\setcounter{equation}{0}

This paper is an attempt to write down field theory generalizations of the monodromy preserving equations.
The latter are non-autonomous Hamiltonian ODE's and our construction is the passage to the non-autonomous Hamiltonian PDE's.
In their definition we start from the isomonodromy problems for infinite rank vector bundles over elliptic curves
(see subsection\,\ref{2ft}).
 The modular parameter $\tau$ of the curves plays the role of "time".
 Here we use the construction proposed for the isospectral problems \cite{KLO,Kr,LOZ}
 \footnote{There was another approach to the  field theory generalizations of the monodromy preserving equations
based on the Painlev{\'e} property of ODE \cite{ARS,WTG}. We don't know interrelations between these two constructions.}.
In general cases we come to non-local equations in dimensions 1+1, 1+2, 1+3 of the form
$\p_\tau\bfS(\bfx,\tau)=P(\bfS,\bfx,\tau)$. Here $P(\bfS,\bfx,\tau)$ is non-linear pseudo differential
operator.
Their non-locality is similar to the non-local equations for vorticities of two-dimensional hydrodynamics of ideal fluid.
 In this case $\bfx=(x_1,x_2)$, the operator $P$ is defined by the Poisson brackets
$\{\bfS,\Delta^{-1}\bfS\}$ where $\Delta$ is the Laplace-Beltrami operator. To come to the isomonodromy problem
we replace $\Delta$ on the operator $\bp^2$, or its "trigonometric" and "elliptic" generalizations.

%%%%%%%%%%%%%%%%%%%%%%%%%%%%%%%%%%%%%%%%%%%%%%%%%%%%%%%%%%%%%%%%%%%%%%%%%%%%%%%%%%%%%%%%%%

\subsection{Painlev\'e VI and non-autonomous elliptic tops}
The six Painlev{\'e}  equations were discovered in the 1900-1910 period \cite{Pa,Fu,Ga} as the second order differential
equations that have only poles in the complex plane as movable singularities \cite{Ince,book,IN2,boalch1}. The Painlev{\'e}
equations
  have a lot of applications
in the contemporary mathematical and theoretical physics \cite{book1,Dubrovin1,Wo}. The most general equation --
 the  Painlev{\'e}  VI (PVI)  has the following
form:
 $$
\frac{d^2X}{dt^2}=\frac{1}{2}\left(\frac{1}{X}+\frac{1}{X-1}+
\frac{1}{X-t}\right)
\left(\frac{dX}{dt}\right)^2-
\left(\frac{1}{t}+\frac{1}{t-1}+\frac{1}{X-t}\right)\frac{dX}{dt}+
 $$
 \beq{p6}
+\frac{X(X-1)(X-t)}{t^2(t-1)^2}\left(\al+\be\frac{t}{X^2}+
\ga\frac{t-1}{(X-1)^2}
+\de\frac{t(t-1)}{(X-t)^2}\right)\,,
 \eq
where $(\al,\be,\ga,\de)$ are arbitrary complex constants.
%This equation is the top of the hierarchy of  Painlev{\'e}  V-I equations. They are the
%second order ordinary differential equations whose singularities have the
%Painlev{\'e}  property: the only movable singularities are poles.
%
%
%It is related to the group ${\rm SL}(2,\mC)$ (see Section 5).
%
%\mo{Zdes' nebolshaja vstavka.}
One of the main goals of this paper is to construct analog of this equation in dimension 1+1.
It takes the form (\ref{Naft}). We consider also some multi-component generalizations of (\ref{p6})
and construct their field theory analogs in higher dimensions. These generalizations share two
main properties of PVI: they are monodromy preserving equations for some linear problems and
they have Hamiltonian form.

Before going to the general case we analyze structures behind PVI.
 The PVI can be represented in the elliptic form \cite{Painleve1906,M}
 \beq{rr1}
 \frac{d^2u}{d\tau^2}  = \sum_{a=0}^{3}\nu_a^2 \wp ' (u
+\omega _a)
 \eq
via the change of variables $X =\frac{\wp (u )-e_1}{e_2 -e_1}$,
  $t=\frac{e_3 -e_1}{e_2 -e_1}$ with $e_k=\wp(\om_k)$
  %\mo{1. $T\to t$?}
  and identification of  constants
  $(\nu_0^2,\nu_1^2,\nu_2^2,\nu_3^2)=\frac{1}{(2\pi i)^2}(\al,-\be,\ga,\frac{1}{2}-\delta)$.
  %\mo{\bf 3}
  Here
  \beq{hpe}
  (\om_0,\om_1,\om_2,\om_3)=\left(0,\frac{1}{2},\frac{\tau+1}{2},\frac{\tau}{2} \right)
  \eq
  are the half-periods of the elliptic curve $\Si_\tau=\mC/(\mZ\!+\!\tau\mZ)$, $Im\,\tau>0$,
 $\wp(x)$ is the Weierstrass $\wp$-function (see
Appendix A).
The equation (\ref{rr1}) is described as the  Hamiltonian system with the Hamiltonian function
 \beq{rr3}
 H  =\frac{1}{2}p^2 -\sum_{a=0}^{3}\nu_a^2 \wp(u+\omega _a)
 \eq
and canonical Poisson bracket $\{p,u\}=1$. The system is {\em non-autonomous} since the potential explicitly depends on the
moduli $\tau$ (of $\Si_\tau$) which is the "time" variable.
%\mo{\bf 4}.
It is non-autonomous version of the Calogero-Inosemtsev system \cite{In1}.
In the case when $\nu_a=\frac{1}{2}\nu$ for all $a$ we come to the elliptic two-particle
non-autonomous Calogero-Moser model
 \beq{rr101}
  \frac{d^2u}{d\tau^2}  = \nu^2 \wp ' (2u)\,.
 \eq
In this paper we deal with another (also elliptic) form of the PVI. It is the non-autonomous version of the Zhukovsky-Volterra
gyrostat \cite{ZVG} (NAZVG). Namely, it was shown in \cite{LOZ1}  that (\ref{rr1}) can be written as dynamics of
three-dimensional complex-valued vector %\mo{\bf 5}
$\vec{S}=(S_1,S_2,S_3)$:
 \beq{eu}
\p_\tau\vec{S}=\vec{S}\times \vec{J}(\vec{S})+\vec{S}\times \vec{\nu'}\,,
 \eq
where $\vec{J}(\vec{S})=(J_1S_1\,,J_2S_2\,,J_3S_3)$, $J_k=\wp(\om_k)$ and $\vec{\nu'}=(\nu'_1,\nu'_2,\nu'_3)$ - vector of
linear combinations of constants $\nu_a$ from (\ref{rr1}) multiplied by some ratios of theta-constants (see \cite{LOZ1}).
%  \beq{rr4}
%\nu'_\al=\exp(2\pi i\om_\al\p_\tau\om_\al)\left(\frac{\vth'(0)}{\vth(\om_\al)}\right)^2\sum\limits_{c=0}^3 \exp(4\pi
%i(\om_\al\p_\tau\om_c-\om_c\p_\tau\om_\al))\nu_c\,.
% \eq
The fourth independent linear combination of the constants $\nu_0'=\frac{1}{2}\sum\limits_{c=0}^3\nu_c$ appears to be the
length of $\vec{S}$: $\nu_0'^2=\sum\limits_{\al=1}^3S_\al^2$.
Equation (\ref{eu}) can be rewritten in terms of ${\rm sl}(2,\mC)$-valued
 $\bfS=\sum\limits_{\ga=1}^3\si_\ga S_\ga$, where
$\si_\ga$ are the Pauli matrices, as
  \beq{eu11}
\p_\tau\bfS=[\bfS,\bfJ(\bfS)]+[\bfS,{\bf{\nu'}}]\,,
 \eq
 where ${\bf\nu'}=\sum\limits_{\ga=1}^3\si_\ga \nu'_\ga$.
It is generated by the quadratic Hamiltonian
  \beq{eu13}
H=\frac{1}{2}\tr\left(\bfS\,\bfJ(\bfS)\right)+\tr(\bfS\,{\bf \nu'})\,.
 \eq
and the linear  Poisson-Lie brackets on ${\rm sl}^*(2,\mC)$: $\{S_\al,S_\be\}=\varepsilon_{\al\be\ga}S_\ga$.\footnote{It is
also shown in \cite{LOZ1} that (\ref{eu11}) can be described by the quadratic Poisson brackets generalizing Sklyanin's ones
\cite{Skl0}.} Explicit change of variables $S_\al(v,u)$ can be found in \cite{LOZ1}.
%
%In all instances the derived equations take the form of
The  equation (\ref{eu11}) reduces to the non-autonomous elliptic ${\rm sl}_2$-top when
$\nu'_{1,2,3}=0$.

In general case we
deal with the non-autonomous Euler-Arnold tops on
 corresponding groups. They have the following description \cite{Ar}.
 Let $\gg$ be the corresponding Lie algebra,  $\gg^*$ its
 Lie coalgebra,
 and $\bfJ$ is a map $\gg^*\to\gg$.
 The conjugate operator
% \mo{I replaced $\bfJ^{-1}$ on $\bfJ^{*}$}
$\bfJ^{*}\,:\gg\to\gg^*$ is called \emph{the inertia tensor }
 of the top. The elements of $\bfS\in\gg^*$ are
 called the angular momenta, while the elements of $\bfF=\bfJ(\bfS)\in \gg$ are the angular velocities.
  The equations assume the form
  \beq{eat}
 \p_\tau\bfS={\rm ad}_{\bfJ(\bfS)}^*\bfS=[\bfS,\bfJ(\bfS)]\,.
  \eq
 The phase space of  the non-autonomous tops are the coadjoint orbits
  \beq{eat14}
\cO=\{\bfS\in\gg^*\,|\,\bfS=Ad^*_g(\cS_0)\,,~g\in G\,,~\cS_0~{\rm is \,a\,fixed\,element\,of}\,\gg^*\}\,.
  \eq
The form (\ref{eat}) is convenient for different generalizations in the sense that $\bfS$ can be generalized to some field
$\bfS(x)$ while $\bfJ$ becomes a differential (or pseudo-differential) operator in $x$. Let us recall that some important
examples of integrable equations in dimensions 1+1 and 2+1
  were interpreted as integrable tops on infinite-dimensional groups \cite{KLO,Se,Wa}.

  In some cases we need to rewrite (\ref{eat}) in terms of the angular velocities. Substituting in (\ref{eat})
$\bfS=\bfJ^{*}\bfF$ we obtain
\beq{3.1a}
\bfJ^{*}\p_\tau\bfF=({\rm ad}_{\bfF}^*\bfJ^{*}(\bfF))-(\p_\tau
\bfJ^{*})(\bfF)\,. \footnote{For a group $G=SDiff(M)$ of the volume preserving diffeomorphisms
 of a Riemann manifold $M$ the equation
takes the form of the Euler-Bernoulli equation (see \cite{Ar}). As it was mentioned above,
in this case $\bfJ^{*}$ is the Laplace-Beltrami operator and
the last term in (\ref{3.1a}) is absent.}
  \eq

The multi-component generalization of PVI (related to the group $\SLN$) is defined by the following form of the operator $J$
in the basis (\ref{3.10})
  \beq{it2}
 \bfS=\sum_{\al \in\mZ^{(2)}_N}S_\al T_\al\,,~~~~\bfJ\,:\,S_\al\to J_\al S_\al\,,~~~~
 J_\al=\wp_\f1{N}(\al):=\wp(\frac{\al_1+\al_2\tau}N|\tau)\,.
  \eq
In the case when the orbit $\cO$ (\ref{eat14}) has the minimal dimension $\dim(\cO)=2N-2$
 \beq{j0}
\cS_0=\nu\,\di(N-1,-1,\ldots,-1)
 \eq
  the corresponding non-autonomous top is equivalent to the non-autonomous version of the
elliptic $\SLN$ Calogero-Moser model \cite{LOZ1} (for ${\rm sl}_2$ case see (\ref{rr101}))
  \beq{it202}
 \frac{d^2u_i}{d\tau^2}=\nu^2\sum\limits_{k\neq i}\wp'(u_i-u_k)\,,\ \ \ i=1,...,N\,.
  \eq
 similarly to the equivalence between (\ref{rr1}) and (\ref{eu}).
Here $\nu$ is the single coupling constant. Its square is proportional to the value of the
quadratic Casimir function of $\bfS$. A
general orbit (\ref{eat14}) corresponds to the "spin" generalizations of Calogero model
\cite{GH,Woj}.%\mo{dobavil ssilki}

Another type of the "multi-component" generalization (which is better to refer as "multi-color") comes from the Schlesinger
systems \cite{Sch}. In this case the phase space consists of $n$ orbits (\ref{eat14}) $\bfS_a$, $a=1,...,n$. The equations of
motion can be written in the following general form:
 \beq{Naft0}
  %\frac{\p\,}{\p\tau}
  \p_\tau\bfS_a=[\bfS_a,\bfJ^{I}_a(\bfS_a)]+\sum\limits_{c\neq a}[\bfS_a,\bfJ^{II}_{ac}(\bfS_c)]\,.
   \eq
Here we have two types of the (inverse) inertia tensors $\bfJ^{I}$ and $\bfJ^{II}$. The model is described by quadratic
Hamiltonian
 \beq{Naft01}
  %\frac{\p\,}{\p\tau}
  H_\tau=
  \sum\limits_{a=1}^n\frac{1}{2}\,\tr(\bfS_a\,\bfJ^{I}_a(\bfS_a))+\sum\limits_{b\neq c}\tr(\bfS_b\,\bfJ^{II}_{bc}(\bfS_c))
   \eq
and direct sum of the linear Poisson-Lie brackets generated by the structure constants (\ref{ta}) for each
$\bfS_a$.\footnote{Similarly to the case of PVI in the form of NAVZG there exists the quadratic Poisson structure describing
the same equations \cite{FO,CLOZ2}.} Originally, the Schlesinger model is given by the Hamiltonians corresponding to $n$
marked points while $H_\tau$ (\ref{Naft01}) appears in the elliptic case and corresponds to the moduli $\tau$ of
$\Sigma_\tau$. In the special ${\rm sl}_2$-case when the marked points on $\Sigma_\tau$ are the half-periods
$\om_a$ (\ref{hpe}) there exists the (reflection) symmetry which
generates constraints
 \beq{rr6}
  \bfS_a^\be=\exp(4\pi i(\om_a\p_\tau\om_\be-\om_\be\p_\tau\om_a))\bfS_a^\be\,,\ \ a=0,..,3\,,\ \
  \bfS_a=\sum\limits_{\ga=1}^3\bfS_a^\ga\si_\ga\,.
   \eq
The constraints save $\bfS_0$ and reduce all other orbits to non-dynamical constant matrices
 \beq{rr61}
  \bfS_a^\be=\delta_{a\be}\ti\nu_\be\si_\be\,,\ \ a=1,2,3\,,
   \eq
where $\ti\nu_be$ differ from $\nu'_\be$ by the factors of theta-constant ratios \cite{LOZ1}. In this way we reproduce the
PVI in the NAVZG form (\ref{eu11}). Explicit equations for the elliptic Schlesinger system can be found in \cite{CLOZ}.

\subsection{Linear problems and monodromy preserving equations}

The described non-autonomous models share another common property. The non-autonomous elliptic equations described above can
be considered as a monodromy preserving conditions for some linear problems. This approach was suggested in
\cite{Ga,Garnier,Sch} and then developed in \cite{FN,JM} (see also \cite{IN2}). Different linear problems are known for
Painlev\'e equations. The scalar examples were found  in \cite{Fu,Garnier}. The matrix-valued linear problems were also
obtained in many different variations. See, for instance, \cite{JM} for ${\rm sl}_2$-valued examples and \cite{JKT07} for
${\rm sl}_3$-valued ones. Our approach is based on consideration of the monodromy preserving equations as non-autonomous
version \cite{LO} of the Hitchin systems \cite{Hi}.  From the  computational point of view it is based on Krichever's ansatz for
the Lax pairs of elliptic integrable systems \cite{K1} and the classical Painlev\'e-Calogero correspondence \cite{LO97} (see
also \cite{ZZ,Takasaki01}). The correspondence claims, in particular, that the Lax pair of the elliptic Calogero model also
satisfies the monodromy preserving condition.

The linear problem for the PVI equation in the form of NAZVG has the following description.
 Let $L=L(\bfS, {\bf\nu}, w,\tau)$ and $M=M(\bfS, {\bf\nu}, w,\tau)$ be $2\times 2$ matrices
 depending on the spectral parameter $w\in\mC$.
 The linear system assumes the form
 \beq{ls1a}
 \left\{
 \begin{array}{l}
  (\p_w+L)\Psi=0\,,\\
   \,\,\p_{\bar w}\Psi=0\,,\\
   (\p_\tau+M)\Psi=0\,,
 \end{array}
  \right.
 \eq
where $\Psi=\Psi(\bfS, {\bf\nu}, w,\tau)$ - $2\times 2$ matrix-valued function of solutions. The first and the second
equations in (\ref{ls1a}) mean that $\p_w+L$ is the component of the flat connection of the $\SL$-bundle
%\mo{delete holomorphic because $\ti w\neq\bar w$.}
over a
complex curve with local coordinates $(w,\bar w)$. The third equation means that the monodromies of $\Psi(w,\tau)$ are
independent of $\tau$.
 The consistency condition of the first and the last  equations %\mo{Not two last!}
  is the monodromy preserving equation (or zero-curvature equation
 \footnote{Sometimes this equation is also referred to as the Lax equation since it appears
 from (\ref{0.1}) by $L\rightarrow\p_w+L$.})
for NAZVG
 \beq{0.3}
\p_\tau L-\p_w M=[L,M]\,.
 \eq

In \cite{LOZ1} the explicit expressions of $L$ and $M$ were obtained:
 \beq{t29}
 \begin{array}{l}
 L^{NAZVG}=\sum\limits_{\al=1}^3 \left(
S^\al\vf_\al(w)+\tilde{\nu}_\al\vf_\al(w-\om_\al) \right)\sigma_\al= \sum\limits_{\al=1}^3\left(
S^\al\vf_\al(w)+{\nu'}_\al\frac{1}{\vf_\al(w)} \right)\sigma_\al\,,\\
 M^{NAZVG}=
 \sum\limits_\al -S^\al\frac{\varphi_1(w)\varphi_2(w)\varphi_3(w)}{\varphi_\al(w)}\sigma_\al +E_1(w)L^{NAZVG}(w)\,,
 \end{array}
  \eq
where  $\vf_\al(w)$ and $E_1(w)$ are defined in the Appendix A. In this paper we obtain
the field-theoretical generalizations of the
linear problems (\ref{ls1a}) with (\ref{t29}). Let us also mention that the
L-M pair (\ref{t29}) was obtained from the
another one describing PVI (\ref{rr1}) suggested in \cite{Zotov04} by a singular gauge transformation.
 The gauge transformation is the Hecke operator (or modification of bundle \cite{AL,LOZ} ).
 Another L-M pair can be derived from the rational Schlesinger system by the change
of variables in the linear problem \cite{Z1}.

 In the general case  $L$ and $M$ are elements of a Lie
algebra acting on a finite-dimensional module. Their matrix elements depend on the local coordinates
on the phase space ${\mathcal M}$
 \beq{rr8}
 {\bf\xi}=(\xi_1,\ldots,\xi_{\dim{\mathcal M}} )
 \eq
and the spectral parameter $w\in\Si$,
where $ \Si$ is $\mC P^1$, or an elliptic curve, or
more generally, any complex curve. Examples of systems of this type are
the Schlesinger system \cite{Sch}, the Painlev{\'e}
equations and their generalizations.

There exists a special limit of (\ref{0.3}) to the classical integrable systems.
 Naively, we replace the derivative $\p_w$
with $\ka\p_w$, where $\ka$ is a small parameter. In the limit $\ka\to 0$ we come from (\ref{0.3}) to the isospectral
problem:
 \beq{0.1}
  \p_tL=[L,M]\,.
  \eq
  It is the Lax equation of a finite-dimensional classical
 autonomous  completely integrable system.\footnote{
This approach  was pursued by Boutroux \cite{Bo} and Garnier \cite{Gar} in their studies of the Painlev\'{e} equations and
the Schlesinger system. Its modern version is called the Whitham quasiclassical limit \cite{Wh}
and developed in number of
works \cite{Du,FN,Kr1,Ta}. In connection with our approach it will be considered in Section \ref{whi}.}
In particular, from
PVI we come in this way to the BC$_1$ Calogero-Inozemtsev model \cite{In1}
and from the Schlesinger system to the Gaudin model \cite{Gar}.

In the autonomous case the wide class of  elliptic integrable systems
(Gaudin models) was obtained in \cite{RSTS,Nekr}. More
general models classified by characteristic classes were described in \cite{LOSZ1,LOSZ2} (see also \cite{LOSZ3}).

Different types of degenerations of the elliptic autonomous and non-autonomous
models to the trigonometric and rational case
including scaling limits of the Inozemtsev type \cite{Inoz0} can be found in
 \cite{ChZ,AA3,AA,Sm}. We consider particular
cases of these reductions in the field theories.

\subsection{1+1 field theories}\label{2ft}

Let us start with 1+1 generalizations of the autonomous mechanical
finite-dimensional integrable systems. The latter systems
are described by the Lax equations (\ref{0.1}). First of all, any 1+1 generalization
implies that the variables (\ref{rr8})
of the phase space $\mathcal M$ become fields on a real line or a circle ${\mathrm S}^1$:
 \beq{rr9}
 {\bf\xi}\ \rightarrow \ {\bf\xi}(x)=(\xi_1(x),\ldots,\xi_{\dim{\mathcal M}}(x))\,,\ \ x\in{\mathrm S}^1\,.
 \eq
 Instead of the Lax equation (\ref{0.1}) the  equations of motion in 1+1 case acquire the form of
the Zakharov-Shabat equation \cite{ZSh}
 \beq{0.2}
\p_tL-\p_x M=[L,M]\,.
 \eq
In general the matrices $L$ and $M$ in this equation can not be  obtained from the mechanical one by a
direct substitution (\ref{rr9}). However, in some cases this approach does work. For example, it happens for the
${\rm sl}_2$ elliptic top and its 1+1 field version - Landau-Lifshitz model \cite{BR,Sk}:
$L^{LL}(z)=\sum\limits_{\al=1}^3
S^\al(x)\vf_\al(z)\si_\al$. The equations of motion
 \beq{rr10}
\bfS_t=[\bfS\,,\bfJ(\bfS)]+[\bfS\,,\bfS_{xx}]
 \eq
 show that in this case the field generalization is achieved by replacing the
  conjugate inertia tensor $\bfJ$ in (\ref{eat})
 with the (local) differential operator
  \beq{rr11}
\bfJ\ \ \stackrel{(\ref{eat})\rightarrow (\ref{rr10})}{\longrightarrow}\ \p_x^2+\bfJ\,.
 \eq
Another two-dimensional version of the integrable Euler-Arnold tops \cite{Ar} are the $N$-wave equations \cite{ZM}.
 The integrable field versions of the interacting particles models
 are the Toda field theory
\cite{LS,Mi,MOP} and  the Calogero-Moser field theory \cite{Kr,LOZ}. It turns out that the two-particle case of the
Calogero-Moser field theory
  is equivalent to the Landau-Lifshitz equation similarly to the relation between PVI and NAVZG.
They are examples of the soliton equations which can be solved by the famous Inverse Scattering Method \cite{ZSh,FT2,DKN,MS}.

More general, the field-theoretical generalization of the Hitchin systems \cite{Hi}
was suggested in \cite{LOZ} (see also \cite{Z,Z11,SZ}). Namely, it was shown that the 1+1 version of integrable systems
appears by changing the structure group $G$ of the underlying Higgs bundle by the centrally extended loop group ${\hat L}(G)$.

To construct multi-dimensional version of the monodromy preserving equations we apply  proposed in
\cite{LO} approach to standard isomonodromy problems. It is similar to the Hitchin derivation
of classical integrable systems  \cite{Hi}, but the Higgs $G$-bundles over Riemann surfaces
are replaced by flat bundles.
It should be noted that
 the quasi-classical (Whitham) limit  leads to the Hitchin integrable systems related to the Higgs $G$-bundles.
To pass to field theory analogs we  replace a simple
finite-dimensional Lie groups $G$ by infinite groups, and in this
way obtain field-theoretical generalizations of the isomonodromy
problems and the Hitchin systems. The latter field theories are
different from described above.

In this paper we  consider two  types of infinite-dimensional groups.
One of them is related to the cocentrally extended loop group $\check L(G)$ of the centrally extended loop group $\hat L(G)$,
$\,G=\SLN$,
and described in detail in Section 3. Another field theories are related to the group of noncommutative torus (NCT) (Section 5)
and its dispersionless limit to the area preserving diffeomorphisms of the two-dimensional torus $SDiff(T^2)$.
The local integrable hierarchies related to this group were considered in \cite{HOT,KLO,TT1}. We come to this point below.

 Before going
to the isomonodromic field theory let us comment on the difference between the construction used in this paper (let us call
it the "$\bar k$-case") and the one described in \cite{LOZ} (the "$k$-case"). The latter leads to integrable hierarchies related to
the Inverse Scattering Problem and, in particular, to (\ref{rr10}), while the $\bar k$-case provides different type of
equations (see below). Comments on NCT-based generalization are given below for the isomonodromic version.

As it will be explained in Section 3, the field generalizations arising (via reduction) from the double extended (centrally and
cocentrally) loop groups in $\bar k$-case have the following property: the L-operator depends on two scalar parameters  $k$
(dual to the central extension) and $\bar k$:
 $$
L=L(k,\bar k)
 $$
In fact, the variable $k$  appears as the coefficient in front of $\p_x$ and exists in the described above
$k$-case as well. In the same time, the dependence on the another variable $\bar k$ is non-trivial ($\bar k$
is absent in $k$-case approach). Moreover, these variables satisfy equations of motion
 \beq{keq1}
\p_t\bar k=k\,,~~\p_t k=0\,.
 \eq
Therefore, the Lax operator depends non-trivially on linear function of time and the field theory is non-autonomous even
in the isospectral deformations. Summarizing, the field theories of $\bar k$-case are originally non-autonomous. As we
will see below, the isomonodromic version adds another type of the non-autonomous dependence. Similarly to mechanics the
dependence on $\tau$ appears from the corresponding definitions of the elliptic functions
as the module of the underlying elliptic curve (say, $\vth(u,\tau)$ for the theta-function), while the
dependence on $\bar k$ enters in another way -- as $\vth(u+\bar k,\tau)$.

There is one more special property of the $\bar k$ case. In contrast to the $k$-case ( LL-like soliton equations (\ref{rr10}))
the term $\p_x M$ does not provide any input into equations of motion from (\ref{0.2}). In this case
 \beq{rr12}
\p_t L=\left.\p_t L\right|_{\bar k={\hbox{\tiny{const}}}}+\p_{\bar k}L\,\p_t\bar k\,\stackrel{(\ref{keq1})}{=}\, \left.\p_t
L\right|_{\bar k={\hbox{\tiny{const}}}}+k\p_{\bar k}L\,.
 \eq
As we will see, the last term $k\p_{\bar k}L$ cancels exactly  $k\p_xM$-term. It happens because the Hamiltonian density can
be computed in this case likewise  in finite-dimensional mechanics:
 \beq{rr13}
H=\frac{1}{2}\oint\tr L^2\,.
 \eq
(see more precise expression (\ref{ham1})). This Hamiltonian is not conserved on its own dynamics due to explicit dependence
on the time-variable.
This situation is different from $k$-case, where there is a whole hierarchy of conservation laws
generated by monodromy operator P$\exp\oint L(x)$. The $\bar k$-case is more like the case of monodromy preserving equations
(\ref{0.3}), where $\p_w M$ is canceled by $\left.\p_t L\right|_{t=\tau}$.
The reason of the difference between the $k$-case and the $\bar k$ case can be explained in the following way.
In the derivation of the both types of systems we use the symplectic reduction procedure. While in
derivation of the $k$-systems we demand that symplectic structure $\om$ and Hamiltonians $H_s$ are invariant,
in the $\bar k$ case  we only assume the invariance of their special combination - the Poincar\'{e}-Cartan form \cite{Ar}
$$
\om^{PC}=\om-\sum\de H_s\wedge\de t_s\,,
$$
where $t_s$ are the times related to the Hamiltonians. It turns out that
derived in this way $\bar k$ systems are nonlocal. The obtained field
theories can be considered as ${\rm gl}_\infty$ Gaudin systems. As it is known from \cite{MMRZZ} the ${\rm gl}_N$ Gaudin
models are spectrally dual to magnetics of Landau-Lifshitz type on $N$ cites in finite-dimensional case. Therefore, we can
await that the spectral duality may be hold on the level of ${\rm gl}_\infty$.

%The field version in $\bar k$-case corresponding
%to the Zakharov-Shabat equation is briefly considered in Section 3.7.

%%\mo{Eto ne sovsem verno (sm. str. 25).}
%From the very
%beginning (in Section 3) we study its isomonodromic version since it is more general.

\subsection{Painlev\'e-Schlesinger field-theoretical generalizations}

%  \beq{keq2}
%\ka\p_\tau\bar k=k\,,~~\ka\p_\tau k=0\,.
% \eq

Our goal is to find a field-theoretical  generalization of the monodromy preserving equations. It means that the mixture of
(\ref{0.2}) and (\ref{0.3}) should appear as the zero-curvature equation\footnote{There is also a different problem, where
the both equations (\ref{0.2}) and (\ref{0.3}) are used \cite{IN}.}, i. e.
 \beq{rr20}
    \begin{array}{|c|}
  \hline\\
\p_tL-\p_xM-\p_wM=[L,M]
\\ \ \\
  \hline
  \end{array}
 \eq
In other words, we replace  $L\to\p_x+L$, where $L$ and $M$ are still finite-dimensional matrices.
 Due to the term $\p_w M$ in (\ref{0.3}) this generalization  is highly non-trivial.
% Here we give only one example of this construction. It is  the field-theoretical
%analog of the rational Schlesinger system (Section \ref{ss}). This system is the isomonodromic
%modification of 1+1 Gaudin Model obtained recently  \cite{Z}
%\mo{AZ must refer the paper of Novokshenov and Its}.
%\footnote{Some modifications of this construction was considered in \cite{IN}.
%}

More generally, we allow $L$ in (\ref{0.3}) to be an integro-differential operator acting in some functional space. In this
way the
 field-theoretical generalizations of the monodromy preserving equations  under consideration
 are still the monodromy preserving conditions for the connection $\p_w+ L$  acting on sections
 of a bundle of infinite rank.
 To write down the systems we assume that:
 \begin{itemize}
 \item some reductions of (\ref{0.3}), (for example, taking  $x$-independent solutions)
 lead
 to PVI (\ref{p6}) or to its finite-dimensional generalizations;
 \item (\ref{rr20}) are equations of motion for some non-autonomous Hamiltonian system.
 \end{itemize}
Let us start from the concrete example.
%
%
%
%We deal with three types.........
%
 Consider the space of twelve fields parameterized by the three-vectors
  $$
 \bfS_b^\al(x)\,,\ \al\in (\mZ_2\times\mZ_2)\setminus(0,0)\,,\ \ b\in\mZ_2\times\mZ_2\,.
 %~\Bigl(\al=((1,0)\,,\,(1,1)\,,\,(0,1))\,,~b=((0,0),\al)\Bigr)\,,
  $$
 and let $\bfJ^I(\be,\p_x,\tau)\,,$ $\bfJ^{II}(\be,b,c,\p_x,\tau)$
 be the pseudo-differential operators defined in (\ref{j1}). Then PVI$^{FT}$ has the form of four
  interacting non-autonomous Euler-Arnold tops \cite{Ar,Ar1}, related to the loop group $L(\SL)$
   \beq{Naft}
   \begin{array}{|c|}
  \hline\\
  \frac{\p\,}{\p\tau}\bfS_b^\al(x)=
  \sum\limits_{\be\neq\al}\Bigl(\bfS_b^{\al-\be}(x)\bfJ^I(\al,\be,\p_x,\tau)
  \bfS_b^\be(x)
  +\sum\limits_{c\neq b}\bfS_b^{\al-\be}(x)\bfJ^{II}(\al,\be,b,c,\p_x,\tau)
  \bfS_c^\be(x)\Bigr)\\ \ \\
  \hline
  \end{array}
   \eq
and subjected  to the constraints
\beq{CZV}
\bfS^\al_b(x)=(-1)^{b\times\al}\bfS^\al_b(-x)\,.
\eq
  The equation (\ref{Naft}) itself can be considered as a field
 generalization of the elliptic ${\rm sl}_2$ Schlesinger system \cite{CLOZ} with four marked points.
% while in particular case of ${\rm sl}_2$-connection and
 There are analogs of the constraints (\ref{CZV}) in the finite-dimensional case that lead to the
 vanishing of Hamiltonians related
 to the marked points located at the half-periods of  elliptic curves. The remaining nontrivial
 Hamiltonian defines an evolution with respect to the modular parameter $\tau$ of the elliptic curves.
  This evolution is  the equation for the non-autonomous Zhukovsky-Volterra gyrostat (NAZVG) (\ref{eu}).
 We show that that the zero modes of the field model (\ref{Naft}) along with the constraints
 (\ref{CZV}) satisfy the equation for the non-autonomous Zhukovsky-Volterra gyrostat (NAZVG) (\ref{eu})
 for three variables $S^\al$ -- zero modes of $S^\al_0$.

 %In the finite-dimensional case assume that
% if the marked points are sitting there are constraints
% on the spin variables that provide vanishing of Hamiltonians corresponding
% For the  marked points
% at half-periods of the elliptic curve and on the constraints it generalizes PVI.
% Namely, it can be shown
 %It was proved in \cite{LOZ1} that NAZVG
 %is equivalent to the PVI (\ref{p6}) (see Section \ref{sv}).
%
%  It is not a sole reason
% to consider (\ref{Naft})
% as a natural field theory analog of the PVI.

%\mo{Zdes' dolzhen bit' +}

% We do not need here the explicit form of $L$ and $M$.
%
In terms of the Lax operators corresponding to (\ref{Naft}) we replace $L$ and $M$ from (\ref{ls1a}) with
 operators, which act on
the space of functions on additional variable $x\in{\rm S}^1$. In this way we come from the ODE PVI
 to pseudo-differential equations for functions on the spaces of dimension 1+1.
 Namely, we define two operators $L(\bfS_b^\al(x),x,w,\tau)$ and $M(\bfS_b^\al(x),x,w,\tau)$
to be two by two matrices whose matrix elements are pseudo-differential operators
in $x$, such that the linear system
 \beq{ls1}
 \left\{\begin{array}{l}
  (\p_w+\p_x +L( \bfS_b^\al(x),x,w,\tau))\Psi(x,w,\tau)=0\,,\\
   \,\,\p_{\bar w}\,\Psi(x,w,\tau)=0\,,\\
    (\p_\tau+M(\bfS_b^\al(x),x,w,\tau))\Psi(x,w,\tau)=0
 \end{array}\right.
 \eq
is consistent. Then (\ref{ls1}) is equivalent to (\ref{Naft}). The general isomonodromy problems
we are going to consider
have the form (\ref{ls1}), where $L,M$ are operators related to infinite-dimensional Lie algebras $L(\SLN)$.
Equivalently, we can look on the system (\ref{ls1}) as the $\SLN$ isomonodromy problem over a two-dimensional surface 
$\ti\Si\subset(\Si_\tau\times\mC^*)$ described by the local coordinates $(w+x,\bar w)$.

%The main difference between (\ref{ls1}) and the standard isomonodromy problem (\ref{ls1a}), (\ref{mop}) is
%that the $L,M$ operators in the latter case are elements of infinite-dimensional algebras.
%As we told NAZVG corresponds to  $\hat L(\SL)$.
%
We will exploit also  another important property of PVI. It is a non-autonomous Hamiltonian system,
where the Hamiltonian is defined by $L$.
In this way all equations we consider here are flows of infinite-dimensional Hamiltonian systems.
%
%
%
%
%The monodromy preserving equations based on the NCT group are equations in dimension 2+1.
% Their derivation is also based  on the Lax equation (\ref{0.3}), where $L$ and $M$ are operators
% that act in
%the space of functions on the torus $S^1\times S^1$.
%
%As by product we derived simultaneously some classes of integrable models in dimension 1+1 and 2+1. As their
%isomonodromic counterparts they are generally nonlocal. Consider first the one-dimensional case.
%

In general case the monodromy preserving equations we derived here are nonlocal and described
by the pseudodifferential operators. But some of their degenerations lead to PDE.
For example, for the  Lie algebra of NCT the dispersionless  limit of the
nonlocal equations leads to the equation in the 2+1 space $(\tau,x_1,x_2)$, depending on two parameters $\ep_1$, $\ep_2$
  \beq{oe}
 \begin{array}{|c|}
  \hline\\
\p_\tau\bp^2_Z\bfF(x,\tau)-\{\bp^2_Z\bfF(x,\tau),\bfF(x,\tau)\}+\ep_2\p_{x_2}\bp_Z\bfF(x,\tau)=0\,,
\\ \ \\
\hline
  \end{array}
 \eq
where
 $$
~\bp_Z=\f1{2\pi\imath}(\ep_1\p_{x_1}+ \ep_2\tau\p_{x_2})\,,~~\{f,g\}= \p_{x_1}f\p_{x_2}g-\p_{x_2}f\p_{x_1}g\,.
 $$

We use the approach which was earlier applied  for the derivation of the classical integrable systems (\ref{0.1}). It was based
on the symplectic reduction of the Higgs bundles over $\Si$ \cite{Hi}.
 To come to (\ref{0.3}) one should replace the Higgs bundles with the space of smooth
connections   \cite{LO}. In this way starting from
     $\SLN$-bundles over elliptic curves we derived earlier  the Painlev{\'e}  VI equation, its
   multi-component generalizations and the elliptic Schlesinger system \cite{CLOZ,LO,LOZ1}.
%The  modular parameter of the
%elliptic curves plays the role of the independent variable .

 The field theory (\ref{0.2}) can be derived similarly to  (\ref{0.1}),
 but in this case  the Higgs bundle has an infinite rank,
 and the group of automorphisms of the bundle is an infinite-dimensional group.
%Implicitly,  this approach was used in \cite{Se,Wa} to derive the KdV equation.
%
% is the centrally extended  group  $\hat{L}(G)$ \cite{LOZ1}.
 Following this idea we construct monodromy preserving equations in higher dimension
 using the symplectic reduction procedure applied to the space of smooth connections
 on the infinite rank bundles.

 In this paper we use a few types of infinite-dimensional groups. First, we consider the bundles
 with  the  loop group  $L(\SLN)$ as  the group of their automorphisms.
 In this case we come to 1+1 theory. In particular, the equations (\ref{Naft})
 are related to the SL$(2,\mC)$ theory. In general case they have the form of
 the so-called non-autonomous elliptic hydrodynamics \cite{KLO}. To derive the trigonometric
 and the rational systems we should extend the loop-group $L(\SLN)$ to the two-loop and three-loop groups.
 In this way we come to models in two-dimensional and three-dimensional spaces.

 For the next type of bundles the group of automorphisms is a specially defined group GL$(\infty)$,
 or more exactly, the group $SIN_\te$  of the noncommutative torus (NCT) ${\cal T}_\theta$,
 where $\te$ is the parameter of non-commutativity.
 In this case the field theories are defined on the
 two-dimensional noncommutative space ${\cal T}_\theta$.
 The  group  $SIN_\te$ for $\te=\oh$ is isomorphic to the two-loop group $LL(\GLT)$ while
 $\GLT$ is a subgroup of $LL(\GLT)$.
 In this sense
 we obtain the field-theoretical generalization of the particular case of PVI (\ref{p6}).
 We consider also
the  classical limit ($\te\to 0)$ of $SIN_\te$ to the group of the volume preserving diffeomorphisms $SDiff(T^2)$ of the
two-dimensional torus $T^2$ (see Appendix B).
 The equation (\ref{oe}) is one of the equations related to this group.

%As we have mentioned, we come in a general case to integro-differential equations.
We also consider  the Whitham quasiclassical limit which reduces  the monodromy preserving equations to integrable systems.
In this way we come to integrable nonlocal field theories. Notice that this construction leads to a new class of integrable
field models. For example, starting with  $\SL$ Euler-Arnold top we come to the nonlocal equation (\ref{nlll}), while the old
construction leads to the Landau-Lifshitz equation \cite{LOZ}. On the other hand, many interesting integrable equations,
related to noncommutative space were analyzed (see, for example \cite{DM,L,LMPPT,Ha}). It would be also interesting to
compare results with the classification suggested in \cite{Fer}. In particular, for the case of NCT the integrable Toda field
theories
 \cite{HOT,SV} and the so-called elliptic hydrodynamic \cite{KLO,Ol} were obtained.

As it was already mentioned, generally we get some nonlinear integro-differential equations. In some approximation they can be
described as a tower of differential equations. For  the loop group there exists a parameter $k$. It is the cocentral charge
corresponding to the term $k\p_x$. Taking perturbations of the equations for small $k$  we come to a tower of equations
related to degrees $k^j$. For $j=0$ the equation is just the original one-dimensional system parameterized by the space
coordinate $x$. On the level $j>0$ the equations are linear differential equations of order $j$ with respect to $\bfS_j$,
where $\bfS=\sum_{j\geq0}\bfS_jk^j$. Its coefficients have quadratic dependence on $\bfS_{j-1},\ldots,\bfS_0$.

 In all instances the derived equations take the form of non-autonomous Euler-Arnold tops on
 corresponding groups.
In a general case the operator $\bfJ$ is a pseudodifferential operator and in this way we come to nonlocal equations. There
are two ways to come to local equations. Generally, we consider degenerations of the systems related to degenerations of the
underlying elliptic curves. Namely, we consider trigonometric, rational and scaling limits. Some of these equations can be
considered as the field-theoretical versions of the PI-PV equations. Notice that the trigonometric degenerations of the
Euler-Arnold tops are rather subtle. We use here the approach from \cite{Sm} that allows us to come to non-trivial equations.
Another way, that works in some cases, is to rewrite (\ref{eat}) in terms of the angular velocities $\bfF$ (see (\ref{3.1a})).
Equations  (\ref{oe}) are of this type.

%\bigskip

The paper is organized as follows. %In Section 2 we present PVI in the form NAZVG following \cite{LOZ}. It will be used as a
%The starting point to go to PVI field theory is NAZVG.
%
In Section 2 we propose the general construction of the monodromy preserving equations from the flat bundles over Riemann
surfaces. The case of elliptic curves is considered in details.
In Section 3 this construction is generalized to the
infinite-dimensional case by consideration of the affine Lie
algebras and infinite rank bundles.
In Section 4 we define the bundles over curves with marked points
with  loop groups as the groups of their automorphisms. Using these
bundles we derive related monodromy preserving equations, their
degenerations and corresponding integrable field theories.
Section 5 is devoted to derivations of the trigonometric and rational equations based on the two and three-loop algebras.
In Section 6 we derive and analyze  field theories related to the noncommutative torus. We consider also
 their classical limits wherein the equations are simplified.

%\paragraph{List of abbreviation:}

\vspace{0.3cm}
%\bigskip
{\small {\bf Acknowledgments.}\\
The work was supported by RFBR grants 12-01-00482 and 12-02-00594,
by the project FASI RF 14.740.11.0347 and by grants for young
scientists (G.A., S.A. and A.Z.) RFBR 12-01-33071 mol$\_$a$\_$ved,
RFBR ”my first grant” 12-01-31385 and by the President fund
MK-1646.2011.1. The work of A.L. was partially supported by AG
Laboratory GU-HSE, RF government grant, ag. 11 11.G34.31.0023. The
work of A.Z. was supported, in part, by the Ministry of Education
and Science of Russian Federation under the contract 8528. The work
of G. Aminov, S. Arthamonov and A. Zotov was also partially
supported by the D. Zimin’s fund ”Dynasty”.}

%%%%%%%%%%%%%%%%%%%%%%%%%%%%%%%%%%%%%%%%%%%%%%%%%%%%
%%%%%%%%%%%%%%%%%%%%%%%%%%%%%%%%%%%%%%%%%%%%%%%%%%%%

\section{Monodromy Preserving Equations as Non-autonomous Version of Hitchin Systems}
\setcounter{equation}{0}

Here we describe a general approach to the monodromy preserving equations, based on the symplectic reduction of the space of
smooth connections on a bundle of any rank (finite or infinite) over a Riemann surface with marked points. The connection
takes values in a complex Lie algebra $\gg$
 (finite-dimensional or infinite-dimensional). The only restriction is that $\gg$ has an invariant
 bilinear form that allows us to identify $\gg$ and its dual. After the symplectic reduction procedure
 one comes to  the monodromy preserving equations. The finite-dimensional algebras lead to
    ODE. This construction was proposed in \cite{LO,LOZ}. It is easily
 generalized to the infinite-dimensional algebras, such as affine algebras and algebras of
 noncommutative torus. Our main goal is to consider equations related to these two types of algebras
 in detail.

%%%%%%%%%%%%%%%%%%%%%%%%%%%%%%%%%%%%%%%%%%%%%%%%%%%%%%%%%
\subsection{Flat bundles}

Let $\Si_g$ be a smooth oriented compact surface of genus $g$. For  a complex Lie group $G$ consider a principal  $G$-bundle
$\cP$ and the associated bundle $E_G=\cP\times_GV$, where $V$ is a representation space of $G$. Let $FConn_{\Si_g,G}$ be the
space of flat connections on $E$
 \beq{fla}
FConn_{\Si_g,G}=\{d+\cA\,|\,d\cA+\oh\cA\wedge\cA=0\}\,.
 \eq
The group of  automorphisms  $\cG=Aut\,(FConn_{\Si_g,G})$ (the gauge group) is the group of smooth maps
$\cG=Map_{C^\infty}\,:\,\Si_g\to G$ acting as $\cA\to f^{-1}df+f^{-1}\cA f$. The moduli space of the flat connections
$FBun_{\Si_g,G}$ is the quotient $FConn_{\Si_g,G}/\cG$.
This space can be also described as a result of the symplectic reduction of the space of all smooth connections
$Conn_{(\Si_g,G)}$ equipped with the symplectic form
 \beq{1}
\om=\oh\int_{\Si_g}(\de\cA\wedge\de\cA)
 \eq
invariant under the action of the gauge group. Here the brackets mean the Killing form in the Lie algebra $\gg=Lie(G)$. The
moduli space of flat bundles is the  coset space
 \beq{dmf}
 FBun_{\Si_g,G}:=FConn_{\Si_g,G}/\cG=Conn_{\Si_g,G}//\cG\,,
  \eq
where the double slash is the symplectic quotient of $\cG$.
In this way $Conn_{(\Si_g,G)}$ will be considered as an unrestricted phase space equipped with the symplectic form (\ref{1}).
The flatness condition (\ref{fla}) plays the role of the moment constraint generating the  automorphisms  $\cG$.

%%%%%%%%%%%%%%%%%%%%%%%%%%%%%%%%%%%%%%%%%%%%%%%%%%%%%%%%%%

\subsection{Deformation of complex structures }

Introduce a complex structure on $\Si_g$. The choice of the complex structure defines
the polarization of $Conn(\Si_g,G)$. Then the connection is decomposed in $(1,0)$ and $(0,1)$
parts $\cA=(A,\bA')$ \footnote{The notation $\bA'$ will be justified
below.}. We can write $(1,0)$ and $(0,1)$ components of the connection in local coordinates  $(z,\bz)$
 $$
(\ka\p+A)\otimes dz\,,~~(\bp+\bA')\otimes d\bz\,,~~(\p=\p_z\,,~\bp=\p_{\bz})\,.
\footnote{We introduce here a new parameter
$\ka$ ($\ka$-connection ) in order to pass later from the monodromy
preserving equations to integrable systems in the limit
$\ka\to 0$.}
 $$
The holomorphic sections of bundles are those which are annihilated by $\bp+\bA'$.
The moduli space of holomorphic bundles is
the quotient
 \beq{bun}
Bun(\Si_g,G)=\{\bp+\bA'\}/\cG\,.
 \eq
In terms of $(A,\bA')$ (\ref{1}) assumes the form
 \beq{1a}
\om=\int_{\Si_g}(\de A\wedge\de\bA')\,.
 \eq
Let $E_G$ be a holomorphic bundle corresponding to a point $b\in Bun(\Si_g,G)$ and
\beq{flb}
 Flat_b=\{\p+A\,|\,F(A,\bA')=\p\bA'-\bp A+[A,\bA']=0\}
\eq
is the space of flat holomorphic connections on $E_G$.
There is a map of the moduli space $FBun_{\Si_g,G}$ (\ref{dmf}) to $Bun(\Si_g,G)$ (\ref{bun}).
The fiber of this projection at the point $b$ is isomorphic to $Flat_b$. These fibers are
Lagrangian with respect to $\om$ (\ref{1a}).

The tangent space to $Bun(\Si_g,G)$ at the point $b$ is isomorphic to the first cohomology group\\
$H^1(\Si_g,End E_G)$. The dual vector space is $H^0(\Si_g,End E_G\otimes\Om^1(\Si_g))$. The space
$Flat_b$ (\ref{flb}) is isomorphic to the principal homogeneous space
over $H^0(\Si_g,End E_G\otimes\Om^1(\Si_g))$.

The vector fields generated by $Lie(\cG)=\{\ep\}$  act on $Conn_{(\Si_g,G)}$. They have
 the form $\de_\ep A=\p\ep+[A,\ep]$,
$\de_\ep\bA'=\bp\ep+[\bA',\ep]$, where $\de_\ep=di_\ep+i_\ep d$ is the Lie derivative.
 The corresponding moment map\\
  $F\,:\,Conn(\Si_g,G)\to Lie^*(\cG)$
 is defined as $i_\ep\om=\int_{\Si_g}(\ep,\de F)$.
The Hamiltonian of the gauge transformations $\de_\ep$ assumes the form
 \beq{1b}
h^{gauge}=\int_{\Si_g}(\ep, F)\,,~~F=\bp A-\p\bA'+[\bA',A]\,.
 \eq

Consider small deformations of the complex structure on $\Si_g$. The complex structure on $\Si_{g}$ is defined
 by the $\bp$ operator. Take the change of variables
 $$
w=z-\ep(z,\bz)\,,~~\bar  w=\bz-\overline{\ep(z,\bz)}\,,
 $$
where $\ep(z,\bz)$ is small. Up to a common multiplier the partial derivatives assume the form
 $$
\left\{
 \begin{array}{l}
\p_w=\p_z+\bar{\mu}\p_{\bz}\,,\\
\p_{\bw}=\p_{\bz}+\mu\p_z\,,
 \end{array}
\right.
 $$
where
 \beq{db}
\mu=\frac{\bp\ep}{1-\p\ep}\sim\bp\ep
 \eq
 is the Beltrami differential $\mu\in\Om^{(-1,1)}(\Si_{g})$.
We pass from $(w,\bw)$ to the chiral coordinates $(w,\ti{w}=\bz)$ because
 the  dependence on $\bar\mu$ is  nonessential in our construction
 \beq{nv}
\left\{
 \begin{array}{l}
\p_w=\p_z\,,\\
\p_{\bw}=\p_{\bz}+\mu\p_z\,.
 \end{array}
\right.
 \eq
 Note that $ \bar{w}$ does not mean the complex conjugation. The pair $(w, \bar{w})$ is just
  a pair of local coordinates on $\Si_g$. What is important is that $\p_{\bw}$ annihilates
  holomorphic functions  $\p_{\bw}f(w)=0$.
We assume that $\mu(w,\bar{w})$ is equivalent to
$\mu(z,\bz)$, if $w(z,\bz)$ is a global diffeomorphism. The classes of equivalence relations in
 $\Om^{(-1,1)}(\Si_{g})$ under the action
 of $Diff_{C^\infty}(\Si_g)$, are the
moduli space $\gM(\Si_{g})$ of complex structures on $\Si_{g}$. The tangent space to the
moduli space is the Teichm\"{u}ller space ${\cal T}_{g}\sim
H^1(\Si_{g},\G)$, where $\G\in T\Si_g$.
From the Riemann-Roch theorem one has
 \beq{2.6}
\dim(\gM(\Si_{g}))=3(g-1)\,.
 \eq
Let $(\mu_1^0,\ldots,\mu_l^0)$ be a basis in the vector space
$H^1(\Si_{g},\G)$. Then
  \beq{3}
 \mu=\sum_{l=1}^{3g-3}\tau_l\mu^0_l\,,
  \eq
  where the local coordinates $\tau_l$ will play the role of times in the isomonodromic
  deformation problem.

The monodromy preserving equations can be derived from the classical integrable systems
(the Hitchin systems) by the so-called Whitham quantization. Here we will consider the inverse procedure - the quasiclassical limit of the
monodromy preserving equations to  integrable systems. The introduced above
constant $\ka$ plays the role of the Planck constant.
Deform $(0,1)$ component of connection $\bA'$ as
  \beq{nba}
 \bA'=\bA-\f1{\ka}\mu A\,,~~ \bp+\mu\p+\bA=\p_{\bar w}+\bA\,.
  \eq
  In other words, $\bA$ is the $(0,1)$ component of the connection in the deformed
 structure. The form $\om$ (\ref{1a}) becomes
 $$
 \om=\int_{\Si_g}(\de A\wedge \de\bA)-\f1{\ka}\int_{\Si_g}(A,\de A)\de\mu\,.
 $$
The connections $(A,\bA)$ play the role of the canonical coordinates in $Conn_{(\Si_g,G)}$.
Taking into account (\ref{3}) we rewrite $ \om$  as the  differential of the
Poincar\'e-Cartan one-form $\vartheta$ $(\om=\de\vartheta)$ \cite{Ar}
  \beq{2}
 \om=\om_0-\f1{\ka}\sum_{l=1}^{3g-3}\de H_l\de \tau_l\,,~~H_l=\oh\int_{\Si_g}(A,A)\mu^0_l\,,~~
 \om_0=\int_{\Si_g}(\de A\wedge \de\bA)\,.
  \eq
We will discuss this form below.
 The  Poincar\'e-Cartan form gives rise to the action functional
  $$
 S=\sum_{l=1}^{3g-3}\int_0^\infty\Bigl(
 \int_{\Si_g}(A,\p_l\bA)-\f1{2\ka}(A,A)\mu^0_l\Bigr)d\tau_l\,,~~
 (\p_l=\p_{\tau_l})\,.
  $$
 The equations of motion following from the action (or from the Hamiltonians) are
  \beq{em}
 \p_l\bA=\f1{\ka}A\mu^0_l\,,~~ \p_lA=0\,.
 \eq
These equations are the compatibility conditions for the following  linear system
 \beq{2.15}
\left\{
 \begin{array}{ll}
  1. & (\ka\p+A)\psi=0\,, \\
  2. & (\bp+(\sum_{l=1}^{3g-3}\tau_l\mu^0_l)\p+\bA)\psi=0\,, \\
  3. & \ka\p_l\psi=0\,,
 \end{array}
\right.
 \eq
where $\psi\in \Om^{(0)}(\Si_{g},{\rm Aut}\, E_G)$. The equations of motion (\ref{em}) for $A$ and $\bA$ are the consistency
conditions of (1.$\&$3.) and (2.$\&$3.) in (\ref{2.15}). The monodromy of $\psi$ is  the transformation
 $$
\psi\rar\psi {\cal Y}, ~~{\cal Y}\in {\rm Rep}(\pi_1(\Si_{g})\to  G).
 $$
The equation (3.\,\ref{2.15}) means that the monodromy is independent on the times. The consistency
 condition of 1. and 2. is the flatness constraint (\ref{fla}).
 $$
\p_{\bar w} A-\ka\p\bA+[\bA,A]=0\,.
 $$
The linear equations (\ref{em}) defined on  $Conn(\Si_g,G)$ become nontrivial on $FBun_{\Si_g,G}$. Before consideration of
(\ref{em}) we extend the phase space by including the quasi-parabolic structures at the marked points on the surfaces
\cite{Si}.

%%%%%%%%%%%%%%%%%%%%%%%%%%%%%%%%%%%%%%%%%%%%%%%%%%%%%%%%%%%

\subsection{Quasi-parabolic structures and deformation of elliptic curves}

\subsubsection*{Gauge groups and flag varieties}

Assume that the algebra $\gg$ has the decomposition into subalgebras
 $$
\gg=\gn^-\oplus\gh\oplus\gn^+\,,~~[\gh,\gn^\pm]=\gn^\pm\,.
 $$
For complex simple $\gg$ it is the decomposition into the positive and negative nilpotent subalgebras
and the Cartan subalgebra. The subalgebra $\gh\oplus\gn^+=\gb$ is the Borel subalgebra.
Let $B$ be the corresponding Borel subgroup.
The quotient $G/B$ is the variety of the  $Fl(G)$.
%For a general $G$ we define the "Borel" subgroup ($\gb=Lie(B)$) and
%$Fl=G/B$ is the variety of the "$G$-flag".

Denote by $\Si_{g,n}$ the surface $\Si_g$ with $n$ marked points
 $\vec x=(x_1,\ldots,x_n)$.
 We fix  $G$-flags at these points and assume that the gauge group preserves $Fl_a$.
 It means that $\cG$ is reduced to the Borel subgroup $B\subset G$ at the marked points.
 In other words, for $f\in\cG$ and $t_a=z-x_a$, $f(t_a,\bar t_a)|_{t_a=0}\in B$. We denote
 this group $\cG_B$.  The quotient $\{\bp+\bA\}/\cG_B$ is the moduli space $Bun(\Si_{g,n},G)$
  (see (\ref{bun})) of the holomorphic bundles with \emph{the quasi-parabolic structures }at the marked points.

 Consider the coadjoint $G$-orbits in the Lie coalgebra $\gg^*$  located at the marked points
  \beq{co1}
 \cO_a=\{S_a=Ad_gS^0_a\,,~g\in G\,,~S^0_a\in\gg^*\}\,, ~~(a=1,\ldots,n)\,.
  \eq
 $ \cO_a$ is  the affine space over the cotangent bundle of the flag variety  $T^*Fl_a$.
 It is a symplectic variety  with the Kirillov-Kostant symplectic forms
  $\om^{KK}=\de\lan S^0,\de gg^{-1}\ran=\lan S, g^{-1}\de g\wedge  g^{-1}\de g\ran$.
  The space of connections
   $Conn(\Si_{g,n},G)$  is equipped with the symplectic form
  \beq{2.2}
 \om+\sum_{a=1}^n\om^{KK}_a\,,~~\om^{KK}_a=(S_ag^{-1}\de g\wedge g^{-1}\de g)\,,
  \eq
 where $\om$ is (\ref{1}). We assume now that
the flat connections have logarithmic singularities at the
 marked points with the residues taking values in $\cO_a$ (\ref{co1}). In this way the flatness
 condition (\ref{fla}) is replaced with
  \beq{fla1}
FConn_{\Si_{g,n},G}=\{d+\cA\,|\,d\cA+\oh\cA\wedge\cA=\sum_{a=1}^nS_a\de(x_a)\}\,.
 \eq
The gauge group $\cG_B$ is the group of symplectic automorphisms of the symplectic space
   $Conn(\Si_{g,n},G)$ and (\ref{fla1}) is the moment constraint condition generated by this action.
   The moduli space of flat connections $FBun_{\Si_g,G}$ is the result of symplectic reduction
    $$
   FBun_{\Si_g,G}=FConn_{\Si_{g,n},G}/\cG_B=Conn(\Si_{g,n},G)//\cG_B\,.
    $$

%%%%%%%%%%%%%%%%%%%%%%%%%%%%%%%%%%%%%%%%%%%%%%%%%%
\subsubsection*{Moving points}

 Consider the moduli space $\gM(\Si_{g,n})$ of complex structures of $\Si_{g,n}$. This space is
 foliated over the moduli space of complex structures of compact curves $\gM(\Si_{g})$ with fibers
 $\cU\subset\mC^n$ corresponding to the moving  marked points.
 The moduli space $\gM(\Si_{g,n})$ is the classes of equivalence relation in
 the space of differentials $\Om^{(-1,1)}(\Si_{g,n})$
 under the action of the group of diffeomorphisms   $Diff_{C^\infty}(\Si_{g,n})$
  vanishing at the marked points.

 Consider local coordinates in a fiber. The basis is defined as follows. Let $(z,\bz)$ be the local coordinates in
a neighborhood of the marked point $x^0_a$ and ${\cal U}_a$ is a neighborhood of $x^0_a$ such that $x^0_b\not\in{\cal U}_a$
if $x^0_b\neq x^0_a$. Define the $C^\infty$ function
 \beq{chi}
\chi(t_a,\bar t_a))=\left\{
 \begin{array}{cl}
1,& z\in {\cal U}_a'\subset {\cal U}_a\\
0,& z\not\in {\cal U}_a.
 \end{array}
\right.
 \eq
The moving points $(x_a^0\to x_a)$ correspond to  the following local deformation
 $$
 \begin{array}{l}
  w=z-\sum_{a=1}^n\ep_a(z,\bz)\,, \\
  \ep_a(z,\bz)=-(t_a+\sum_{j=1}t_a^{(j)}\f1{j!}(z-x^0_a)^j)\chi(x^0_a,\bar x^0_a))\,.
 \end{array}
 $$
In particular,
 \beq{t1}
t_a=x_a-x^0_a\,,~~t_a^{(1)}=\p_zw|_{z=x_a^0}-1\,,~~t_a^{(j)}=\p^j_zw|_{z=x_a^0}\,,~(j>1)\,.
 \eq
Under the action of  $Diff_{C^\infty}(\Si_{g,n})$ we can kill all the terms except the first. Thus, in general we have only
time $t_a$.  The part of the Beltrami differential related to the marked points assumes the form
 \beq{mp}
\mu=\sum_{a=1}^nt_a\mu_a^{(0)}\,,~~~
\mu_a^{(0)}=\bp\chi_a(z,\bz)\,.
 \eq
From (\ref{2.6}) we find
 \beq{dmp}
\dim(\gM(\Si_{g,n}))=3(g-1)+n\,.
 \eq

%%%%%%%%%%%%%%%%%%%%%%%%%%%%%%%%%%%%%%%%%%%%%%%%%%5

 \subsubsection*{Deformation of elliptic curve}
Let $\ti{T}^2=\{(x,y)\in\mR^2\,|\,x,y\in\mR/\mZ\}$ be a torus.
%\footnote{Don't mix it with the torus $ T^2$ (\ref{T})}.
 Complex structure on $\ti T^2$ is defined by the complex coordinate
$z=x+\tau_0y\,$, $\Im m\,\tau_0>0$. In this way we define the elliptic curve
$\Si_{\tau_0}\sim\mC/(\mZ+\tau_0\mZ)$.
%or by the operator $\p_{\bz}$ annihilating the one form $dz$.
It follows from (\ref{dmp}) that $\ti{T}^2$ should have at least one moving point. Only this case ($n=1$) will be considered
here. Since there is $\mC$ action on $\Si_{\tau_0}$ $z\to z+c$, it is possible to put the one marked point at $z=0$ and we
are left with one module related to the deformation $\Si_{\tau_0}\to\Si_{\tau}\sim\mC/(\mZ+\tau\mZ)$. We assume that
$\tau-\tau_0$ is small.
According to the general prescription introduce the new coordinate $ w=z-\frac{\tau-\tau_0}{\rho_0}(\bz-z)$, where
$\rho_0=\tau_0-\bar\tau_0$.
 If we shift $z$ as $z+1$ and $z+\tau_0$, then $w$ is transformed as $w+1$ and $w+\tau$.
Thereby $w$ is a well defined holomorphic coordinate on $\Si_\tau$.
 But this deformation moves the points $x_a$.
Instead we should use the following transformation
 $$
 w=z-\frac{\tau-\tau_0}{\rho_0}(\bz-z)(1-\sum_{a=1}^n\chi_a(z,\bz))\,,
  $$
where $\chi(z,\bz)$ is the characteristic function (\ref{chi}) and $\chi_a(z,\bz)=\chi(z-x^0_a,\bz-\bar x^0_a)$. As in the
general case, we  use the coordinates
 \beq{dec}
\left\{
 \begin{array}{l}
  w=z-\frac{\tau-\tau_0}{\rho_0}(\bz-z)(1-\sum_{a=1}^n\chi_a(z,\bz))\,,\\
\ti w=\bz\,.
 \end{array}
\right.
 \eq
Notice that
 $$
 (z+\tau_0,\bz+\bar\tau_0)\rightarrow (w+\tau\,,~\ti w+\bar\tau_0)\,.
 $$
Taking into account (\ref{mp}) define
 $$
\ep(z,\bz)=\frac{t_\tau}{\rho_0}(\bz-z)(1-\sum_{a=1}^n\chi_a(z,\bz))+
\sum_{a=1}^nt_a\chi_a(z,\bz)\,.
 $$
The deformed operator assumes the form $\p_{\ti{w}}=\p_{\bz}+\mu\p_z$, where (see (\ref{mp}))
$\mu$ is represented as the sum
 \beq{dmu}
\mu=t_\tau\mu^{(0)}_\tau+
\sum_{a=1}^nt_a\mu_a^{(0)}\,.
 \eq
From (\ref{db}) and (\ref{dec}) we find the form of $\mu^{(0)}_\tau$
  \beq{bel}
 \mu^{(0)}_\tau=\f1{\rho_0}\bp(\bz-z)(1-\sum_{a=1}^n\chi_a(z,\bz)),
 ~t_\tau=\tau-\tau_0\,,
  \eq
 and from (\ref{t1}) $\mu_a^{(0)}$
  $$
 \mu_a^{(0)}=\bp\chi_a(z,\bz)\,,~~t_a=x_a-x_a^0\,.
 $$
The dual to the Beltrami-differentials basis $\mu^{(0)}_\tau$,
 $\{\mu_a^{(0)}\,,a=1,\ldots n\}$ with respect
to the integration over $\Si_\tau$ is given by the first Eisenstein functions (\ref{A.1}) $E_1(z-x_a)$ and $1$. There is only one time
$t_\tau$  for the one
 marked point case (see (\ref{dmp})).

%%%%%%%%%%%%%%%%%%%%%%%%%%%%%%%%%%%%%%%%%%%%%%%%%%
\subsection{Equations of motion and the isomonodromy problem}

The choice of the complex structure defines the polarization $(A,\bA)$ of $Conn(\Si_{g,n},G)$.
We define the bundle ${\cal P}(G)$ over the Teichm\"{u}ller space
${\cal T}_{g,n}$ with the local
coordinates
 $$
(A,\bA,\bfS,\bft)\,,~~Res A_{z=x_a}=S_a\,,~~\bfS=(S_1,\ldots,S_n)\,,~~
\bft=(\tau_1,\ldots,\tau_{3g-3};t_1,\dots,t_n)\,,
 $$
 $$
 \begin{array}{cc}
{\cal P}(G)& \\
\downarrow& Conn(\Si_{g,n},G)\\
{\cal T}_{g,n} &
 \end{array}
 $$
The bundle ${\cal P}(G)$ plays the role of the extended phase space while $Conn(\Si_{g,n},G)$ is the standard phase space
with degenerate form (\ref{2.2}) and $\bft$ is the set of times. The  form (\ref{2}) acquires additional terms
 \beq{4}
\om=\om_0+\sum_{a=1}^n\om_a-
\f1{\ka}\Bigl(\sum_{l=1}^{3g-3}\de H_l\de \tau_l+\sum_{a=1}^{n}\de H_a\de t_a\Bigr)\,,~~
H_a=\oh\int_{\cU_a}(A,A)\bp\chi_a(z,\bz)\,.
  \eq
The symplectic form $\om$ is  defined on the total space of ${\cal P}(G)$.
It is degenerate on $3g-3+n$ vector fields $D_s$:~ $\om(D_s,\cdot)=0$, where
 $$
D_l=
\p_{\tau_l}+\f1{\ka}\{H_l,\cdot\}_{\om_0}\,,~~(l=1,\dots,3g-3)\,,
~~~
D_a=
\p_{t_a}+\f1{\ka}\{H_a,\cdot\}_{\om_a}\,,~(~a=1,\dots,n)\,.
 $$
The Poisson brackets corresponding to $\om_0$ are the Darboux brackets,  and those corresponding to $\om_a$ are the Lie
brackets. They are non-degenerate on the fibers.
 The vector fields $D_s$ define the
equations of motion for any function $f$ on ${\cal P}(G)$:
 $$
\frac{df}{dr_s}=\p_{r_s}f +\f1{\ka}\{H_s,f\}\,,~~r_s=\tau_s\,,~{\rm or}~r_s=t_s.
 $$
The compatibility conditions are the so-called Whitham equations \cite{Kr}:
 \beq{WE}
\ka\p_sH_r-\ka\p_rH_s+\{H_r,H_s\}=0.
 \eq
The Hamiltonians are the Poisson commuting quadratic Hitchin Hamiltonians. It means that  there exists the generating
function (the tau-function)
 $$
H_s=\frac{\p}{\p t_s}\log \tau, ~~
\tau=\exp\oh\sum_{s=1}^l\int_{\Si_{g,n}}<A^2>\mu_s.
 $$

%%%%%%%%%%%%%%%%%%%%%%%%%%%%%%%%%%%%%%%%%%%%%%%%%%%%%%%%%%

\subsection{Symplectic reduction}\label{whi}

Up to now the equations of motion, the linear problem, and the tau-function are
trivial.
The meaningful equations arise after imposing the corresponding constraints (\ref{fla1}) and
 the gauge fixing.
 Here we explain in detail the structure of the moduli space of the flat bundles upon the choice
 of the polarization.
 The form $\om$ (\ref{4}) is invariant under the action of  the gauge group $\cG_B$
 $$
A\rar f^{-1}\ka\p f+f^{-1}Af,~~\bA\rar f^{-1}\bp f+f^{-1}\bA f\,.
 $$
% $$
%S_a\rar f_a^{-1}S_a f_a,~~g_a\rar g_af_a,~~f_a=f(z,\bz)|_{z=x_a}\,.
% $$
%
Let us fix $\bA$ such that $\bL$ parameterized orbits of the $\cG_B$ action
 \beq{2.30}
\bA=f(\bp+\mu\p)f^{-1} +f\bL f^{-1}\,.
 \eq
Then the dual field is obtained from $A$ by the same element $f$
 \beq{2.31}
L=f^{-1}\ka\p f+f^{-1}Af\,,
 \eq
Thus, in local coordinates  the moment equation takes the form (see (\ref{fla1}))
 \beq{2.18}
(\bp+\p\mu)L-\ka\p\bL+[\bL,L]=2\pi i\sum_{a=1}^nS_a\de(x_a)\,.
 \eq
 Let $U_a$ be a small   neighborhood of the marked point $x_a$, where $f(x_a,\bar x_a)\in B$.
 We fix the gauge as  $\bL=0$ locally on $U_a$. Then the Hamiltonian $h^{gauge}$ (\ref{1b}) takes
 the form
  $$
 h^{gauge}=\int_{U_a}(\ep, \bp L)\,, ~~\ep|_{t_a=0}=\gb+O(|z|)\,,~~\gb\in Lie(B)\,.
 $$
 The zero moment map condition with respect to the $\cG_B$ action is
 \beq{mm}
\bp L|_{U_a}=\sum_{a=1}^nS_a\de^{(2)}(t_a,\bar t_a)\,,~~S_a\in \cO_a\,,~{\rm~}\hbox{Res}_{x_a} L=S_a\,.
 \eq
If $G$ is a finite-dimensional group, the gauge fixing (\ref{2.30}) and the moment constraint (\ref{mm}) kill almost all
degrees of freedom. The fibers $FBun(\Si_{g,n},G)=\{L,\bL,{\bf S}\}$ become finite-dimensional as well as the bundle ${\cal
P}(G)$:
 $$
\dim\,(FBun(\Si_{g,n},G))=2\dim\,(G)(g-1)+ n\dim\cO\,,
 $$
 $$
\dim{\cal P}(G)=\dim\,(FBun(\Si_{g,n},G))+
3g-3+n\,.
 $$
Due to the invariance of $\om$ (\ref{4}) it preserves its form  on $FBun(\Si_{g,n},G)$
 \beq{red}
\om_0=\int_{\Si_{g,n}}\lan dL,d\bL\ran+\sum_{a=1}^n\om^{KK}_a\,,
~~H_s(L)=\oh\int_{\Si_{g,n}}\lan L^2\ran\mu_s^0\,.
 \eq
But now, due to (\ref{2.18}), the system is no longer free, because $L$ depends on $\bL$ and ${\bf S}$. Moreover, since $L$
depends explicitly on $\bft$, the system (\ref{red}) is non-autonomous.
Let $M_s=\p_sff^{-1}$.  It follows from (\ref{2.30}) and (\ref{2.31}) that the equations of motion (\ref{em}) on the space
$FBun(\Si_{g,n},G)$  take the form
 \beq{2.19}
\left\{
 \begin{array}{ll}
  1. & \ka\p_sL-\ka\p M_s+[M_s,L]=0,~~s=1,\ldots,l\,, \\
  2. & \ka\p_s\bL-(\bp+\p\mu)M_s+[M_s,\bL]=L\mu_s^0\,. \\
 \end{array}
\right.
 \eq
The equations  1.(\ref{2.19}) are the Lax equations. The essential
difference with the integrable systems is the differentiation $\p$ with respect
to the spectral parameter.
On  $FBun(\Si_{g,n},G)$ the linear system (\ref{2.15}) assumes the form
 \beq{4.36}
\left\{
 \begin{array}{ll}
1.&(\ka\p+L)\psi=0\,,\\
2.&(\bp+\sum_st_s\mu_s^0\p+\bL)\psi=0\,,\\
3.&(\ka\p_s+M_s)\psi=0,~~(s=1,\ldots,l_2)\,.
 \end{array}
\right.
 \eq
 The equations 1. and  2. (\ref{2.19})  are consistency conditions for (1. 3.) and (2. 3.),
  and (1. 2.) is the flatness condition (\ref{2.18}).
The equations 3.(\ref{4.36}) provide the isomonodromy property of the system 1.,2.(\ref{4.36}), with respect to variations
of the times $t_s$. For this  we refer to the nonlinear equations (\ref{2.19}) as the Hierarchy of the Isomonodromic
Deformations.

%%%%%%%%%%%%%%%%%%%%%%%%%%%%%%%%%%%%%%%%%%%%%%%%%%%

\subsection{Isomonodromic deformations and integrable systems}\label{idi}

We can consider the isomonodromy preserving equations as a deformation
({\it Whitham quantization}) of integrable equations \cite{Ta,LO}.
The level $\ka$ plays the role of the deformation parameter.
Introduce the independent times $t^0_s=(\tau_l^0\,,x_a^0)$ as $\tau_l=\tau_l^0+\ka t_l$, $t_a=\ka\ti t_a$ for $\ka\to 0$. It
means that $t_s=(t_l,\ka\ti t_a)$ plays the role of a local coordinate in a neighborhood of the point $(t_l^0,x_a^0)$ on the
moduli space $\gM(\Si_{g,n})$. In this limit  the equations of motion  1.(\ref{2.19}) are the standard Lax equation (the
Zakharov-Shabat equation in the 1+1 case)
 \beq{lae}
\p_sL^{(0)}+[M^{(0)}_s,L^{(0)}]=0,~~s=1,\ldots,l\,,
 \eq
where $L^{(0)}=L(t^0_s)$, $(M^{(0)}_s=M_s(t^0_s))$.
The linear problem for this system is obtained from the linear problem for the isomonodromy problem (\ref{4.36}) by the
analog of the quasiclassical limit in quantum mechanics. Represent the Baker-Akhiezer function in the WKB form
 \beq{WKB}
\psi=\Phi\exp\bigl(\frac{{\cal S}^{(0)}}{\ka}+{\cal S}^{(1)}\bigr)
 \end{equation}
and substitute (\ref{WKB})
 into the linear system (\ref{4.36}). If
$\p_{\bar z}{\cal S}^{(0)}=0$ and $\p _t{\cal S}^{(0)}=0$, then the terms of order $\ka^{-1}$ vanish. In the quasiclassical
limit we put $\p{\cal S}^{(0)}=\la$. In the zero order approximation we come to the linear system %for CM
 \beq{lis}
\left\{
 \begin{array}{ll}
 i. & (\la+L^{(0)}(z,\tau_0))Y=0\,,\\
  ii. & \p_{\bz}Y=0\,,\\
iii. & (\p_{t_s}+M_s^{(0)}(z,\tau_0))Y=0\,.
 \end{array}
\right.
 \eq
The Baker-Akhiezer function $Y$ takes the form
 $$
Y=\Phi e^{\sum_s t_s\frac{\p}{\p t^0_s}{\cal S}^{(0)}}\,.
 $$
The consistency condition of $i$ and $ii$  is the Lax equation (\ref{lae}).

%The Baker-Akhiezer function $Y$ takes the form
% $$
%Y=\Phi e^{ t\frac{\p}{\p \tau_0}{\cal S}^{(0)}}\,.
% $$

%%%%%%%%%%%%%%%%%%%%%%%%%%%%%%%%%%%%%%%%%%%%%%%%%%%%
%%%%%%%%%%%%%%%%%%%%%%%%%%%%%%%%%%%%%%%%%%%%%%%%%%%%

\section{Affine Algebras and Isomonodromic Deformations}\label{lgb}
\setcounter{equation}{0}

We apply the general scheme to the bundles related to affine algebras, and
 briefly describe here their central and the cocentral extensions
\cite{Ka}.

\subsection{Affine Lie algebras}

Let $\gg$ be a simple complex Lie algebra and $L(\gg)=\gg\otimes\mC[x,x^{-1}]$ $(x\in \mC^*)$
 is the loop algebra of Laurent polynomials.
Consider its central extension  $\hat{L}(\gg)=\{(\cX(x),\la)\}$.
 The commutator in  $\hat{L}(\gg)$ assumes the form
 $$
[(\cX_1,\la_1),(\cX_2,\la_2)]=([(\cX_1,\cX_2)]_0,\oint\lan\cX_1,\p_x\cX_2\ran)\,,
 $$
where $[(\cX_1,\cX_2)]_0$ is a commutator in $\gg$ and $\lan\,,\,\ran$ is the Killing metric on $\gg$. The  cocentral
extension $\check{L}(\gg)$ of $\hat L(\gg)$ is the algebra
 \beq{lcc}
\check{L}=\{\cW=(k,Y,\la)=(k\p_x+Y,\la)\,,~Y\in L(\gg)\,,~~k\in\mC\,,~\la\in\mC\}\,.
 \eq
with the commutator
 \beq{lc}
[\cW_1,\cW_2]=[(k_1,\cY_1,\la_1),(k_2,\cY_2,\la_2)]= (0,k_1\p_x\cY_2-k_2\p_x\cY_1+[\cY_1,\cY_2]_0,\oint\lan
\cY_1,\p_x\cY_2\ran )
 \eq
and the invariant non-degenerate form
 \beq{inm}
(\cW_1,\cW_2)=\oh\Bigl(\oint\lan \cY_1,\cY_2\ran+k_1\la_2+k_2\la_1\Bigr)\,.
 \eq
 Let $L(G)$ be the loop group corresponding to the Lie algebra $L(\gg)$.
 The following statement can be checked directly.
 \begin{lem}
The group $L(G)$ acts as Aut($\check{L}(\gg)$) preserving the metric (\ref{inm})
 \beq{gtr}
\cW\to (k\,,kf^{-1}\p_xf+f^{-1}\cY f\,,
\la-\oint\lan \cY,f^{-1}\p_xf\ran-k\oint\lan f^{-1}\p_xf,f^{-1}\p_xf\ran)\,.
 \eq
 \end{lem}
The Lie algebra End($\check{L}(\gg))=L(\gg)$ is
 generated by the  vector fields
 $\ep\in L(\gg)$
 \beq{igt}
\de_\ep \cY=k\p_x\ep+[\ep,\cY] \,,~\de_\ep k=0\,,~\de_\ep \la=-\oint\lan \cY,\p_x\ep\ran\,.
 \eq

 %%%%%%%%%%%%%%%%%%%%%%%%%%%%%%%%%%%%%%%%%%%%%%%%

\subsection{Vector bundles of infinite rank}\label{inf}

Let $\cP$ be a principal $L(G)$ bundle over the complex curve $\Si_g$.
Consider the adjoint bundle $E=\cP\otimes_\cG\check{L}(\gg)$.
Its local sections are the maps $\Si_g\to \cW$ (\ref{lcc}).  The space of connections  on $E$
 is
$$
 Conn(\Si_g,L(G))=\{d_\cA\}\,,~~ d_{\cA}\,:\, \G(E)\to\G(E)\otimes\Om^1(\Si_g)\,,~~d_{\cA}=d+\cA\,,~
%  \cA\in\Om^1(\Si,\check{L}(\gg) )\,,
   $$
 where  $d_{\cA}$ has a component description (\ref{lcc}), and locally
  \beq{svi}
d_{\cA}= d+\cA=(d+\emph{k}\p_x+\cY\,,d+\emph{$\lambda$})\,.
  \eq
 The component $d+\emph{$\lambda$}$ is a connection in a trivial line bundle over $\Si_g$.

%Let $\{\cU_\al\}$ be a covering of $\Si_g$ by contractible sets,
%and $\cU_{\al\be}=\cU_\al\cap\cU_\be$.
The gauge group of the bundle $\cG$ is the group $Map_{C^\infty}\,(\Si_g\to  L(G))$.
 %It means that the transition functions are the maps
%  $$
% \cG_{\al\be}=Map_{C^\infty}(\cU_{\al\be}\to L(G))\,.
%  $$
It acts on sections as (\ref{gtr}).
%Consider the space of smooth connections $Conn(\Si_g,L(G))=\{d_\cA\}$  on $E$
%   $$
%  d_{\cA}\,:\, \G(E)\to\G(E)\otimes     \Om^1(\Si)\,,~~d_{\cA}=d+\cA\,,~
%%  \cA\in\Om^1(\Si,\check{L}(\gg) )\,,
%   $$
% where  $\cA$ has a component description (\ref{lcc}), and locally
%  \beq{svi}
%d_{\cA}= d+\cA=(d+\emph{k}\p_x+\cY\,,d+\emph{$\lambda$})\,.
%  \eq
%
%%
Using the form (\ref{inm}) we introduce the symplectic structure in
the space $Conn(\Si_g,L(G))$
 \beq{ss23}
\om=\oh\int_{\Si_{g,n}}\oint\lan\de\cA\wedge\de\cA\ran=
\oh\int_{\Si_{g,n}}\oint\lan\de \cY\wedge\de \cY\ran
+\de k\wedge\de\la\,.
 \eq
 The form is invariant under the gauge action.

%%%%%%%%%%%%%%%%%%%%%%%%%%%%%%%%%%%%%%%%%%%%%%%%%%%%%%
\subsection{The gauge group}

In what follows we use the central extension $\hat\cG$ of the gauge group $\cG$.
%Let ${\mathcal G}$ be the group of automorphisms of $\mathcal{R}$
%(the gauge group).
It is based on
the central extension  of the loop group
$$
1\longrightarrow\mC^*\longrightarrow\hat{L}(G)\longrightarrow L(G)\longrightarrow 1
$$
and defined as
 \beq{gg}
\hat {\cal G}={\rm Map}_{C^\infty}(\Si_{g}\to \hat{L}(G))=
\{f(z,\bz,x),s(z,\bz)\}\,,
 \eq
where $f(z,\bz,x)$ takes values in $G$,
and $s$ is the map $\Si_{g}$  to the central element of $\hat{L}(G))$.
The multiplication is pointwise with respect to $\Sigma_g$
 \beq{cex}
(f_1,s_1)\times(f_2,s_2)=\left(f_1f_2,s_1s_2\exp {\cal C}(f_1,f_2)\right)\,.
 \eq
Here $\exp {\cal C}(f_1,f_2)$ is a map from $\Si_{g}$ to the central element of
$\hat{L}(G)$ defined by the Wess-Zumino-Witten action \cite{Fel,PS}.

%
%The group $ \cG=Map_{C^\infty}(  \Si_{g,n}\to L(G))$
%is the group of symplectomorphisms of $Conn(\Si_g,L(G))$,
 %  with the action (\ref{gtr}).
   Consider the Lie algebra $Lie(\hat\cG)=\{(\ep,\varepsilon)\}$, where $\ep$ is a smooth map
   of  $\Si_{g}$ to $\gg$ and $\varepsilon$ is a smooth map
   of  $\Si_{g}$ to $\mC$. $Lie(\hat\cG)$
   is represented by the Hamiltonian  vector fields:
   (see (\ref{igt}))
    \beq{hac}
    \begin{array}{lll}
   \de_\ep\cY=  (d+k\p_x)\ep+[\cY,\ep]\,, & ~\de_\ep k=0\,, & \de_\ep\la=-\oint\lan \cY,\p_x\ep\ran\,, \\
      \de_\varepsilon\cY=0\,, & \de_\varepsilon k=0\,,  & \de_\varepsilon\la=d\varepsilon\,.
    \end{array}
    \eq
   It defines the moment map $F\,:\,Conn(\Si_{g},L(G))\to Lie^*(\cG)$, with
    $$
   \cF=(F,dk)\,,~~F=(d+k\p_x)\cY+\oh[\cY,\cY]\,.
 $$
Rewrite $\cA$ in terms of a complex structure on $\Si_g$. Let $(z,\bz)$ be
local coordinates  on $\Si_g$. Then
 $$
\cA=(A,\bA')=\bigl((\ka\p_z+k\p_x+Y,\p_z+\la)\otimes dz\,,~
(\p_{\bz}+\bar{k}'\p_x+\bar{Y}',\p_{\bz}+\bar{\la}')\otimes d\bz\bigr)\,,
 $$
In these terms $\om$ (\ref{ss23}) and the action of $Lie\,(\hat\cG)$ (\ref{hac}) assume the form
 \beq{ss1}
\om=\oint\int_{\Si_{g}}\lan\de A\wedge\de\bA'\ran=\oint\int_{\Si_{g,n}}\lan\de Y\wedge\de\bar{Y}'\ran
+\de k\wedge\de\bar{\la}'+\de\la\wedge\de\bar{k}'
 \eq
 \beq{vf}
 \begin{array}{l}
\de_{\ep} Y=(\ka\p_z+k\p_x){\ep}+[Y,{\ep}]\,,~~
\de_{\ep}\la=-\oint(Y,\p_x{\ep})\,,~~\de_{\ep} k=0\,,\\
\de_{\ep} \bar{Y}'=(\p_{\bz}+\bar{k}'\p_x){\ep}+[\bar{Y}',{\ep}]\,,~~
\de_{\ep}=-\oint(\bar{Y}',\p_x{\ep})\,,~~\de_{\ep} \bar{k}'=0\,,\\
\de_\varepsilon\la=\ka\p_z\varepsilon\,,~~\de_\varepsilon\bar{\la}'=\p_{\bz}\varepsilon\,.
 \end{array}
 \eq
Then we come to the following expression for the moment
 \beq{F1}
\cF=\Bigl((\ka\p_{z}+k\p_x)\bar{Y}'-(\p_{\bz}+\bar{k}'\p_x)Y+[Y,\bar{Y}']),
(\ka\p_{z}\bar k'-\p_{\bz}k)\Bigr)\,.
 \eq
The flatness condition $\cF=0$ is the moment constraint.

As in (\ref{nba}) we pass to the new components
 $$
\bar{Y}=\bar{Y}'+\f1{\ka}\mu Y\,,~~ \bar{k}= \bar{k}'+\f1{\ka}\mu k\,,~~
\bar{\la}=\bar{\la}'+\f1{\ka}\mu\la\,.
 $$
In this way using (\ref{nv}) we get the flat connection in the coordinates $(w,\ti w)$:
 \beq{nfl}
\left\{
 \begin{array}{l}
 ((\ka\p_{w}+k\p_x+Y),(\p_{ w}+\la))\otimes dw\,, \\
  ((\p_{\ti w}+\bar{k}\p_x+\bar{Y}),(\p_{\ti w}+\bar{\la}))\otimes d\ti w \,.
 \end{array}
\right.
 \eq
Recall that $\mu=\sum_st_s\mu^0_s$. Then $\om$ (\ref{ss1}) becomes
 \beq{ss2}
\om=\om_0-\f1{\ka}\de H_s\de t_s\,,
 \eq
 $$
\om_0=\oint\int_{\Si_{g,n}}\lan\de Y\wedge\de\bar Y\ran+
\int_{\Si_{g,n}}(\de k\wedge\de\bar\la+\de \la\wedge\de\bar k)\,,
 $$
 \beq{ham}
H_s=\int_{\Si_{g,n}}\Bigl(\oint\lan Y,Y\ran+k\la\Bigr)\mu_s^0\,.
 \eq
The equations of motion corresponding to these Hamiltonians
 take the form:
 \beq{eqom}
 \begin{array}{lll}
  \p_sY=0\,, & \p_s\la=0\,, & \p_sk=0\,, \\
  \ka\p_s\bar{Y}=Y\mu_s^0\,,~ & \ka\p_s\bar{k}=k\mu_s^0\,, & \ka\p_s\bar\la=\la\mu_s^0\,.
 \end{array}
 \eq

%%%%%%%%%%%%%%%%%%%%%%%%%%%%%%%%%%%%%%%%%%%%%%%%%%%%

\subsection{Quasi-parabolic structure}

Let $B\subset G$ be a Borel subgroup of $G$. Define $L^+(G)=(B+G \otimes t\mC[x])$ the Borel subgroup of $L(G)$. The quotient
space $L(G)/L^+(G)$ is the affine flag variety  $Fl^{Aff}$ \cite{PS}. It is the loop analog of the finite-dimensional flag
variety. Assume that $\cG$ is restricted to $L^+(G)$ at the marked points on $\Si_{g,n}$. As in the general case we denote this
subgroup $\cG_B$. The existence of the quasi-parabolic structure for these bundles means that we fix affine flags
$Gr_a^{Aff}$ at the marked points $x_a$ $(a=1,\dots,n)$.
To define the moduli space of flat connections we attach the coadjoint orbits (\ref{co1}) of $\hat L(\SLN)$ to the marked
points
 \beq{orbi}
 {\mathcal O}(\bfS^{(0)}_a,r_a)=
\{ \bfS_a=-g^{-1}\bfS_a^{(0)}g-r_ag^{-1}\p_xg,~r_a\,,~(\bfS_a^{(0)},r_a)\in \hat L^*(\gg)\}\,.
 \eq
The coadjoint orbits are infinite-dimensional symplectic manifolds with the
 Kirillov-Kostant form
 \beq{1.6a}
\om_a^{KK}=\oint \lan\de (\bfS_a g^{-1})\wedge\de g\ran
+\frac{r_a}{2}\oint \lan g^{-1}\de g\wedge\de  (g^{-1}\p_xg \ran\,.
 \eq
Notice, that due to the residue theorem
 \beq{rt}
\sum_{a=1}^nr_a=0\,
 \eq
%In what follows we put $r_a=0$.
the orbits are the affine spaces over the cotangent bundle $T^*Fl_a^{Aff}$  of the affine
 flag varieties $ Fl_a^{Aff}$.
 For the bundles with the quasi-parabolic structure we replace $\om_0$ in (\ref{ss2}) with
 $\om_0+\sum_{a=1}^n\om_a^{KK}$.
 The flatness condition for these bundles takes the form (see (\ref{F1})):
 $$
 (\ka\p_{z}+k\p_x)\bar{Y}'-(\p_{\bz}+\bar{k}\p_x)Y+[Y,\bar{Y}']=-\sum_{a=1}^n\bfS_a\de(x_a)\,,
 $$
  $$
 \ka\p_{z}\bar k'-\p_{\bz}k=-\sum_{a=1}^nr_a\de(x_a)\,,
 $$
%Let us fix the gauge $\bL=f^{-1}(\p_{\bz}+\mu\p_z)f+f^{-1}\bar Yf$ then
% we come to the moment constraint (\ref{2.18}).
or from (\ref{nfl}) we get:
 \beq{flbq}
 (\ka\p_{w}+k\p_x)\bar{Y}-(\p_{\ti w}+\bar{k}\p_x)Y+[Y,\bar{Y}]=-\sum_{a=1}^n\bfS_a\de(x_a)\,,
 \eq
  $$
 \ka\p_{w}\bar k-\p_{\ti w}k=-\sum_{a=1}^nr_a\de(x_a)\,.
 $$

 %%%%%%%%%%%%%%%%%%%%%%%%%%%%%%%%%%%%%%%%%%%%%%%%%%%%%%%
 %%%%%%%%%%%%%%%%%%%%%%%%%%%%%%%%%%%%%%%%%%%%%%%%%%%%%%%%%%%%%%%

\section{Painlev\'e type field theory related to $L(\SLN)$ bundles}
\setcounter{equation}{0}

 Here we consider a particular example of general theory and assume that $G=\SLN$ and the base
 of the loop group bundle is taken over elliptic curve $\Si_{\tau}=\mC/(\mZ\oplus\tau\mZ)$
 with coordinates $(z,\bz)$. We consider the case of one marked point $z=0$.

\subsection{Bundles over  $\Si_\tau$}

%Let $\cP$ be a principle infinite rank $\hat L(\GLN)$-bundle over  the elliptic curve
%$\Si_{1,1}=\Si_{\tau}=\mC/(\mZ\oplus\tau\mZ)$
%with the complex coordinate  $w$ (\ref{bel}).
%Consider
 %an integrable representation $V$ of $\hat{L}(\sln) $ \cite{Ka}.

 Let $V$ be an integrable representation of $\hat{L}(\sln)$ of level $l$ \cite{Ka}.
It means that $\la$ acts on $V$ as a scalar $l$.
 Consider the adjoint bundle $E=\cP\times_{\hat L(\SLN)} V$.
The sections $s(w,\ti w,x)\in\G(E)$ satisfy the quasi-periodicity conditions
 \beq{saa}
\left\{
 \begin{array}{ll}
1.&  s(w+1,\ti w+1,x)=Q^{-1}s(w,\ti w,x)\,, ~~(\ti\La=-\bfe(\frac{-w-\tau/2}N)\La\,,~(\ref{la})) \\
 2.& s(w+\tau,\ti w+\bar\tau_0,x)=
 \ti\La^{-1}\bfe(-\frac{\bar k}{2\pi\imath}\p_x)s(w,\ti w,x)=\ti\La^{-1}s(w,\ti w,x-\bar k) \,,
 \end{array}
\right.\,.
 \eq
To describe the  sections of the bundles of this type  it is convenient to work with the sin-basis $T^\al$ in $\sln$
(\ref{ta2}) $s(w)=\sum_{\al\in\mZ^{(2)}_N} s^\al(w)T^\al$. The sections $ s^\al(w)$ are defined by
 the quasi-periodicities (\ref{saa}).
It follows from (\ref{nfl}) and (\ref{saa}) that we pass to the connections
%\mo{How the quasi-periodicity condition
%(\ref{saa} 2.) "kill" the derivative $\bar k\p_x$
% in the $(0,1)$ component?}
 \beq{cb}
\left\{
 \begin{array}{l}
  \ka\p_w+(k\p_x+Y,\la)\,, \\
  \p_{\ti w}+(\bar Y,\bar\la)\,.
 \end{array}
\right.
 \eq
It follows from (\ref{vf}) that $k$, $\bar k$, $\la$ and $\bar\la$ can be chosen to be double-periodic,
while
%\footnote{In fact $Y$ and $\bar Y$ depend also on $\ti w$, where $\ti w$ is
%defined up to the shifts $\ti w\to\ti w+1$, $\ti w\to\ti w+\tau_0$.}
 \beq{yqp}
 \begin{array}{ll}
  Y(w+\tau)=-\ka\frac{2\pi\imath}N Id_N+\La Y(w)\La^{-1}\bfe(-\frac{\bar k}{2\pi\imath}\p_x)\,, &
   Y(w+1)=Q Y(w) Q^{-1}\,, \\
   \bar Y(w+\tau)=\La\bar Y(w)\La^{-1}\bfe(-\frac{\bar k}{2\pi\imath}\p_x)\,, &
 \bar Y(w+1)=Q \bar Y(w) Q^{-1}\,.
 \end{array}
 \eq
Let $\hat\cG=C^\infty(Map)\,:\, \Si_\tau\to\hat L(\GLN)$ be the gauge group (\ref{gg}).
Its elements satisfy the quasi-periodicity conditions (\ref{saa}).
The gauge transformation (\ref{vf}) allows one to choose $k$ and $\bar k$ to be $(z,\bz)$-independent,
 $\la$ to be antiholomorphic, and $\bar\la$ holomorphic. Since $\la$ and $\bar\la$ are double-periodic,
 then they are also  $(z,\bz)$-independent.
 For  the bundles (\ref{saa})  almost all  $\bar Y$ can be represented as a pure gauge
 $\bar Y=-f^{-1}(\p_{\ti w})f$.
Then the gauge transformed connections (\ref{cb}) assume the form
 \beq{ca12}
\left\{
 \begin{array}{l}
\ka\p_w+(k\p_x+L,\la)\,,\\
\p_{\ti w}+(0,\bar\la)\,,
 \end{array}
\right.
 \eq
where $k$, $\la$, $\bar k$, $\bar\la$ are constants.
% \mo{This statement is valid for $\bar k=0$. In this case we have for any value $0\leq x<1$
 %a bundle of degree one over $\Si_\tau$. In fact, we deal here with a bundle $\cE$  over a
 %four-dimensional manifold $W=C^*\times\Si_\tau$.
 %$Y$ and $\bar Y$ are components of connection on $\cE$.
  %To define $\cE$ we assume that in addition to (\ref{yqp}) $Y$ and $\bar Y$ are $x$-periodic.
   %It can be suggested that $\bar Y$ can be gauged away if $\oint Y=0$, where integral
   %is taken over a non-contractible contour in $\mC^*$.}
%It follows from the definition of $A$ and $\bA$ (\ref{svi})  that its components
%$k$, $\la$, $\bar{k}$ and $\bar{\la}$ are double-periodic on $\Si_\tau$, while
%
%
Let us consider $\Si_\tau$ with a single marked point $z=0$. Then $L$ is fixed by the following conditions:
 \vskip3mm
\noindent{$\bf 1$.}\emph{ The flatness} of the connections (\ref{ca12})
 \beq{fl}
\p_{\bar{w}}L=0\,.
 \eq
$\bf 2.$ \emph{The quasi-periodicity
conditions}
 \beq{guy}
  L(x,w+\tau)=-\ka\frac{2\pi\imath}N Id_N+
  \La L(w)\La^{-1}\bfe(-\frac{\bar k}{2\pi\imath}\p_x)\,,~~
   L(w+1)=Q L(w) Q^{-1}\,,
 \eq
$\bf 3.$ \emph{The quasi-parabolic structure:}
 \beq{res}
Res_{w=0}L=\bfS-\ka\frac{2\pi\imath}N Id_N\,,~~\bfS\in{\mathcal O}(\bfS^{(0)},r=0)\,.
\footnote{$r=0$ follows from (\ref{rt}).}
 \eq
Since $\det(\ti{\La}(z,\tau))=(-1)^N\bfe(-z-\oh\tau)$ we see that   $\ti{\La}$
takes value in $\GLN$, but not in $\SLN$. Thus, for fixed $x$ (\ref{guy})
defines  $\GLN$ bundles. These  $\GLN$-bundles have degree one. In this way, the $\hat{L}(\SLN)$-bundles which we consider are the analogues of the $\GLN$-bundles of degree one (see \cite{LOSZ1,LOSZ2} for the bundles of arbitrary
characteristic class).
%It is not important for $\cP\otimes_{L(\GLN)}ad_{\hat{L}\sln}$
%bundles, because it coincides with the $\SLN$-bundles.
\begin{rem}\label{2d}
The construction of the infinite rank bundle $E=\cP\times_{\hat L(\SLN)} V$ (\ref{saa})
can be reformulated in terms of the bundle of rank $N$ over two-dimensional complex surface
$\Si_\tau\times \mC^*$, where the second component is defined by the coordinate
$\bfe(x)$, $x\in mC$. Its structure group is $\mC^*\ltimes\GLN$ (see (\ref{gg}), (\ref{cex})).
It has degree one and the corresponding nontrivial cycle is $\Si_\tau$.
\end{rem}

We rewrite the Hamiltonians (\ref{ham}) in terms of $L$
 \beq{ham1}
H_\tau=\int_{\Si_{\tau}}\Bigl(\oint\lan L,L\ran+k\la\Bigr)\mu_\tau^0\,,~~
\mu_\tau^0=\f1{\rho_0}\,,
 \eq
where we preserve the notion $\la$ after the gauge transformation.

%%%%%%%%%%%%%%%%%%%%%%%%%%%%%%%%%%%%%%%%%%%%%%%%%%%%%%%%%%

\subsection{Spin variables}
We refer to the elements  $\bfS=\sum_{\al\in\mZ^{(2)}_N}S^\al(x) T^\al$
of the orbit ${\mathcal O}(\bfS^{(0)},r=0)\subset
L^*(\gg) $ (\ref{orbi}) as the spin variables. To be more precise, we consider
 the loop  algebra $L(\gg)$ of trigonometric
polynomials in the basis $T_n^\al=\bfe(nx)T^\al$. In other words,
 $$
L(\gg)=\{\bfW(x)=\sum_{n\in\mZ}\sum_{\al\in\mZ^{(2)}_N}W_n^\al T_n^\al\,,~~ W_n^\al=0~{\rm for ~almost ~all }~n\}\,.
 $$
 The dual space $L^*(\gg)$ is the space of distributions on
$L(\gg))$. For the functionals
\beq{ixv}
\bfS^\al(x)=\lan\bfS(x),T^{-\al}\ran\,,~~\bfS^\al(x)=\sum_nS^\al_nT_n^\al\,,~~(T_n^\al=\bfe(nx)T^\al)
\eq
 on $L^*(\gg)$ we have the  Poisson-Lie brackets
 \beq{pbo}
\{S^\al(x),S^\be(y)\}=S^{\al+\be}(x)\bfC(\al,\be)\de(x-y)\,,
 \eq
or, in terms of the Fourier expansion
 \beq{pbs}
\{S_n^{\al},S_m^\ga\}=S_{n+m}^{\al+\ga}\bfC(\al,\ga)\,.
 \eq
The brackets  are non-degenerate on the orbits ${\mathcal O}$. We can identify the algebra $L(\gg)$ of the trigonometric
polynomials with the subspace in $L^*(\gg)$ by means of the pairing via the contour integral and the Killing form on $\gg$.
The element
 \beq{ca}
C_2=\sum_{n\in\mZ}\sum_{\al\in\mZ^{(2)}_{N}}T_n^\al T_{-n}^{-\al}
 \eq
is the central element of the algebra (\ref{ta}). Though it is not an element of the universal enveloping algebra
$\cU(L(\gg))$ it is well defined on some representations of $L(\gg)$ \cite{Ka}. The Casimir function of the brackets
(\ref{pbo}),  (\ref{pbs}) is given by
 \beq{ca11}
H_2=\sum_{n\in\mZ}\sum_{\al\in\mZ^{(2)}_{N}}S_n^\al S_{-n}^{-\al}=\oint\lan\bfS(x),\bfS(x)\ran\,.
 \eq
In addition to (\ref{vf}) the gauge transformation of $\bfS$ takes the form:
 \beq{1.8a}
\de_{\ep}\bfS= [\bfS,\ep^0]\,,~~\ep^0=\ep(x,z,\bz)|_{z=0}\,.
 \eq

%%%%%%%%%%%%%%%%%%%%%%%%%%%%%%%%%%%%%%%%%%%%%%%%%%%%

\subsection{Lax operator and equations of motion}\label{hami}

Represent the Lax operator in the form of the Fourier expansion
 $$
L(x,w)=\frac{\ka}NE_1(w|\tau) Id_N+\sum_{n\in\mZ} \sum_{\al\in\mZ^{(2)}_{N}}L^{\al}_n(w)T^\al_n\,,~~
T^\al_n=\bfe(nx)T^\al\,,~\bfe(~)=\exp\,2\pi\imath(~)\,.
 $$
%\mo{I have add the term $\frac{\ka}NE_1(w|\tau) Id_N$ to satisfy (\ref{guy}).
%Otherwise $L(x,w,\bw)$ is the Lax
%operator for the Hitchin system, but not for the Isomonodromy problem.}
Taking into account
  the flatness condition (\ref{fl}) and (\ref{res}) $L^{\al}_n(w)$ is $\ti w$ independent
  and
 \beq{mc2}
L^{\al}_n(w)|_{w\to 0}\sim\frac{S^{\al}_n}w\,.
 \eq
From the quasi-periodicities of $E_1$ (\ref{A.12}) and $\phi\Bigl(-\al_\tau+\bar{k}n,w\Bigr)$ (\ref{A.14}) we find that
 \beq{ln}
 \begin{array}{|c|}
  \hline\\ L^{\al}_n(w)=S_n^{\al}X_n^{\al}\,,~~X_n^{\al}=\bfe \Bigl( -\frac{w\al_2}N\Bigr) \phi\Bigl(-\al_\tau+\bar{k}n,w\Bigr)
\\ \ \\
\hline
  \end{array}
 \eq
with $\al_\tau=\frac{\al_1+\tau\al_2}N$ has the quasi-periodicity conditions (\ref{guy}). In addition, from (\ref{A.3a}) we
come to (\ref{mc2}). Therefore, in terms of the functions $\bfS^{\al}(x)=\sum_{n\in\mZ} S_n^{\al}\bfe(nx)$
 \beq{lop}
L(x,w)=\frac{\ka}NE_1(w|\tau) Id_N+\sum_{\al\in\mZ^{(2)}_{N}}\bfS^{\al}(x)\bfe
\Bigl(-\frac{w\al_2}N\Bigr)
\phi\Bigl(-\al_\tau+\frac{\bar{k}\p_x}{2\pi\imath},w\Bigr)T^\al\,.
 \eq

%%%%%%%%%%%%%%%%%%%%%%%%%%%%%%%%%%%%%%%%%%%%%%%%%%%%%

\subsubsection*{Hamiltonian, phase space and equations of motion}
Recall that for the one marked point case we have only one time $t_\tau$ (\ref{bel}). In order to calculate the Hamiltonian
we consider the integral
 $$
\oint\tr L^2(x,w)\mu^{(0)}_\tau=\left(\frac{\ka^2}NE^2_1(w)+
\sum_{n\in\mZ}\sum_{\al\in\mZ^{(2)}_{N}}L^{\al}_n,L^{-\al}_{-n}\right)\mu^{(0)}_\tau
%\sum_{n\in\mZ}\sum_{\al\in\mZ^{(2)}_{N}}(\rho\bar{k}n+L_n^{\al})(-\rho\bar{k}n+L_{-n}^{-\al})=
 $$
 $$
=\Bigl(\frac{\ka^2}NE^2_1(w)+
\sum_{n\in\mZ}\sum_{\al\in\mZ^{(2)}_{N}}S^{\al}_nS^{-\al}_{-n}\phi(-\al_\tau+\bar{k}n,w)
\phi(\al_\tau-\bar{k}n,w)\Bigr)\mu^{(0)}_\tau
 $$
 \beq{tra}
\stackrel{(\ref{d3})}{=}\Bigl(\frac{\ka^2}NE^2_1(w)+\sum_{\al\in\mZ^{(2)}_{N}}
\sum_{n\in\mZ}S^{\al}_nS^{-\al}_{-n}
(E_2(w)-E_2(-\al_\tau+\bar{k}n)\Bigr)\mu^{(0)}_\tau\,.
 \eq
Notice that  the first term of  this expression is not double periodic (see (\ref{A.12})). It reflects the fact that
 $L$ is not the Higgs field but a component of the connection of the flat bundle over $\Si_\tau$.
It follows from the definition of $\mu^{(0)}=\mu_\tau^{(0)}$ (\ref{bel}) that
 \beq{di}
 \begin{array}{lll}
  \int_{\Si_\tau}E_2(w)\mu_\tau^{(0)}=-2\pi\imath\,, &  \int_{\Si_\tau}\mu_\tau^{(0)}=1\,,  &
  \int_{\Si_\tau}E_1^2(w)\mu_\tau^{(0)}=const\,.
 \end{array}
 \eq
Then integrating (\ref{tra}) over $\Si_\tau$ and using (\ref{di}) we obtain the Casimir function
(\ref{ca})
 $$
\int_{\Si_\tau}E_2(w)\sum_{n\in\mZ}S^{\al}_nS^{-\al}_{-n}\mu_\tau^{(0)}=H_2\,,
 $$
 the Hamiltonian $H_\tau$
 \beq{hta0}
 \begin{array}{|c|}
  \hline\\
H_\tau=k\la+ \sum_{\al\in\mZ^{(2)}_{N}}\sum_{n\in\mZ} S^{\al}_nS^{-\al}_{-n}E_2\bigl(-\al_\tau+\rho_0\bar{k}n\bigr)
\\ \ \\
\hline
  \end{array}
 \eq
 and the
constant term proportional to $\frac{\ka^2}N$
 coming from the last integral  in (\ref{di}). It does not depends on the dynamical variables
 $(S_n^\al,k,\la)$ and we ignore it.
In terms of fields the Hamiltonian takes the form
 \beq{hta1}
 \begin{array}{|c|}
  \hline\\
H_\tau=k\la+ \sum_{\al\in\mZ^{(2)}_{N}}\oint \bfS^{\al}(x)
\Bigl(E_2\bigl(-\al_\tau+\frac{\rho_0\bar{k}}{2\pi\imath}\p_x\bigr)\bfS^{-\al}(x) \Bigr)
\\ \ \\
\hline
  \end{array}
 \eq
The Hamiltonian can be written in the form of the Euler-Arnold top on $\check{L}(\gg)$
 $$
H_\tau=k\la+\int_{\Si_\tau}\lan\bfS(x),\bfJ(\bfS(x))\ran\,,
 $$
where  $\bfJ\,:\,L^*(\gg)\to L(\gg)$ is the conjugate inertia tensor
 $$
\bfJ\,:\,\bfS^\al(x)\to E_2\bigl(-\al_\tau+\frac{\bar{k}}{2\pi\imath}\p_x\bigr)\bfS^{\al}(x)\,.
 $$
The phase space $\cM$  of the dynamical system is defined by means of the functionals (\ref{ca11}), (\ref{hta1})
 $$
\cM=\{\bfS^\al(x)\,|\, H_2(\bfS)<\infty\,,~H_\tau(\bfS)<\infty\,.\}
 $$
Taking into account the Poisson brackets (\ref{pbo}) we find that the evolution on $\cM$ generated by $H_\tau$ is given as
follows:
 \beq{eom}
\ka\p_\tau \bfS^\al(x)=\sum_{\ga\in\mZ^{(2)}_{N},\ga\neq\al}\Bigl(
E_2\bigl(-\ga_\tau+\frac{\bar{k}}{2\pi\imath}\p_x|\tau\bigr)\bfS^{-\ga}(x)\Bigr)\bfS^{\al+\ga}(x)
\bfC(\al,\ga)\,.
 \eq
In terms of the Fourier modes the equations of motion assume the form
 \beq{efm}
 \begin{array}{|c|}
  \hline\\
\ka\p_\tau S_n^\al= \sum_{\ga}\sum_{m}S^{-\ga}_{-m}S^{\al+\ga}_{n+m}E_2\bigl(-\ga_\tau+\bar{k}m|\tau\bigr) \bfC(\al,\ga)
\\ \ \\
\hline
  \end{array}
 \eq
The equation (\ref{eom}) can be written in the top like form (\ref{eat})
 $$
\ka\p_\tau \bfS(x)=ad^*_{J\cdot\bfS(x)}\bfS(x)\,,~~\ka\p_\tau \bfS(x)=[J\cdot\bfS(x),\bfS(x)]\,.
 $$
From (\ref{eqom}) we find
 \beq{keq}
\ka\p_\tau\bar k=k\,,~~\ka\p_\tau k=0\,.
 \eq
 \begin{rem}
Consider the subalgebra $\gg \subset L(\gg)$ of the $x$-independent loops.
  Then (\ref{efm}) coincides with the monodromy preserving
equation related to $\sln$ \cite{LO, LOZ}
%The quotient $L(\gg)/I$ is isomorphic to
 for $\gg=\sln$ . For $\slt$  (\ref{efm}) is the particular case
of the Painlev\'e VI  (\ref{p6}) (or, to be exact (\ref{rr101})). In this way   (\ref{efm})  is the field-theoretical
generalization of the Painlev\'e VI equation.
 \end{rem}
It is useful to rewrite  (\ref{eom}) in terms of velocities:
 $$
\bfF^\al(x)=E_2\bigl(-\al_\tau+\frac{\bar{k}}{2\pi\imath}\p_x|\tau\bigr)\bfS^\al(x)\,.
 $$
They are the functionals on $L(\gg)$. Then (\ref{eom}) takes the form:
 \beq{bee}
\ka\p_\tau\wp^{-1}\bigl(\al_\tau+\frac{\bar{k}}{2\pi\imath}\p_x|\tau\bigr)\bfF^\al(x)=
\sum_{\ga\in\mZ^{(2)}_{N},\ga\neq\al}\bfF^{-\ga}(x)
\wp^{-1}\bigl(\al_\tau+\ga_\tau+\frac{\bar{k}}{2\pi\imath}\p_x|\tau\bigr)\bfF^{\al+\ga}(x)\,.
 \eq
In some limiting cases considered below (\ref{eom}) is simplified to some local equation.

%%%%%%%%%%%%%%%%%%%%%%%%%%%%%%%%%%%%%%%%%%%%%%%%%%%%%%%%

\subsubsection*{Perturbation with respect to $k$}
Assume that $k$ is a small parameter and consider expansion of the equation of motion in degrees  $k^j$. Using (\ref{keq}) we
can take $\bar{k}=\tau k$. Let
 $$
\bfS^{\al}(x,k)=\sum_{j\geq 0}\bfS_j^{\al}k^j\,,~~
\bfJ(k)=E_2(-\ga_\tau+ \frac{\tau k}{2\pi\imath}\p_x)=\sum_{j\geq 0}J_j\p_x^j\,,~
J_j=\frac{\p^jE_2(-\ga_\tau)(\tau k)^j}{(2\pi\imath)^jj!}\,.
 $$
Then we come to the system of equations
 \beq{peq}
 \begin{array}{ll}
0)&\ka\p_\tau\bfS_0(x)=[J_0\bfS(x),\bfS_0(x)]\,,     \\
1)&\ka\p_\tau\bfS_1(x)= [J_1\p_x\bfS_0(x),\bfS_0(x)]+[J_0\bfS_1(x),\bfS_0(x)]    +[J_0\bfS_0(x),\bfS_1(x)]\,,    \\
\ldots\\
j)&\ka\p_\tau\bfS_j(x)= [J_j\p_x^j\bfS_0(x),\bfS_0(x)]+[J_{j-1}\p_x^{j-1}\bfS_1(x),\bfS_0(x)]  +\ldots
  +[J_0\bfS_0(x),\bfS_j(x)]\,.    \\
  \end{array}
 \eq
The equation 0) describes non-autonomous $L(\SLN)$  Euler-Arnold top. The next equation is linear on the background of
solutions $\bfS_0(x)$. In this way we come to the tower of linear equations. The equation on level j) is linear on the background of solutions of the lower equations 0),1),...,(j-1))
 $\bfS_j=\bfS_j(\bfS_0,\bfS_1,\ldots,\bfS_{j-1})$.
For SL$(2,\mC)$ $J_{2i+1}=0$ (\ref{pv}) and $\p_x$ appears first time on the second level
 \beq{pert}
 \begin{array}{l}
  \ka\p_\tau\bfS_0(x)=[J_0\bfS_0(x),\bfS_0(x)]\,,~~J_0=(E_2(\tau/2),E_2((1+\tau)/2),E_2(1/2)) \,,\\
  \ka\p_\tau\bfS_1(x)=[J_0\bfS_1(x),\bfS_0(x)]    +[J_0\bfS_0(x),\bfS_1(x)]\,,   \\
   \ka\p_\tau\bfS_2(x)=[J_2\p_x^2\bfS_0(x),\bfS_0(x)]
      +[J_0\bfS_2(x),\bfS_0(x)] +[J_0\bfS_0(x),\bfS_2(x)]
   +[J_0\bfS_1(x),\bfS_1(x)]\,.
 \end{array}
 \eq

%%%%%%%%%%%%%%%%%%%%%%%%%%%%%%%%%%%%%%%%%%%%%%%%%%%%%

\subsubsection*{Lax equations}
In addition to the  connection $A$ we define the connection $\ka\p_\tau+M$. Thus, we consider the pair of operators
 $$
\left\{
 \begin{array}{l}
\ka\p_w+ k\p_x+L\,,\\
\ka\p_\tau+M\,.
   \end{array}
  \right.
 $$
%\aag{Here should be no  $\rho_0$.}
 \begin{predl}\label{leq}
The Lax equation
 \beq{lm}
\Bigl[\ka\p_\tau +M\,,\,
\ka\p_w+k\p_x+L\Bigr]=0
 \eq
%\aag{and here.}
is equivalent to the equations of motion (\ref{eom})-(\ref{keq}) for
 \beq{M}
M(x,w)=\frac{\ka}N\p_\tau\vartheta(w|\tau)+
\sum_{n\in\mZ}\sum_{\al\in\mZ^{(2)}_{N}}M^{\al}_n(w)T_n^\al\,,
 \eq
where
 $$
M^\al_n=S_n^{\al}Y_n^\al\,,~~Y_n^\al=\f1{2\pi\imath}
\bfe\Bigl( -\frac{w\al_2}N\Bigr)
f\Bigl(-\al_\tau+\bar{k}n,w\Bigr)\,,
 $$
and $f$ is defined in (\ref{A3c}).
 \end{predl}
\emph{Proof}.\vskip2mm \noindent{From (\ref{lm}) we have}
 $$
\overbrace{\ka\p_\tau L_n^\al}^{\underline{a}}-
\overbrace{(\ka\p_w+ kn)M^{\al}_{n}}^{\underline{b}}+
+ \overbrace{\frac{\ka}N\p_\tau\p_w\vartheta(w|\tau)
- \frac{\ka}N\p_\tau\p_w\vartheta(w|\tau)}^{\underline{c}}+
 $$
%\mo{In fact, the term $\underline{a}$ is $(\ka\p_\tau +2\pi\imath(w-\ti w)\frac{kn}{\rho_0})L_n^\al$.
%But in this case $\underline{2}=\underline{7}$ and these terms are not cancels.}
 \beq{lmn}
+
\overbrace{\sum_{\ga}\sum_{m}M^{\ga}_{m}L_{n-m}^{\al-\ga}\bfC(\ga,\al)}^{\underline{e}}=0\,,
 \eq
%\aag{Here should be no  $\rho_0$.}
where $L_n^\al=S_n^{\al}X_n^\al $ is defined in (\ref{ln}).
First of all, notice that $\underline{c}=0$. % and we consider the last terms.
Let us substitute $L^{\al}_n(w)$  and $M^{\al}_n(w)$ in (\ref{lmn}).
Taking into account the equation of motion (\ref{keq}) %\aag{Without  $\rho_0$.}
we find
 $$
\underline{a}=\ka\p_\tau L_n^\al=
\ka\p_\tau S^{\al}_nX_n^\al(w,\ti w)+\ka S^{\al}_n\p_\tau X_n^\al(w,\ti w)=
 $$
 $$
\underbrace{\ka\p_\tau S^{\al}_nX_n^\al(w,\ti w)}_{\underline 1}+
(\underbrace{-\ka\frac{\al_2}N}_{\underline{2}}+
\underbrace{kn}_{\underline{3}})S_n^\al Y_n^\al(w,\ti w)+
 $$
 $$
\underbrace{S^{\al}_n\bfe\Bigl(
-\frac{w\al_2}N\Bigr)\ka\p_\tau\phi\Bigl(-\al_\tau+\bar{k}n,w\Bigr)}_{\underline{4}}\,.
 $$
 On the other hand
  $$
\underline{b}=( \ka\p_w+kn )M^{\al}_{n}=
  $$
  $$
S^{\al}_n\Biggl( \underbrace{-\ka
\frac{\al_2}NY_n^\al(w,\ti w)}_{\underline{5}}+
 \underbrace{\frac{\ka}{2\pi\imath}\bfe\Bigl(
-\frac{w\al_2}N\Bigr)\p_wf\Bigl(-\al_\tau+\bar{k}n,w\Bigr)}_{\underline{6}}\Biggr)
  $$
  $$
 + \underbrace{knS^{\al}_{n}Y_n^\al(w,\ti w)}_{\underline{7}}\,.
  $$
Finally,
 $$
\underline{e}=\sum_{\ga}\sum_{m}M^{\ga}_{m}L_{n-m}^{\al-\ga}\bfC(\ga,\al)=
\sum_{\ga}\sum_{m}S^{-\ga}_{-m}S^{\al+\ga}_{n+m}X^{-\ga}_{-m}Y^{\al+\ga}_{n+m}\bfC(\al,\ga)=
 $$
 $$
\bfe\Bigl(-\frac{w\al_2}N\Bigr)
\sum_{\ga}\sum_{m}S^{-\ga}_{-m}S^{\al+\ga}_{n+m}\phi(\ga_\tau-\bar{k}m,w)
f\Bigl(-(\ga+\al)_\tau+\bar{k}(n+m),w\Bigr)
\bfC(\al,\ga)\stackrel{(\ref{d2})}{=}
 $$
 $$
\underbrace{X_n^\al\sum_{\ga}\sum_{m}S^{-\ga}_{-m}S^{\al+\ga}_{n+m}\Biggl (E_2
\Bigl(-\al_\tau+\ga_\tau+\,\bar{k}(n-m)\Bigr)-E_2\bigl(-\ga_\tau+\,\bar{k}m\bigr)\Bigr)
\Biggr)\bfC(\al,\ga)}_{\underline{8}}\,.
 $$
Comparing the terms in these expressions we get
 $$
 \begin{array}{ll}
   \underline{1}+\underline{8}=0 ~({\rm is~eq.~of~motion}~(\ref{efm}))\,,
   &\underline{4}+\underline{6}=0  ~({\rm the ~heat ~eq. ~} (\ref{A.4b}))\,,\\
     \underline{2}+\underline{5}=0\,, &
   \underline{3}+\underline{7}=0\,.\\
   \end{array}
   $$
In this way we come to the equations of motion (\ref{efm}). $\blacksquare$

%%%%%%%%%%%%%%%%%%%%%%%%%%%%%%%%%%%%%%%%%%%%%%%%%%%%%

\subsubsection*{Limit to integrable systems}\label{idi1}
Applying the general construction %of Section 3.6
we obtain  the integrable equations from (\ref{eom})
 $$
\p_t \bfS^\al(x)=\sum_{\ga}\Bigl(E_2\bigl(-\ga_\tau+\frac{\bar{k}}
{2\pi\imath}\p_x\bigr)\bfS^{-\ga}\Bigr)\bfS^{\al+\ga}\bfC(\al,\ga)\,, ~~\p_t\bar{k}=k/\rho_0\,,
 $$
 $$
\p_t \bfS(x)=[\bfJ\cdot\bfS,\bfS]\,,~~
\bfJ=\sum_{\ga}E_2\bigl(-\ga_\tau+\frac{kt}{2\pi\imath}\p_x\bigr)T^\ga\,,
 $$
%\aag{Again the question about $\rho_0$}
or in terms of the Fourier modes
 \beq{nlll}
\p_t S_n^\al=
\sum_{\ga}\sum_{m}S^{-\ga}_{-m}S^{\al+\ga}_{n+m}E_2\bigl(-\ga_\tau+kt m\bigr)
\bfC(\al,\ga)\,.
 \eq
These equations are the two-dimensional (nonlocal) version of the integrable Euler-Arnold $\SLN$-top \cite{RSTS,KLO}.
As above consider the perturbation of these equations for small $k$. We come to the system (\ref{peq}), where $\ka\p_\tau$ is
replaced with $\p_t$. Again, for $\slt$ we obtain
 \beq{nleat}
 \begin{array}{ll}
1)&  \p_t\bfS_0(x)=[J_0\bfS_0(x),\bfS_0(x)]\,,~~J_0=(E_2(\tau/2),E_2((1+\tau)/2),E_2(1/2)) \,,\\
2)&  \p_t\bfS_1(x)=[J_0\bfS_1(x),\bfS_0(x)]    +[J_0\bfS_0(x),\bfS_1(x)]\,,   \\
 3)&  \p_t\bfS_2(x)=[J_2\p_x^2\bfS_0(x),\bfS_0(x)]
      +[J_0\bfS_2(x),\bfS_0(x)] +[J_0\bfS_0(x),\bfS_2(x)]
   +[J_0\bfS_1(x),\bfS_1(x)]\,,\\
   j)& \p_t\bfS_j(x)= [J_j\p_x^j\bfS_0(x),\bfS_0(x)]+[J_{j-1}\p_x^{j-1}\bfS_1(x),\bfS_0(x)]
    +\ldots
  +[J_0\bfS_0(x),\bfS_j(x)]\,,
 \end{array}
 \eq
where $J_j=0$ for $j$ odd. There exists a local two-dimensional version of the
SL$(\mC,2)$ Euler top. It is the famous Landau-Lifshitz equation \cite{Sk}
 $$
 \p_t\bfS_0(x)=[J_0\bfS_0(x),\bfS_0(x)]+[\bfS_0(x),\p^2\bfS_0]\,,
 $$
The advantage of (\ref{nleat}) in compare with the Landau-Lifshitz
equation is the existence of the  isomonodromic version (\ref{pert})
of the former system while for the latter system such a version is
unknown.

 In the integrable case we come to the Lax equation (\ref{lae}). The linear system
 (\ref{lis})  assumes the
form
 $$
\left\{
 \begin{array}{l}
(k\p_x+L)Y=0\,,\\
(\p_t+M)Y=0\,,
   \end{array}
  \right.
 $$
where $L(x,z,\bz)=\sum_{n\in\mZ}\sum_{\al\in\mZ^{(2)}_{N}}L^{\al}_n(z,\bz)T_n^\al$,
 $$
L^{\al}_n(z,\bz)=\bfS^{\al}_nX_n^\al\,,~X_n^\al=\bfe
\Bigl(-\frac{z\al_2}N\Bigr)
\phi\Bigl(-\al_\tau+t kn,z\Bigr)\,,
 $$
$M(x,z,\bz)=\sum_{n\in\mZ}\sum_{\al\in\mZ^{(2)}_{N}} M^{\al}_n(z,\bz)T_n^\al$,
 $$
M^\al_n=S_n^{\al}Y_n^\al\,,~~Y_n^\al=\f1{2\pi\imath}
\bfe\Bigl(-\frac{z\al_2}N\Bigr)f\Bigl(-\al_\tau+t kn,z\Bigr)\,.
 $$
The Lax equation assumes the form:
 $$
\p_tL- k\p_xM+[M,L]=0\,.
 $$
%\aag{$\rho_0$?}
It is a simplified version of Proposition \ref{leq}.

%%%%%%%%%%%%%%%%%%%%%%%%%%%%%%%%%%%%%%%%%%%%%%%%%%%%%%%%%%%%%%%%%%%%%%%%%%%%%%%%%%

\subsection{Trigonometric and rational limits}
Let $\tau=\tau_1+\tau_2$ and $\Im m\tau_2\to+\infty$. Then using (\ref{c5}) we derive
the trigonometric degeneration of the equations of motion (\ref{efm})
 \beq{eomt}
\ka\p_\tau \bfS^\al(x)=\sum_{\ga\in\mZ^{(2)}_{N},\ga\neq\al}\Bigl(
\pi^2\sin^{-2}\pi\bigl(
-\ga_\tau+\frac{\bar{k}}{2\pi\imath}\p_x\bigr)\bfS^{-\ga}(x)\Bigr)\bfS^{\al+\ga}(x)
\bfC(\al,\ga)\,,
 \eq
 where $\tau$ stands for $\tau_1$.
 This equation is the compatibility condition for the isomonodromy problem
  \beq{itr}
 \left\{
  \begin{array}{l}
   (\ka\p_w+k\p_x+L^{tr}(\bfS,w))\Psi(w)=0\,, \\
    (\ka\p\tau+M^{tr}(\bfS,w))\Psi(w)=0\,,
  \end{array}
 \right.
  \eq
 Let
  $$
 T_{n,j}^\al=\bfe(nx+jy)T^\al\,,
  $$
 where $T^\al$ is defined in (\ref{ta2}). Then the trigonometric operators take the form
  $$
 L^{tr}=\sum_{n\in\mZ}\sum_{\al\in\mZ^{(2)}_{N},j\in\mZ} L^{tr,\al}_{n,j}T_{n,j}^\al\,,~~
 M^{tr}=\sum_{n\in\mZ}\sum_{\al\in\mZ^{(2)}_{N},j\in\mZ} M^{tr,\al}_{n,j}T_{n,j}^\al\,,
  $$
  $$
 \begin{array}{ll}
L^{tr,\al}_{n,j}=S_{n,j}^\al X_{n,j}^\al &X_{n,j}^\al=\bfe
\Bigl(
(-\frac{w\al_2}N-jw\Bigr)
\pi\Bigl(\cot\pi(-\al_\tau-j\tau+\bar{k}n)+\cot(\pi w)\Bigr)  \,, \\
 M^{tr,\al}_{n,j}=S_{n,j}^\al Y_{n,j}^\al & Y_{n,j}^\al=  -\bfe\Bigl( -\frac{w\al_2}N-jw\Bigr)
\pi^2\sin^{-2}\pi\Bigl(-\al_\tau-j\tau+\bar{k}n\Bigr)\,,
                 \end{array}
  $$
 where we use the forms of $\phi^{tr}$ and $f^{tr}$  (\ref{c4}), (\ref{c4a}). The compatibility
of (\ref{itr}) can be checked directly.
Rewrite (\ref{eomt}) in the form (\ref{bee}). Define the second order difference operator
 \beq{dtr}
\cD^{tr}_x(\tau) \bfF^\al(x)=\f1{4}\left(\bfe(-\al_\tau)\bfF^\al(x+\bar k)+
\bfe(\al_\tau)\bfF^\al(x-\bar k)-\bfF^\al(x)\right)\,.
 \eq
It plays the role of the inertia operator.
Then (\ref{eomt}) assumes the form of the difference equation in $x$
 \beq{emtf}
\ka\p_\tau \cD^{tr,\al}_x(\tau) \bfF^\al(x)=\sum_{\ga\in\mZ^{(2)}_{N},\ga\neq\al}
\bfF^{-\ga}(x)\cD^{tr,\al}_x(\tau)\bfF^{\al+\ga}(x)
\bfC(\al,\ga)\,.
 \eq
For the $\slt$ case we have three fields (see (\ref{100})) arranged in the three vector
 $$
\vec\bfF=(\bfF^1=\bfF^{(1,0)}\,,\bfF^1=\bfF^{(1,1)}\,,\bfF^3=\bfF^{(0,1)})\,,
 $$
and the vector $\overrightarrow{D^{tr}\bfF}=(\cD^{tr,\al}_x(\tau)\bfF^\al)$. Then  the $\slt$ commutator becomes the wedge
product and (\ref{emtf}) is simplified to
 \beq{emtf1}
 \begin{array}{|c|}
  \hline\\ \ka\p_\tau (\overrightarrow{D^{tr}\bfF})(x,\tau)=
\vec\bfF(x,\tau)\wedge(\overrightarrow{D^{tr}\bfF})(x,\tau)\\ \ \\
\hline
  \end{array}
 \eq

\bigskip

In the rational case we come to the equations
 \beq{eomr}
\ka\p_\tau \bfS^\al(x)=\sum_{\ga\in\mZ^{(2)}_{N},\ga\neq\al}\Bigl(\bigl(
-\ga_\tau+\frac{\bar{k}}{2\pi\imath}\p_x\bigr)^{-2}\bfS^{-\ga}(x)\Bigr)\bfS^{\al+\ga}(x)
\bfC(\al,\ga)\,,
 \eq
The corresponding isomonodromy problem is defined by the operators
 $$
 \begin{array}{ll}
L^{r,\al}_n=S_n^\al X_n^\al &X_n^\al=\bfe
\Bigl(-\frac{w\al_2}N\Bigr)
\Bigl(\f1{-\al_\tau+\bar{k}n}+\f1{ w}\Bigr)  \,, \\
 M^{r,\al}_n=S_n^\al Y_n^\al & Y_n^\al=  -\bfe\Bigl( -\frac{w\al_2}N\Bigr)
\Bigl(-\al_\tau+\bar{k}n\Bigr)^{-2}\,,
                 \end{array}
  $$
The rational inertia operator takes the form:
 \beq{dr}
\cD^{r,\al}_x(\tau) \bfF^\al(x)=
\left(-\al_\tau+\frac{\bar k}{2\pi\imath}\p_x\right)^2\bfF^\al(x)\,.
 \eq
Then (\ref{eomr}) can be rewritten in terms of the angular velocities
 \beq{emtfr}
\ka\p_\tau \cD^{r,\al}_x(\tau) \bfF^\al(x)=\sum_{\ga\in\mZ^{(2)}_{N},\ga\neq\al}
\bfF^{-\ga}(x)\cD^{r,\al}_x(\tau)\bfF^{\al+\ga}(x) \bfC(\al,\ga)\,.
 \eq
For $\slt$ case we introduce as above two three-dimensional vectors $\vec{\bfF}=(\bfF^\al)$ and
$\overrightarrow{D^{r}\bfF}=(\cD^{r,\al}_x(\tau)\bfF^\al)$. Then as in the trigonometric case we have
 \beq{emtfr2}
 \begin{array}{|c|}
  \hline\\ \ka\p_\tau\overrightarrow{D^{r}\bfF} (x,\tau)=
\vec\bfF(x)\wedge\overrightarrow{D^{r}\bfF} (x,\tau)\\ \ \\
\hline
  \end{array}
 \eq

%%%%%%%%%%%%%%%%%%%%%%%%%%%%%%%%%%%%%%%%%%%%%%%%%%%%%%%%%%%%%%%%%%%%%%%%%%%%%%%%%5

\subsubsection{Special trigonometric and rational limits for the $\slt$ case}

In the $\slt$ case another interesting limit can be described explicitly using the results of  \cite{Sm}. We consider $\slt$
Lax operators in the basis of the sigma matrices
 \begin{equation}
\nonumber
L=\sum_{n\in\mZ}\bfe(nx)
\mat{S^3_n\,\varphi_{10}\brc{\bar{k}n,w}}{S^1_n\,\varphi_{01}\brc{\bar{k}n,w}-\imath S^2_n\,\varphi_{11}
\brc{\bar{k}n,w}}{S^1_n\,\varphi_{01}\brc{\bar{k}n,w}+\imath S^2_n\,\varphi_{11}
\brc{\bar{k}n,w}}{-S^3_n\,\varphi_{10}\brc{\bar{k}n,w}}+
 \end{equation}
 \begin{equation}
\label{eq:sl2Lax}
+\frac{\ka}2E_1(w|\tau) Id,
 \end{equation}
 \begin{equation}
\nonumber
M=\frac1{2\pi\imath}\sum_{n\in\mZ}\bfe(nx)
\mat{S^3_n\,f_{10}\brc{\bar{k}n,w}}{S^1_n\,f_{01}\brc{\bar{k}n,w}-\imath S^2_n\,f_{11}
\brc{\bar{k}n,w}}{S^1_n\,f_{01}\brc{\bar{k}n,w}+\imath S^2_n\,f_{11}\brc{\bar{k}n,w}}{-S^3_n\,f_{10}\brc{\bar{k}n,w}}+
 \end{equation}
 \begin{equation}
\label{eq:sl2Lax2} +\frac{\ka}2\p_\tau\vartheta(w|\tau) Id,
 \end{equation}
where
 \begin{equation}
\nonumber
\varphi_{10}\brc{\bar{k}n,w}=\varphi\brc{-\frac12+\bar{k}n,w},\quad
\varphi_{01}\brc{\bar{k}n,w}=\varphi\brc{-\frac{\tau}2+\bar{k}n,w},
 \end{equation}
 \begin{equation}
\nonumber
\varphi_{11}\brc{\bar{k}n,w}=\varphi\brc{-\frac{1+\tau}2+\bar{k}n,w},
 \end{equation}
and functions $f_{ij}$ are related to the function $f$ in the same way.
Hamiltonian for the $\slt$ case can be written as
 \begin{equation}
\label{eq:sl2Ham}
H=k\lambda -\frac1{4\pi\imath} \sum_{\al=1,2,3}\sum_{n\in\mZ}S^{\al}_nS^{\al}_{-n}E_2\brc{-\al_{\tau}+\bar{k}n}.
 \end{equation}
The Poisson brackets are of the form
 \begin{equation}
\label{eq:sl2bra}
\bfi{S^i_n,S^j_m}=2\imath\epsilon_{ijk}S^k_{n+m},\quad \bfi{\lambda,\,\bar{k}}=1.
 \end{equation}
In what follows we are going to use the Chevalley basis:
 \begin{equation}
\nonumber
S^+_{n}=\frac12\brc{S^1_{n}+\imath S^2_{n}},\quad S^-_{n}=\frac12\brc{S^1_{n}-\imath S^2_{n}},
 \end{equation}
 \begin{equation}
\label{eq:sl2chev}
\bfi{S^3_n,S^{\pm}_m}=\pm2S^{\pm}_{n+m},\quad \bfi{S^+_{n},S^{-}_{m}}=S^3_{n+m}.
 \end{equation}
One can also write down the Poisson brackets in terms of the field variables:
 \begin{equation}
\nonumber
\bfi{\textbf{S}^3(x),\textbf{S}^{\pm}(y)}=\pm2\textbf{S}^{\pm}(y)\delta(x-y),\quad \bfi{\textbf{S}^+(x),\textbf{S}^{-}(y)}=\textbf{S}^3(y)\delta(x-y).
 \end{equation}
To perform the non-autonomous trigonometric limit, we decompose the parameter of the elliptic curve as
 \begin{equation}
\nonumber
\tau=\tau_1+\tau_2,
 \end{equation}
where $\tau_1$ plays the role of time of the limiting system and $\tau_2$ gives the
trigonometric limit $\Im m\tau_2\rightarrow+\infty$.
Then we introduce time-independent change of variables
 \begin{equation}
\nonumber
S^3_n\rightarrow S^3_n,\quad S^+_n\rightarrow q_2^{1/4} S^+_n,\quad
S^-_n\rightarrow q_2^{-1/4} S^-_n,
 \end{equation}
where $q_2 \equiv\bfe(\tau_2)$. Notice that the Poisson structure (\ref{eq:sl2chev}) is preserved under this transformation.

The trigonometric Lax operator can be derived as the limit of the gauge transformed elliptic Lax operator
 \begin{equation}
\nonumber
L^T=\lim_{q_2\rightarrow0}A^T(q_2)\,L\,A^T(q_2)^{-1},
 \end{equation}
where
 \begin{equation}
\nonumber
A^T(q_2)=\mat{q_2^{1/8}}{0}{0}{q_2^{-1/8}}.
 \end{equation}
This gives
 \begin{equation}
\nonumber
L^T=\pi\sum_{n\in\mZ}\mat{\brc{\cot(\pi w)-\tan(\pi\bar{k}n)}S^3_n}{\dfrac{2}{\sin(\pi w)}S^-_n} {\dfrac{2}{\sin(\pi w)}S^+_n+8\sqrt{q_1}\sin\brc{\pi w+2\pi\bar{k}n}S^-_n}{-\brc{\cot(\pi w)-\tan(\pi\bar{k}n)}S^3_n}\bfe\brc{nx}+
 \end{equation}
 \begin{equation}
\nonumber
+\frac{\pi\ka}2\cot(\pi w) Id,
 \end{equation}
where $q_1 \equiv\bfe(\tau_1)$.
Trigonometric Hamiltonian is defined via
 \begin{equation}
\nonumber
H^T=\lim_{q_2\rightarrow0}H=k\lambda+\frac{\imath\pi}4\sum_{n\in\mZ}\brc{\frac{S^3_nS^3_{-n}}{\cos^2\brc{\pi\bar{k}n}}- 16\sqrt{q_1}\cos\brc{2\pi\bar{k}n}S^-_nS^-_{-n}},
 \end{equation}
or in terms of the field variables
 \begin{equation}
\nonumber
H^T=k\lambda+\frac{\imath\pi}4\oint\brc{\textbf{S}^3(x)\cos^{-2}\brc{\frac{\imath\bar{k}}2\partial_x}\textbf{S}^3(x)- 16\sqrt{q_1}\textbf{S}^-(x)\cos\brc{\imath\bar{k}\partial_x}\textbf{S}^-(x)}.
 \end{equation}
Applying the same gauge transformation as for the $L$-operator, we come to the trigonometric $M$-operator
 \begin{equation}
\nonumber
M^T=\lim_{q_2\rightarrow0}A^T(q_2)\,M\,A^T(q_2)^{-1}=\frac{\pi}{2\imath}\sum_{n\in\mZ}\mat{-\dfrac1{\cos^2\brc{\pi\bar{k}n}}S^3_n}{0}{16\sqrt{q_1}\cos\brc{\pi w+2\pi\bar{k}n}S^-_n}{\dfrac1{\cos^2\brc{\pi\bar{k}n}}S^3_n}\bfe\brc{nx}.
 \end{equation}
The Lax equation
 \begin{equation}
\nonumber
\kappa\partial_{\tau_1}L^T-\kappa\partial_w M^T-k\partial_x M^T=\left[L^T,M^T\right]
 \end{equation}
in the trigonometric case is equivalent to the Hamiltonian equations of motion
 \begin{equation}
\nonumber
\kappa\partial_{\tau_1}\bar{k}=k,\qquad
\kappa\partial_{\tau_1}S^3_n=\frac{16\pi}\imath\sqrt{q_1}\sum_{m\in\mZ}\cos\brc{2\pi\bar{k}m}S^-_{n-m}S^-_{m},
 \end{equation}
 \begin{equation}
\nonumber
\kappa\partial_{\tau_1}S^+_n=\pi\imath\sum_{m\in\mZ}\brc{\frac{S^+_{n-m}S^3_{m}} {\cos^2\brc{\pi\bar{k}m}}-8\sqrt{q_1}\cos\brc{2\pi\bar{k}m}S^3_{n-m}S^-_{m}},
 \end{equation}
 \begin{equation}
\nonumber
\kappa\partial_{\tau_1}S^-_n=\frac\pi\imath\sum_{m\in\mZ}\frac{S^-_{n-m}S^3_{m}} {\cos^2\brc{\pi\bar{k}m}}.
 \end{equation}
Using the field variables, one can rewrite the  equations of motion as follows:
 \begin{equation}
\nonumber
\kappa\partial_{\tau_1}\bar{k}=k,\qquad
\kappa\partial_{\tau_1}\textbf{S}^3(x)=\frac{16\pi}\imath\sqrt{q_1}\textbf{S}^-(x) \cos\brc{\imath\bar{k}\partial_x}\textbf{S}^-(x),
 \end{equation}
 \begin{equation}
\nonumber
\kappa\partial_{\tau_1}\textbf{S}^+(x)=\pi\imath\brc{\textbf{S}^+(x)\cos^{-2}\brc{\frac{\imath\bar{k}}2\partial_x}\textbf{S}^3(x) -8\sqrt{q_1}\textbf{S}^3(x)\cos\brc{\imath\bar{k}\partial_x}\textbf{S}^-(x)},
 \end{equation}
 \begin{equation}
\nonumber
\kappa\partial_{\tau_1}\textbf{S}^-(x)=\frac\pi\imath\textbf{S}^-(x)\cos^{-2} \brc{\frac{\imath\bar{k}}2\partial_x}\textbf{S}^3(x).
 \end{equation}
The zero Fourier modes are related to the canonical coordinates $u,\,v$ by the so-called bosonization formulas:
 \begin{equation}
\nonumber
S^3_0=-\frac{v}{\pi\tan\brc{2\pi u}}-\frac{\nu}{\sin^2\brc{2\pi u}},\quad
S^-_0=-\frac{v}{4\pi\sin\brc{2\pi u}}-\frac{\nu\cos\brc{2\pi u}}{4\sin^2\brc{2\pi u}},
 \end{equation}
 \begin{equation}
\nonumber
S^+_0=\frac{v\cos^2\brc{2\pi u}}{\pi\sin\brc{2\pi u}}+
\frac{\nu\cos\brc{2\pi u}\brc{1+\sin^2\brc{2\pi u}}} {\sin^2\brc{2\pi u}}.
 \end{equation}
Thus, in the zero modes we obtain a non-autonomous version of the trigonometric Calogero-Moser system with the following Hamiltonian:
 \begin{equation}
\nonumber
H^T_0=k\lambda +\frac{\imath}{4\pi}\brc{\frac{v^2}{\sin^2\brc{2\pi u}}\brc{\cos^2\brc{2\pi u}-\sqrt{q_1}}+\frac{\pi^2\nu^2}{\sin^4\brc{2\pi u}}\brc{1-\sqrt{q_1}\cos^2\brc{2\pi u}}}+
 \end{equation}
 \begin{equation}
\nonumber
+\frac{\imath\nu}{2}\frac{v\cot\brc{2\pi u}}{\sin^2\brc{2\pi u}}\brc{1-\sqrt{q_1}}.
 \end{equation}
The equations of motion for this system are equivalent to the following second-order differential equation:
 \begin{equation}
\nonumber
\frac{\rmd^2u}{\rmd\tau_1^2}\brc{\cos^2\brc{2\pi u}-\sqrt{q_1}}+2\pi
\brc{\frac{\rmd u}{\rmd \tau_1}}^2\cot\brc{2\pi u}\brc{1-\sqrt{q_1}}+
 \end{equation}
 \begin{equation}
\nonumber
+\pi\imath\frac{\rmd u}{\rmd\tau_1}\sqrt{q_1}+\frac{\pi}2\nu\brc{\nu+1}\cot\brc{2\pi u}\sqrt{q_1}=0.
 \end{equation}
The rational limit can be described in terms of the small parameter  $y\rightarrow0$ and the following transformations:
 \begin{equation}
\nonumber
S^3_n\rightarrow S^3_n+\frac4{y^2}S^-_n,\quad S^-_n\rightarrow\frac{S^-_n}{y^2},\quad
S^+_n\rightarrow y^2S^+_n -2S^3_n-\frac4{y^2}S^-_n,
 \end{equation}
 \begin{equation}
\nonumber
w\rightarrow\frac{y}{\pi}w,\quad \bar{k}\rightarrow\frac{y}{\pi}\bar{k},\quad \tau_1\rightarrow\frac{y^2}{\imath\pi}t_1,\quad k\rightarrow\frac{\pi}{y}k,\quad \lambda\rightarrow\frac{\pi}{y}\lambda.
 \end{equation}
Again, the Poisson structure (\ref{eq:sl2chev}) remains unchanged along with the canonical bracket
 \begin{equation}
\nonumber \bfi{\lambda,\,\bar{k}}=1\,.
 \end{equation}

The rational Lax operator is defined via
 \begin{equation}
\nonumber
L^R=\lim_{y\rightarrow0}\frac{y}{\pi}A^R(y)\,L^T\,A^R(y)^{-1},
 \end{equation}
where
 \begin{equation}
\nonumber
A^R(y)=\mat{y}{0}{2/y}{1/y}.
 \end{equation}
This gives
 \begin{equation}
\nonumber
L^R=\sum_{n\in\mZ}\mat{\dfrac1w S^3_n-2\brc{w+2\bar{k}n}S^-_n}{\dfrac2w S^-_n}{\dfrac2w S^+_n-2\brc{w+2\bar{k}n}S^3_n+\varphi_R\brc{\bar{k}n,w,t_1}S^-_n}{-\dfrac1w S^3_n+2\brc{w+2\bar{k}n}S^-_n}\bfe(nx)+
 \end{equation}
 \begin{equation}
\nonumber
+\frac{\ka}{2w}Id,
 \end{equation}
where
 \begin{equation}
\nonumber
\varphi_R\brc{\bar{k}n,w,t_1}=8\brc{w+2\bar{k}n}t_1-2\brc{8\bar{k}^3n^3+8\bar{k}^2n^2w +4\bar{k}nw^2+w^3}.
 \end{equation}
The rational Hamiltonian assumes the following form:
 \begin{equation}
\nonumber
H^R=\lim_{y\rightarrow0}\frac{y^2}{\imath\pi}\,H^T=-\pi\imath k\lambda+2\sum_{n\in\mZ}\brc{S^3_n S^-_{-n}+ \brc{5\bar{k}^2n^2-2t_1}S^-_{n}S^-_{-n}},
 \end{equation}
or in terms of the field variables
 \begin{equation}
\nonumber
H^R=-\pi\imath k\lambda+2\oint\brc{\textbf{S}^3(x)\textbf{S}^-(x)- \textbf{S}^-(x)\brc{2t_1+\frac{5\bar{k}^2}{4\pi^2}\,\partial^2_x}\textbf{S}^-(x)},
 \end{equation}
The rational $M$ operator is related to the trigonometric one as
 \begin{equation}
\nonumber
M^R=\lim_{y\rightarrow0}\frac{y^2}{\imath\pi}A^R(y)\,M^T\,A^R(y)^{-1},
 \end{equation}
which gives
 \begin{equation}
\nonumber
M^R=2\sum_{n\in\mZ}\mat{S^-_n}{0}{S^3_n-2\brc{2 t_1-w^2-4w\bar{k}n-6\bar{k}^2n^2}S^-_n}{-S^-_n}\bfe\brc{nx}.
 \end{equation}
The Lax equation
 \begin{equation}
\nonumber
\kappa\partial_{t_1}L^R-\kappa\partial_w M^R-k\partial_x M^R=\left[L^R,M^R\right]
 \end{equation}
is equivalent to the equations of motion
% \begin{equation}
%\nonumber
%\kappa\partial_{t_1}\bar{k}=-\pi\imath k,\quad
%\kappa\partial_{t_1}S^3_n=4\sum_{m\in\mZ}\brc{S^-_{n-m} S^3_{m}+ 2\brc{5\bar{k}^2m^2-2t_1}S^-_{n-m}S^-_{m}},
% \end{equation}
% \begin{equation}
%\nonumber
%\kappa\partial_{t_1}S^+_n=2\sum_{m\in\mZ}\brc{2S^+_{n-m} S^-_{m}-S^3_{n-m} S^3_{m}-2\brc{5\bar{k}^2m^2-2t_1}S^3_{n-m}S^-_{m}},
% \end{equation}
% \begin{equation}
%\nonumber
%\kappa\partial_{t_1}S^-_n=-4\sum_{m\in\mZ}S^-_{n-m} S^-_{m}.
% \end{equation}
 \beq{erl}
\left\{
 \begin{array}{l}
  \kappa\partial_{t_1}\textbf{S}^3(x)=4\textbf{S}^-(x) \textbf{S}^3(x)-
8\textbf{S}^-(x)\brc{2t_1+\frac{5\bar{k}^2}{4\pi^2}\,\partial^2_x}\textbf{S}^-(x)\,, \\
  \kappa\partial_{t_1}\textbf{S}^+(x)=4\textbf{S}^+(x)\textbf{S}^-(x)-
2\brc{\textbf{S}^3(x)}^2 +4\textbf{S}^3(x)\brc{2t_1+\frac{5\bar{k}^2}{4\pi^2}\,\partial^2_x} \textbf{S}^-(x), \\
\kappa\partial_{t_1}\textbf{S}^-(x)=-4\brc{\textbf{S}^-(x)}^2\,,~~
\kappa\partial_{t_1}\bar{k}=-\pi\imath k\,.
 \end{array}
\right.
 \eq
The Hamiltonian for the zero Fourier modes can be written in canonical coordinates $u,\,v$ as follows:
 \begin{equation}
\nonumber
H^R_0=v^2-\frac{t_1}{u^3}\brc{\nu+uv}v-\frac{\nu^2}{4u^2}\brc{1+\frac{t_1}{u^2}},
 \end{equation}
where $\nu=const$ and we use the following bosonization formulas:
 \begin{equation}
\nonumber
S^3_0=uv-\frac{\nu}2,\quad S^+_0=-\frac{u^3v}2+\frac{3\nu u^2}4,\quad S^-_0=\frac{v}{2u}+\frac{\nu}{4u^2}.
 \end{equation}
The equations of motion can be represented as the second-order differential equation
 \begin{equation}
\nonumber
\frac{\rmd^2u}{\rmd t_1^2}\brc{u^2-t_1}+\frac{\rmd u}{\rmd t_1}-
\brc{\frac{\rmd u}{\rmd t_1}}^2\frac{t_1}{u}+\frac{\nu}{u}\brc{\nu+1}=0
 \end{equation}
with the solution of the form
 \begin{equation}
\nonumber
4u^2=-C_1-4 C_1 \nu\brc{\nu+1}+16 \frac{C_2^2}{C_1} +4 t_1+16\frac{C_2}{C_1} t_1+\frac{4}{C_1}t_1^2.
 \end{equation}

\subsubsection{Scaling limit}

%General description
In this subsection we construct a limiting procedure based on generalizations \cite{AA3,AA}
 of the Inozemtsev limit \cite{Inoz0}. To define the procedure we decompose the parameter
 $\tau$ of the elliptic curve:
 \begin{equation*}
\tau=\tau_1+\tau_2,
 \end{equation*}
where $\tau_1$ plays the role of time of the limiting system and $\tau_2$ gives the
trigonometric limit $\Im m\tau_2\rightarrow+\infty$. Before taking the trigonometric limit
we shift the spectral parameter and coordinate on $S^1$
 \begin{equation}
\label{eq:LoopShifts}
w=\wt w+\frac{\tau}2,\qquad x=\wt x+\frac{\bar k}{2}
 \end{equation}
and scale the coordinates in the following way:
 \begin{equation}
\label{eq:LoopScalings} S_n^{\alpha_1,\alpha_2}=\wt S_n^{\alpha_1,\alpha_2} q_2^{-g(\alpha_2)},\quad q_2
\equiv\bfe(\tau_2),\quad g(\alpha_2)=\frac{1-\delta_{\alpha_2,0}}{2N}.
 \end{equation}
%
%
%Algebra
After scalings (\ref{eq:LoopScalings}) we obtain the contraction of the Poisson algebra in the limit $\Im
m\tau_2\rightarrow+\infty$
 \begin{equation*}
\bfi{\wt S_n^{\alpha_1,\alpha_2},\wt S_m^{\beta_1,\beta_2}}=
\frac N{\pi}\sin\brc{\frac{\pi}N\brc{\alpha_1\beta_2-\alpha_2\beta_1}}
\wt S_{n+m}^{\alpha_1+\beta_1,\alpha_2+\beta_2}\,q_2^{g(\al_2)+g(\be_2)-g(\al_2+\be_2)},
 \end{equation*}
where $\wt S_\alpha \equiv\wt S_{\alpha_1,\alpha_2}$, $\alpha\in\mZ^{(2)}_{N}$. Scaled  coordinates $\wt S_\alpha$ with the
Poisson brackets form the algebra in the limit of $\Im m\tau_2\rightarrow+\infty$ provided that
 \begin{equation}
\label{eq:loopscalingcond}
\forall\alpha_2,\beta_2\in\mZ:\quad g(\al_2)+g(\be_2)-g(\al_2+\be_2)\geqslant0.
 \end{equation}
For $g(\alpha_2)=\brc{1-\delta_{\alpha_20}}/(2N)$ the condition (\ref{eq:loopscalingcond}) is trivial and we can write down
all nonzero brackets corresponding to the equality
 in (\ref{eq:loopscalingcond})
 \begin{equation}
\label{eq:LoopLimitingAlgebra}
\bfi{\wt S_n^{\alpha_1,0},\wt S_m^{\beta_1,\beta_2}}=
\frac N{\pi}\sin\brc{\frac{\pi}N\alpha_1\beta_2}\wt S_{n+m}^{\alpha_1+\beta_1,\beta_2}.
 \end{equation}
We start from the $\slt$ case. The scalings of coordinates in the basis of the Pauli matrices can be written as follows:
 \begin{equation}
S^1_n\rightarrow S^1_n\,q_2^{-1/4},\quad S^2_n\rightarrow S^2_n\,q_2^{-1/4},\quad S^3_n\rightarrow S^3_n.
 \end{equation}
Then the Lax operator (\ref{eq:sl2Lax}) acquires the following form in the limit $\Im m\tau_2\rightarrow+\infty$:
 \begin{equation}
\nonumber
L^I=4\pi\sum_{n\in\mZ}L^I_n\bfe\brc{n\wt x},
 \end{equation}
where $L^I_n$ is $2\times 2$ matrix of the form
 {\small{\begin{equation}
\nonumber \mat{-\dfrac{\imath\,S^3_n}{4\cos\brc{\pi\bar{k}n}}} {\brc{\sin\brc{\pi\wt w+\pi\bar{k}n}S^1_n-\cos\brc{\pi\wt
w+\pi\bar{k}n}S^2_n}q_1^{1/4}} {\brc{\sin\brc{\pi\wt w+\pi\bar{k}n}S^1_n+\cos\brc{\pi\wt w+\pi\bar{k}n}S^2_n}q_1^{1/4}}
{\dfrac{\imath\,S^3_n}{4\cos\brc{\pi\bar{k}n}}}
 \end{equation}}}
and $q_1 \equiv\bfe\brc{\tau_1}$.
For the limit of the Hamiltonian (\ref{eq:sl2Ham}) we have
 \begin{equation}
\label{eq:sl2InHam}
H^I=k\lambda+\frac{\imath\pi}4\sum_{n\in\mZ}\brc{\frac{S^3_nS^3_{-n}}{\cos^2\brc{\pi\bar{k}n}}- 8\sqrt{q_1}\cos\brc{2\pi\bar{k}n}\brc{S^1_nS^1_{-n}-S^2_nS^2_{-n}}}.
 \end{equation}
The Poisson structure  (\ref{eq:sl2bra}) transforms as well. In addition to the other zero brackets we get
 \begin{equation}
\nonumber
\bfi{S^1_n,S^2_m}=0,\quad \bfi{S^1_n,S^3_m}=-2\imath S^2_{n+m},\quad \bfi{S^2_n,S^3_m}=2\imath S^1_{n+m},\quad \bfi{\lambda,\,\bar{k}}=1,
 \end{equation}
or in terms of the field variables
 \begin{equation}
\nonumber
\bfi{\textbf{S}^1(x),\textbf{S}^2(y)}=0,\quad \bfi{\textbf{S}^1(x),\textbf{S}^3(y)}=-2\imath\textbf{S}^2(y)\delta(x-y),
 \end{equation}
 \begin{equation}
\nonumber
\bfi{\textbf{S}^2(x),\textbf{S}^3(y)}=2\imath\textbf{S}^1(y)\delta(x-y).
 \end{equation}
The equations of motion
 \begin{equation}
\label{eq:Insl2eqm0}
\kappa\partial_{\tau_1}\bar{k}=k,\qquad
\kappa\partial_{\tau_1}S^3_n=-8\pi\sqrt{q_1}\sum_{m\in\mZ}\cos\brc{2\pi\bar{k}m} \brc{S^1_{n-m}S^2_{m}+S^1_{m}S^2_{n-m}},
 \end{equation}
 \begin{equation}
\label{eq:Insl2eqm}
\kappa\partial_{\tau_1}S^1_n=-\pi\sum_{m\in\mZ}\frac{S^2_{n-m}S^3_{m}} {\cos^2\brc{\pi\bar{k}m}},\qquad
\kappa\partial_{\tau_1}S^2_n=\pi\sum_{m\in\mZ}\frac{S^1_{n-m}S^3_{m}}{\cos^2\brc{\pi\bar{k}m}}
 \end{equation}
can be represented in the form of the Lax equation
 \begin{equation}
\nonumber
\kappa\partial_{\tau}L^I-\kappa\partial_{\wt w}M^I-k\partial_{\wt x} M^I=\left[L^I,M^I\right],
 \end{equation}
with
 {\small{\begin{equation}
\nonumber M^I=\pi^2\sum_{n\in\mZ} \mat{-\dfrac{S^3_n}{\cos^2\brc{\pi\bar{k}n}}} {4\,\bfe\brc{-\dfrac{\wt
w}2-\bar{k}n}\brc{S^1_n+\imath S^2_n}q_1^{1/4}} {4\,\bfe\brc{-\dfrac{\wt w}2-\bar{k}n}\brc{S^1_n-\imath S^2_n}q_1^{1/4}}
{\dfrac{S^3_n}{\cos^2\brc{\pi\bar{k}n}}}\bfe\brc{n\wt x}.
 \end{equation}}}
Equations (\ref{eq:Insl2eqm}) can be simplified by the following change of coordinate:
 \begin{equation}
\nonumber
S^3_n=\cos^2\brc{\pi\bar{k}n}\wt S^3_n.
 \end{equation}
Then  in terms of the field variables the Hamiltonian (\ref{eq:sl2InHam}) and equations
of motion (\ref{eq:Insl2eqm0}), (\ref{eq:Insl2eqm}) assume the following form:
 \begin{equation}
\nonumber
H^I=k\lambda-2\imath\pi\sqrt{q_1}\oint\brc{\textbf{S}^1(x)\cos\brc{\imath\bar{k}\partial_{\wt x}}\textbf{S}^1(x) -\textbf{S}^2(x)\cos\brc{\imath\bar{k}\partial_{\wt x}}\textbf{S}^2(x)}+
 \end{equation}
 \begin{equation}
\nonumber
+\frac{\imath\pi}4\oint\widetilde{\textbf{S}}^3(x) \cos^2\brc{\frac{\imath\bar{k}}2\partial_{\wt x}}\widetilde{\textbf{S}}^3(x),
 \end{equation}
 \begin{equation}
\nonumber
\kappa\partial_{\tau_1}\brc{1+\cos\brc{\imath\bar{k}\partial_{\wt x}}}\widetilde{\textbf{S}}^3 =-16\pi\sqrt{q_1}\brc{\textbf{S}^1\cos\brc{\imath\bar{k}\partial_{\wt x}}\textbf{S}^2+ \textbf{S}^2\cos\brc{\imath\bar{k}\partial_{\wt x}}\textbf{S}^1},
 \end{equation}
 \begin{equation}
\nonumber
\kappa\partial_{\tau_1}\textbf{S}^1=-\pi\textbf{S}^2_{n-m}\widetilde{\textbf{S}}^3_{m},\qquad
\kappa\partial_{\tau_1}\textbf{S}^2=\pi\textbf{S}^1_{n-m}\widetilde{\textbf{S}}^3_{m}.
 \end{equation}
The Hamiltonian for the zero Fourier modes can be written in canonical coordinates $u,v$:
 \begin{equation}
\nonumber
H^I_0=k\lambda-\frac{\imath}{2\pi}\brc{\frac{v^2}2+4M^2\pi^2\sqrt{q_1}\cos\brc{4\pi u}},
 \end{equation}
where we use the bosonization formulas of the form
 \begin{equation}
\nonumber
S^1_0=M\cos\brc{2\pi u},\quad S^2_0=-M\sin\brc{2\pi u},\quad S^3_0=\frac{\imath v}\pi.
 \end{equation}
Equations of motion for the coordinates $u,v$ in the form of the second-order differential equation
 \begin{equation}
\nonumber
\frac{\rmd^2 u}{\rmd\tau_1^2}=-4M^2\pi\sqrt{q_1}\sin\brc{4\pi u}
 \end{equation}
are equivalent to a particular case of the Painlev\'{e} III equation:
 \begin{equation}
\nonumber
\frac1t\frac{\rmd}{\rmd t}t \frac{\rmd\brc{4\pi u}}{\rmd t}=\sin\brc{4\pi u},
 \end{equation}
where $t=8M\bfe\brc{\tau_1/4}$.
%
%Hamiltonian
The Hamiltonian of the limiting system in the case of $\textrm{sl}\brc{N>2,\mathbb C}$ has the following form:
 \begin{equation}
\label{eq:LoopLimitingHam}
H_{\tau_1}=k\lambda+\pi^2\sum_{\alpha_1\neq 0,n}\dfrac{\wt S^{\alpha_1,0}_n
\wt S^{-\alpha_1,0}_{-n}}{\sin^2\brc{\pi\brc{\frac{\alpha_1}N-\bar{k}n}}}-
 8\pi^2q_1^{\frac 1N} \sum_{\alpha_1,n}\rme\brc{\dfrac{\alpha_1}N-
 \bar{k}n}\wt S^{\alpha_1,1}_n \wt S^{-\alpha_1,-1}_{-n}.
 \end{equation}
Its worth noting that the Hamiltonian (\ref{eq:LoopLimitingHam}) contains only coordinates of the form
$S_n^{\alpha_1,\alpha_2},\;\alpha_2=0,\pm1$. Other elements have simple linear dynamics which can be integrated once we know
the dynamics of the coordinates included in the Hamiltonian.
%Equations of motion
The equations of motion of the limiting system can be presented as the Hamiltonian equations with respect to the brackets
(\ref{eq:LoopLimitingAlgebra}):
 \begin{subequations}
 \begin{eqnarray}
\kappa\partial_{\tau}\wt S^{\gamma_1,0}_m&=&8\pi Nq_1^{\frac 1N}
\sum_{\alpha_1,n} \rme\brc{\dfrac{\alpha_1}N-\bar{k}n}
\sin\brc{\dfrac{\pi \gamma_1}N}\brc{\wt S^{\alpha_1+\gamma_1,1}_{n+m}
\wt S^{-\alpha_1,-1}_{-n}-\wt S^{\alpha_1,1}_n\wt S^{\gamma_1-\alpha_1,-1}_{m-n}},\\
\kappa\partial_{\tau}\wt S^{\gamma_1,1}_m&=&2\pi N\sum_{\alpha_1,n}
\dfrac{\sin\brc{\frac{\pi\alpha_1}N}} {\sin^2\brc{\pi
\brc{\frac{\alpha_1}N-\bar{k}n}}}\wt S^{\alpha_1+\gamma_1,1}_{n+m}\wt S^{-\alpha_1,0}_{-n},\\
\kappa\partial_{\tau}\wt S^{\gamma_1,-1}_m&=&-2\pi N
\sum_{\alpha_1,n}\dfrac{\sin\brc{\frac{\pi\alpha_1}N}}
 {\sin^2\brc{\pi\brc{\frac{\alpha_1}N-\bar{k}n}}}\wt S^{\alpha_1+\gamma_1,-1}_{n+m}\wt S^{-\alpha_1,0}_{-n}.
 \end{eqnarray}
 \end{subequations}
%
%Lax representation
The equations of motion of the original system admit the Lax representation (\ref{lm}) which is equivalent to
 \begin{equation}
\label{eq:LoopOriginalLax}
\kappa\partial_{\tau}L-\kappa\partial_w M-k\partial_x M=\left[L,M\right].
 \end{equation}
Since substitutions (\ref{eq:LoopShifts}) are time-dependent, equation (\ref{eq:LoopOriginalLax}) transforms as follows
% \begin{subequations}
 \begin{equation}
\kappa\partial_{\tau}L-\kappa\partial_{\wt w}(M+\frac12L)-k\partial_{\wt x} (M+\frac12L)=\left[L,M\right],
 \end{equation}
where $L=L\brc{\textbf{S}(\wt x+\bar k/2),\wt w+\tau/2,\tau}$,
$M=M\brc{\textbf{S}(\wt x+\bar k/2),\wt w+\tau/2,\tau}$, and we use the following properties:
 \begin{eqnarray*}
&&\left(\partial_{\wt x}M\right)_{\tau,\wt w}=\left(\partial_xM\right)_{\tau,w},\quad
\left(\partial_{\wt w}L\right)_{\tau,\wt x}=\left(\partial_wL\right)_{\tau,x},\\
&&\left(\partial_{\tau}L\right)_{\wt w,\wt x}=
\left(\partial_{\tau}L\right)_{w,x}+\frac12\left(\partial_w L\right)_{\tau,x}+\frac1{4\pi\imath}
\left(\partial_x L\right)_{\tau,w}\partial_{\tau}\bar k.\\
 \end{eqnarray*}
% \end{subequations}
Thus, the Lax pair of the limiting system is defined via
 \beqnl
\wt L=\lim_{\Im m\tau_2\rightarrow+\infty}L,\quad \wt M=\lim_{\Im m\tau_2\rightarrow+\infty} \brc{M+\dfrac12 L}
 \eq
and the Lax equation assumes the form
 \begin{equation*}
\kappa\partial_{\tau}\wt L-\kappa\partial_{\wt w}\wt M-k\partial_{\wt x}\wt M=\left[\wt L,\wt M\right],
 \end{equation*}
where
 \begin{eqnarray*}
\wt L&=&-\pi\sum_{n,\in\mathbb Z}\sum_{\alpha_1=1}^{N-1}\wt S_n^{\alpha_1,0}
\frac{\rme\left(\frac{\alpha_1}{2N}\right)} {\sin\brc{\pi\brc{\frac{\alpha_1}N-\bar kn}}}\rme(n\wt x)T^{\alpha_1,0}+\\
&&+2\pi\imath\sum_{n\in\mathbb Z}\sum_{\alpha_1=0}^{N-1}\wt S_n^{\alpha_1,1}
\rme\left(\frac{\alpha_1}N-\frac{\wt w}N-\frac{\bar kn}{2}\right) q_1^{\frac1{2N}}\rme(n\wt x)T^{\alpha_1,1}-\\
&&-2\pi\imath\sum_{n\in\mathbb Z}\sum_{\alpha_1=0}^{N-1}\wt S_n^{\alpha_1,-1}
\rme\brc{\frac{\bar kn}{2}+\frac{\wt w}N}q_1^{\frac1{2N}}\rme(n\wt x)T^{\alpha_1,-1},
 \end{eqnarray*}
 \begin{eqnarray*}
\wt M&=&\frac{\imath\pi}4\sum_{n\in\mathbb Z}\sum_{\alpha_1=1}^{N-1}\wt S_n^{\alpha_1,0}
\frac{\rme\brc{\frac{\bar kn}{2}}+\rme\brc{\frac{\alpha_1}N-\frac{\bar kn}2}}
{\sin^2\brc{\pi\brc{\frac{\alpha_1}N-\bar kn}}}\rme\brc{n\wt x}T^{\alpha_1,0}-\\
&&-\imath\pi\sum_{n\in\mathbb Z}\sum_{\alpha_1=0}^{N-1}\wt S_n^{\alpha_1,1}
\rme\left(\frac{\alpha_1}N-\frac{\wt w}N-\rme\brc{\frac{\bar kn}{2}}\right) q_1^{\frac1{2N}}\rme(n\wt x)T^{\alpha_1,1}-\\
&&-\imath\pi\sum_{n\in\mathbb Z}\sum_{\alpha_1=0}^{N-1}\wt S_n^{\alpha_1,-1}
\rme\brc{\frac{\bar kn}{2}+\frac{\wt w}N}q_1^{\frac1{2N}}\rme(n\wt x)T^{\alpha_1,-1}.
 \end{eqnarray*}

%%%%%%%%%%%%%%%%%%%%%%%%%%%%%%%%%%%%%%%%%%%%%%%%%%%%%%%%%%%%%%%%%%%%%%%%%%%%

\subsection{Non-autonomous Zhukovsky-Volterra gyrostat field theory}
Up to now we considered the Painlev\'e VI field theory depending essentially on one constant corresponding to the
finite-dimensional case (\ref{rr101}) and its $\SLN$ generalization (\ref{eat}), (\ref{it202}). Here we give the
field-theoretical generalization of the complete Painlev\'e VI  equation (\ref{rr1}) in the form (\ref{eu}) .
We consider the bundle  $E=\cP\otimes_\cG\check{L}(\gg)$ with $\gg=\slt$
 over elliptic curves with four marked points $\Si_{\tau,4}$ identified with
  the half-periods $\om_a$ (\ref{ome}). Similarly to the finite-dimensional case \cite{LOZ1}
  consider the involution $\varsigma$ of $E$ ($\varsigma^2=1$). It acts on the basic spectral curve $\Si_\tau$
  and on the fields as
   \beq{invo}
  \varsigma\,:\,(w,x)=(-w,-x)\,.
   \eq
  The crucial point is that the half-periods $w=\om_a$ and $w=0$  are the  fixed points of $ \varsigma$ under the
  action of it on $\Si_\tau$.
There are two eigenspaces of the operator $\varsigma$ acting on the space of sections
 \beq{ine}
\G({\rm End}E)=\G^+({\rm End}E)\oplus\G^-({\rm End}E)\,,
 \eq
where $\varsigma\G^+=\G^+$ and $\varsigma\G^-=-\G^-$.
The Lax operators corresponding to the bundle over  $\Si_{\tau,4}$ can be decomposed in the basis of the Pauli matrices
(\ref{100})
 \begin{equation}
\nonumber L(\bar k,x,w)=\frac{\ka}2E_1(w|\tau) \si_0+\sum_{n\in\mZ}\sum_{b=0}^3\sum_{\al=1}^3
S^{\alpha}_{b,n}\varphi_{\alpha}\brc{\bar{k}n,w-\omega_b}\bfe\brc{nx-\frac{n\bar{k}}2b_2} \sigma_{\alpha},
 \end{equation}
 \begin{equation}
\nonumber M(\bar k,x,w)=\frac{\ka}2\p_\tau\vartheta(w|\tau) \si_0+
\frac{1}{2\pi\imath}\sum_{n\in\mZ}\sum_{b=0}^3\sum_{\al=1}^3
S^{\alpha}_{b,n}f_{\alpha}\brc{\bar{k}n,w-\omega_b}\bfe(nx-\frac{n\bar{k}}2b_2)\sigma_{\alpha}-
 \end{equation}
 \begin{equation}
\nonumber -\sum_{n\in\mZ}\sum_{b=0}^3\sum_{\al=1}^3 S^{\alpha}_{b,n} \frac{\partial\omega_b}{\partial\tau}\varphi_{\alpha}
\brc{\bar{k}n,w-\omega_b} \bfe\brc{nx-\frac{n\bar{k}}2b_2}\sigma_{\alpha},
 \end{equation}
where $\varphi_\al(\bar kn,w)$ and $f_{\alpha}\brc{\bar{k}n,w}$ are defined
 in (\ref{var}) and (\ref{eq:fadef}) and $\si_\al$ are the Pauli matrices (\ref{100}).
  Coordinates $S^{\alpha}_{b,n}$
 are the Fourier modes of the field variables $\textbf{S}^{\alpha}_{b}$:
 \begin{equation}
\nonumber
\textbf{S}^{\al}_{b}(x)=\sum_{n\in\mZ}S^{\alpha}_{b,n}\bfe(nx).
 \end{equation}
The Lax pair can be rewritten in terms of these field variables as
 \begin{equation}
\nonumber L(\bar k,x,w)=\frac{\ka}2E_1(w|\tau) \si_0+\sum_{b=0}^3\sum_{\al=1}^3
 \varphi_{\alpha}\brc{\frac{\bar{k}}{2\pi\imath}\partial_x,w-\omega_b}
 \textbf{S}^{\alpha}_{b}\brc{x-\bar{k}b_2/2}\sigma_{\alpha},
 \end{equation}
 \begin{equation}
\nonumber M(\bar k,x,w)=\frac{\ka}2\p_\tau\vartheta(w|\tau) \si_0+ \frac{1}{2\pi\imath}\sum_{b=0}^3\sum_{\al=1}^3f_{\alpha}
\brc{\frac{\bar{k}}{2\pi\imath}\partial_x,w-\omega_b} \textbf{S}^{\alpha}_{b} \brc{x-\bar{k}b_2/2}\sigma_{\alpha}-
 \end{equation}
 \begin{equation}
\nonumber -\sum_{b=0}^3\sum_{\al=1}^3\frac{\partial\omega_b}{\partial\tau}
 \varphi_{\alpha}\brc{\frac{\bar{k}}{2\pi\imath}\partial_x,w-\omega_b}
  \textbf{S}^{\alpha}_{b}\brc{x-\bar{k}b_2/2}\sigma_{\alpha}.
 \end{equation}
The latter operators give the Lax equation
 \begin{equation}
\nonumber
\kappa\partial_{\tau}L-\kappa\partial_w M-\frac k{\rho_0}\partial_x M=\left[L,M\right],
 \end{equation}
which is equivalent to the equation of the elliptic top
 \beq{eat1}
\frac{\partial S^{\ga}_{b,n}}{\partial\tau}= \sum_{m\in\mZ}\sum_{c=0}^{3}\sum_{\alpha,\beta=1}^3
\ep_{\al\be\ga}S^{\alpha-\beta}_{b,n-m}S^{\beta}_{c,m} J^{\beta}_{b,c,m}\,,
 \eq
where
 $$
J^{\beta}_{b,c,m}=
\frac1{\pi}\brc{1-\delta_{bc}}\brc{f_{\beta}\brc{\bar{k}m,\omega_b-\omega_c} +2\pi\imath\frac{\partial (\omega_b-\omega_c)}{\partial\tau} \varphi_{\beta}
 \brc{\bar{k}m,\omega_b-\omega_c}}\bfe\brc{\frac{m\bar{k}}2(b_2-c_2)}-
 $$
 $$
-\frac1{\pi}\de_{bc}E_2\brc{-\omega_{\beta}+\bar{k}m}.
 $$
% and
% \begin{equation}
%\nonumber
%\left[\sigma_{\alpha},\sigma_{\beta}\right]=\ep_{\al,\be,\ga}
%\sigma_{\alpha+\beta}.
% \end{equation}
In terms of the field variables equation (\ref{eat1}) acquires the following form:
 \begin{equation}
\label{eq:ZVGFeq} \frac{\partial \textbf{S}^{\ga}_{b}(x)}{\partial\tau}=\sum_{c=0}^{3}\sum_{\alpha,\beta=1}^3
\ep_{\al\be\ga}\textbf{S}^{\alpha}_{b}(x)\bfJ^{\beta}_{b,c}(\p_x)\textbf{S}^{\beta}_{c}(x)\,,
 \end{equation}
where the conjugate inertia tensor is $\Psi$DO
 $$
\bfJ^{\beta}_{b,c}(\p_x)=\frac1{\pi}\brc{1-\delta_{bc}}f_{\beta}\brc{\frac{\bar{k}}{2\pi\imath}\p_x,\omega_b-\omega_c} \bfe\brc{\frac{\bar{k}(b_2-c_2)}{4\pi\imath}\p_x}
-\frac1{\pi}\de_{bc}E_2\brc{-\omega_{\beta}+\frac{\bar{k}}{2\pi\imath}\p_x}+
 $$
 $$
+2\imath\frac{\partial (\omega_b-\omega_c)}{\partial\tau} \varphi_{\beta}
 \brc{\frac{\bar{k}}{2\pi\imath}\p_x,\omega_b-\omega_c}\bfe\brc{\frac{\bar{k}(b_2-c_2)}{4\pi\imath}\p_x}.
 $$
Now, in accordance with (\ref{invo}) we take
 \beq{ina}
\varsigma L(x,w)=-L(-x,-w)\,,\quad \varsigma M(x,w)=M(-x,-w)\,,
 \eq
and define
 $$
L^{\,\pm}=\oh\bigl(L\pm\varsigma (L)\bigr)\,,~~L^{\,\pm}\in\G^{\,\pm}({\rm End}E)\,.
 $$
 $$
M^{\,\pm}=\oh\bigl(M\pm\varsigma (M)\bigr)\,,~~M^{\,\pm}\in\G^{\,\pm}({\rm End}E)\,.
 $$
The constraint
 \beq{cons11}
L^{\,-}= 0
 \eq
is consistent with the involutions (\ref{invo}), defined by $M^{+}$
 \begin{equation}
\label{eq:LaxZVG}
\ka\p_\tau L^{\,+}-\ka\p_w M^+ -\frac k{\rho_0}\partial_x M^+ =[L^{\,+},M^{+}]\,.
 \end{equation}
Consider the action of $\varsigma$ on the arguments of $\varphi_\al(\bar kn,w)$ :
%$\varsigma\varphi_\al(\bar kn,w)=\varphi_\al(-\bar kn,-w)$
 $$
\varsigma\varphi_\al(\bar kn,w-\om_b)=-\varsigma\Bigl(\bfe(-(w-\om_b)\frac{\al_2}2)
\phi(-\al_\tau+\bar kn,w-\om_b)\Bigr)=-
\bfe((w+\om_b)\frac{\al_2}2)\phi(-\al_\tau+\bar kn,-w-\om_b)\,.
 $$
we remind that $\om_b=\frac{b_1+b_2\tau}2$ $(b_1,b_2=0,1)$, and
$\al_\tau=\frac{\al_1+\al_2\tau}2$ $(\al_1,\al_2=0,1)$.
Then from (\ref{A.300}) and (\ref{A.14}) we find
 \beq{cma}
\varphi_\al(\bar kn,-w-\om_b)=-(-1)^{\vec b\times\vec\al}\bfe(b_2\bar kn)
\varphi_\al(-\bar kn,w-\om_b)\,,
 \eq
where $\vec b\times\vec\al=b_1\al_2-b_2\al_1$. In this way following (\ref{ina})
 we come to the expression
 $$
\varsigma\,L(\bar k,x,w)=\frac{\ka}2E_1(w|\tau) \si_0+\sum_{n\in\mZ} \sum_{b=0}^3\sum_{\al=1}^3S_{a,n}^{\al}(-1)^{\vec
b\times\vec\al} \varphi_\al(-\bar kn,w-\om_b)\bfe\brc{\frac{\bar kn}2b_2-nx}\si_\al\,.
 $$
Then the constraints (\ref{cons11}) and (\ref{cma})  imply
 \begin{equation}
\label{con12}
S_{b,n}^{\al}=(-1)^{\vec b\times\vec\al}S_{b,-n}^{\al}\,,
 \end{equation}
which in terms of the field variables can be written as
 \begin{equation}
\nonumber
\textbf{S}^{\alpha}_{b}(x)=(-1)^{\vec b\times\vec\al}\textbf{S}^{\alpha}_{b}(-x).
 \end{equation}
Using the relation (\ref{con12}), we get that the constraint (\ref{cons11}) also implies
 \begin{equation}
M^{\,-}= 0,
 \end{equation}
or, equivalently,
 \begin{equation}
\nonumber
M(x,w)=M(-x,-w).
 \end{equation}
Thus, the equations of motion of the Zhukovsky-Volterra gyrostat are defined by
 the Lax equation (\ref{eq:LaxZVG}).
From (\ref{con12}) we find the invariant part of the Lax operator
 $$
L^+(x,w)= \frac{\ka}2E_1(w|\tau) \si_0+ \sum_{\al=1}^3\sum_{b=0}^3\left(\frac12S_{b,0}^{\al}\varphi_\al(w-\om_b)\bigl(1
+(-1)^{\vec b\times\vec\al}\bigr)+\right.
 $$
 $$
+\left.\sum_{n> 0}S_{b,n}^{\al}
\brc{\varphi_\al(\bar kn,w-\om_b)
\bfe\brc{nx-\frac{\bar{k}n}2b_2}+(-1)^{\vec b\times\vec\al}
\varphi_\al(-\bar kn,w-\om_b)\bfe\brc{\frac{\bar{k}n}2b_2-nx}}\right)\si_\al\, .
 $$
Now we can write down the equations of motion of the Zhukovsky-Volterra gyrostat, using the Lax equation (\ref{eq:LaxZVG}).
For the positive Fourier modes we obtain:
 \begin{equation}
\nonumber n>0:\;\frac{\partial S^{\gamma}_{b,n}}{\partial\tau}=
-\frac1{\pi}\sum_{m\in\mZ}\sum_{\alpha,\beta=1}^3\epsilon_{\alpha\beta\gamma}
 S^{\alpha}_{b,n-m}S^{\beta}_{b,m}J^I(\be,m) +
 \end{equation}
 \begin{equation}
\label{eq:ZVGeqm} +\frac1{\pi}\sum_{m\in\mZ}\sum_{c\neq b}\sum_{\alpha,\beta=1}^3 \epsilon_{\alpha\beta\gamma}
S^{\alpha}_{b,n-m}S^{\beta}_{c,m}J^{II}(b,c,\be,m)\,,
 \end{equation}
where $J^I(\be,m)=E_2\brc{-\omega_{\beta}+\bar{k}m}$ and
 $$
J^{II}(b,c,\be,m)=\left(f_{\beta}\brc{\bar{k}m,\omega_b-\omega_c}+2\imath
 \frac{\partial (\omega_b-\omega_c)}{\partial\tau}
 \varphi_{\beta}\brc{\bar{k}m,\omega_b-\omega_c}\right)
\bfe\brc{\frac{m\bar{k}}2(b_2-c_2)}\,.
 $$
If we put $n$ to be zero, the last term in the right hand
 side of (\ref{eq:ZVGeqm}) vanishes on the constraint (\ref{con12}) and we get the
 equations for the zero Fourier modes in the form:
 \begin{equation}
\nonumber \vec b\times\vec\gamma=0:\;\frac{\partial S^{\gamma}_{b,0}}{\partial\tau}=
-\frac1{\pi}\sum_{m\in\mZ}\sum_{\alpha,\beta=1}^3\epsilon_{\alpha\beta\gamma}
S^{\alpha}_{b,-m}S^{\beta}_{b,m}E_2\brc{-\omega_{\beta}+\bar{k}m} +
 \end{equation}
 \begin{equation}
\label{eq:ZVGeqm0} +\frac1{\pi}\sum_{m\in\mZ}\sum_{c\neq b}\sum_{\alpha,\beta=1}^3\epsilon_{\alpha\beta\gamma}
S^{\alpha}_{b,-m}S^{\beta}_{c,m}f_{\beta}\brc{\bar{k}m,\omega_b-\omega_c} \bfe\brc{\frac{m\bar{k}}2(b_2-c_2)}.
 \end{equation}
In order to compute the Hamiltonians of the system, we consider the following expansion in the basis of the Eisenstein
functions:
 \beq{trl2}
const\oint\tr(L(x,w))^2=\sum_b\Bigl(H_{2,b}E_2(w-\om_b)+H_{1,b}E_1(w-\om_b)\Bigr)+H_\tau\,,
 \eq
where
 \begin{equation}
\nonumber H_{2,b}=\frac1{4\pi\imath}\sum_{n\in\mZ}\sum_{\alpha=1}^3S^{\alpha}_{b,n}S^{\alpha}_{b,-n},
 \end{equation}
 \begin{equation}
\nonumber H_{1,b}=-\frac1{2\pi\imath}\sum_{n\in\mZ}\sum_{c\neq b}\sum_{\alpha=1}^3 S^{\alpha}_{b,n}S^{\alpha}_{c,-n}
\varphi_{\alpha}\brc{\bar{k}n,\omega_c-\omega_b}
 \bfe\brc{\frac{n\bar{k}}2(c_2-b_2)},
 \end{equation}
 \begin{equation}
\nonumber H_\tau=\frac1{4\pi\imath}\sum_{n\in\mZ}\sum_{c\neq b}\sum_{\alpha=1}^3S^{\alpha}_{b,n}S^{\alpha}_{c,-n}
f_{\alpha}\brc{\bar{k}n,\omega_c-\omega_b} \bfe\brc{\frac{n\bar{k}}2(c_2-b_2)}-
 \end{equation}
 \begin{equation}
\nonumber -\frac1{4\pi\imath}\sum_{n\in\mZ}\sum_{b=0}^3\sum_{\alpha=1}^3 S^{\alpha}_{b,n}S^{\alpha}_{b,-n}
E_2\brc{\bar{k}n-\omega_\alpha}.
 \end{equation}
In terms of the field variables $H_\tau$ assumes the following form:
 \begin{equation}
\nonumber H_\tau=\frac1{4\pi\imath}\sum_{c\neq b}\sum_{\alpha=1}^3 \oint
\textbf{S}^{\alpha}_{c}(x-\bar{k}c_2/2)\bfJ^{II}(\al,b,c,\p_x,\tau) \textbf{S}^{\alpha}_{b}(x-\bar{k}b_2/2)-
 \end{equation}
 \begin{equation}
\nonumber -\frac1{4\pi\imath}\sum_{b=0}^3\sum_{\alpha=1}^3\oint\textbf{S}^{\alpha}_{b}(x)
\bfJ^{I}(\al,b,c,\p_x,\tau)\textbf{S}^{\alpha}_{b}(x)\,,
 \end{equation}
where
 \beq{j1}
\bfJ^{I}(\al,b,c,\p_x,\tau)=E_2\brc{\omega_\alpha-\frac{\bar{k}}{2\pi\imath}\partial_x,\tau}\,,~~
\bfJ^{II}(\al,b,c,\p_x,\tau)=f_{\alpha}\brc{\frac{\bar{k}}{2\pi\imath}\partial_x,\omega_c-\omega_b}\,,
 \eq
Then we impose the constraint (\ref{con12}), which yields
 $$
H_{1,b}=0.
 $$
Other functions from the expansion (\ref{trl2}) can be represented as
 \begin{equation}
\nonumber H_{2,b}=\frac1{8\pi\imath}\sum_{\alpha=1}^3S^{\alpha}_{b,0}S^{\alpha}_{b,0}\brc{1+(-1)^{\vec
b\times\vec\al}}+\frac1{2\pi\imath}\sum_{n>0}\sum_{\alpha=1}^3S^{\alpha}_{b,n}S^{\alpha}_{b,-n},
 \end{equation}
 \begin{equation}
\nonumber H_\tau=-\frac1{8\pi\imath}\sum_{b=0}^3\sum_{\alpha=1}^3S^{\alpha}_{b,0}S^{\alpha}_{b,0}\brc{1+(-1)^{\vec
b\times\vec\al}}E_2\brc{\omega_\alpha}-\frac1{2\pi\imath}\sum_{n>0}\sum_{b=0}^3\sum_{\alpha=1}^3S^{\alpha}_{b,n}S^{\alpha}_{b,-n}
E_2\brc{\bar{k}n-\omega_\alpha}+
 \end{equation}
 \begin{equation}
\nonumber +\frac1{16\pi\imath}\sum_{c\neq b}\sum_{\alpha=1}^3S^{\alpha}_{b,0}S^{\alpha}_{c,0} \brc{1+(-1)^{\vec
b\times\vec\al}}\brc{1+(-1)^{\vec c\times\vec\al}} f_{\alpha}\brc{\omega_c-\omega_b}+
 \end{equation}
 \begin{equation}
\nonumber +\frac1{2\pi\imath}\sum_{n>0}\sum_{c\neq
b}\sum_{\alpha=1}^3S^{\alpha}_{b,n}S^{\alpha}_{c,-n}f_{\alpha}\brc{\bar{k}n,\omega_c-\omega_b}
\bfe\brc{\frac{n\bar{k}}2(c_2-b_2)}+
 \end{equation}
 \begin{equation}
+\sum_{n>0}\sum_{c\neq b}\sum_{\alpha=1}^3S^{\alpha}_{b,n}S^{\alpha}_{c,-n}\frac{\partial
(\omega_c-\omega_b)}{\partial\tau}\varphi_{\alpha}\brc{\bar{k}n,\omega_c-\omega_b} \bfe\brc{\frac{n\bar{k}}2(c_2-b_2)}.
 \end{equation}
The latter function gives the equations of motion (\ref{eq:ZVGeqm}), (\ref{eq:ZVGeqm0})
 with respect to the Poisson structure
 \begin{equation}
\label{eq:pbZV} \bfi{S^{\alpha}_{b,n},S^{\beta}_{c,m}}=\imath\,\delta_{bc}\sum_{\gamma=1}^3\epsilon_{\alpha\beta\gamma}
\brc{S^{\gamma}_{b,n+m}+(-1)^{\vec{b}\times\vec{\alpha}}S^{\gamma}_{b,m-n}}.
 \end{equation}
Thus, taking into the account (\ref{ham}) we get the Hamiltonian of the Zhukovsky-Volterra gyrostat  field theory
 \beq{hzvt}
   \begin{array}{|c|}
  \hline\\
H^{ZVFT}=H_\tau+k\lambda. \\ \ \\
  \hline
  \end{array}
   \eq
Notice that the other hypothetical Hamiltonians $H_{2,b}$ are the Casimir functions with respect to the Poisson brackets
(\ref{eq:pbZV}). Also, the Poisson structure (\ref{eq:pbZV}) can be reformulated in terms of the field variables as follows:
 \begin{equation}
\label{eq:fpbZV}
\bfi{\textbf{S}^{\alpha}_{b}(x),\textbf{S}^{\beta}_{c}(y)}=\imath\,\delta_{bc}\sum_{\gamma=1}^3\epsilon_{\alpha\beta\gamma}
\textbf{S}^{\gamma}_{b}(y)\brc{\delta(x-y)+(-1)^{\vec{b}\times\vec{\alpha}}\delta(x+y)}.
 \end{equation}
This gives the Hamilton equations of motion (\ref{eq:ZVGFeq}) for the Zhukovsky-Volterra gyrostat field variables.

%%%%%%%%%%%%%%%%%%%%%%%%%%%%%%%%%%%%%%%%%%%%%%%%%%%%%%%% %%
%%%%%%%%%%%%%%%%%%%%%%%%%%%%%%%%%%%%%%%%%%%%%%%%%%%%%%%%%%%%
\section{Trigonometric and Rational Systems and Multiloop Algebras}

\setcounter{equation}{0}

%\subsection{Preliminaries}

Our goal here is to derive trigonometric and rational versions of the monodromy
preserving field theories based on the
centrally and cocentrally extended Lie algebra $\check{L}(\sln)$.
To this end we replace the elliptic curve
$\Si_\tau=\mC/(\mZ\oplus\tau\mZ)$ by the cylinder
 $\cC=\mC/\mZ$ (trigonometric case) and the complex plane $\mC$ (rational case).
In the elliptic case we used the sine-basis $\{T^\al,~\al=(\al_1,\al_2)\}$ of the
Lie algebra $\sln$ and put in correspondence to it a finite lattice
 \beq{elll}
\cL^{ell}=
\frac{(\mZ_N\oplus\tau\mZ_N)\backslash(0,0)}N=\left\{\al_\tau=\frac{\al_1+\tau\al_2}N\right\}
\subset\Si_\tau\,.
 \eq
The elements $\al_\tau$ are arguments of elliptic functions $\phi$ and $f$, which are
matrix elements of the Lax operators. In the trigonometric and rational cases we
consider their degenerated versions (\ref{c3})-(\ref{c5}) and (\ref{c6b})-(\ref{c5}).
The trigonometric and rational functions are well defined on $\cC$ and $\mC$.
Their arguments are elements of the lattices
 \beq{etr}
 \cL^{rat}=\frac{(\mZ\oplus\tau\mZ)\backslash(0,0)}N\subset\mC\,,~~
 \cL^{tr}=\frac{(\mZ_N\oplus\tau\mZ)\backslash(0,0)}N\subset \cC=\mC/(\mZ\oplus 0)\,.
  \eq
%Moreover, they satisfy the addition theorems (\ref{d2}) and (\ref{d3}).

In the previous Section we considered bundles with connections taking values in the Lie algebra
$\gga^{ell}=\check{L}(\sln)$ (\ref{lcc}). For trigonometric and rational systems
 we replace this algebra by the algebras
 $$
\gga^{tr}=\gga^{ell}\otimes P(y)\,,~~\gga^{rat}=\gga^{ell}\otimes P(y_1,y_2)\,,
 $$
where $P(\bfe(y))$ and $P(\bfe(y_1),\bfe(y_2))$ are trigonometric polynomials in one and
two variables. In this way we will establish interrelations between
the following classes of objects

 \begin{center}\vspace{5mm}
 \begin{tabular}{|c|c|c|c|}
  \hline
  &&&\\
              &I elliptic&II trigonometric &III rational\\
              &&&\\
              \hline
              \hline
   %Equations in &  &  &  \\
   Dimension & 2 & 3 & 4 \\
   \hline
   Surfaces & $\Si_\tau=\mC/(\mZ\oplus\tau\mZ)$ & $\cC=\mC/\mZ$ & $\mC$\\
   \hline
   Lattices  & $\cL^{ell}$
    & $\cL^{tr}$ & $\cL^{rat}$ \\
    \hline
  Algebras  & $\gga^{ell}=\check{L}(\sln)$& $\gga^{tr}=\gga^{ell}\otimes P(y)$ &
  $\gga^{rat}=\gga^{ell}\otimes P(y_1,y_2)$  \\
  \hline
 \end{tabular}
\bigskip
%\texttt{ Table 1}
 \end{center}

 %%%%%%%%%%%%%%%%%%%%%%%%%%%%%%%%%%%%%%%%%%%%%%%%%%%%%%%%%%%%%%%%%%%%%%%%%%%%%%%%%%

%%%%%%%%%%%%%%%%%%%%%%%%%%%%%%%%%%%%%%%%%%%%%%%%%%%%%%%%%%%%%%%%%%%%%%%%%%%%%%%%%%

\subsection{Loop algebras $L(\gln)$ and $LL(\gln)$ }

\underline{$L(\gln)$}\vskip2mm

\noindent Introduce the following notations. Let
 \beq{rtr}
\mZ_N^{tr,(2)}=(\mZ_N\oplus\mZ)\setminus(0,0)=\{ (\ti\al_1,\al_2)\}\,,~~
0\leq\ti\al_1<N\,,~
\al_2=\ti\al_2+mN\,,~0\leq\ti\al_2<N\,.
 \eq
For the loop algebra $L(\gln)$ we introduce the basis corresponding to
$\mZ_N^{tr,(2)}$ (\ref{tlg1}).
% \beq
%T_N^\al=T_N^{\ti\al_1,\al_2}\sim T_N^{\ti\al_1,\ti\al_2}\bfe(\al_2y)\,,~~
%y\in\mC\,,~~(\ti\al_1,\ti\al_2)\in\mZ_N^{(2)}\,,~~\al_2=\ti\al_2+mN\,.
% \eq
The Poisson algebra of functions on $L^*(\sln)$ has the following generators:
 \beq{pgtr}
S^{\ti\al_1,\al_2}=\sum_{(\be_1,\be_2)\in\mZ_N^{tr,(2)}}\tr\left(
S^{-\ti\be_1,-\be_2}T_N^{-\ti\be_1,-\be_2} T_N^{\ti\al_1,\al_2} \right)\,.
 \eq
The Poisson brackets follow from (\ref{ctr})
 \beq{pbtr}
\{S^{\ti\al_1,\al_2},S^{\ti\be_1,\be_2}\}=
\bfC^{tr}_{1/N}(\al,\be)S^{\widetilde{\ti\al_1+\ti\be_1},\al_2+\be_2}\,.
 \eq
Introduce the currents
 \beq{cur}
\bfS^{\ti\al_1}(y)=\sum_{\al_2\in\mZ}S^{\ti\al_1,\al_2}\bfe(\al_2y)\,.
 \eq
 In terms of the currents the Poison structure
(\ref{pbtr}) takes the form
 \beq{pbtr1}
\{\bfS^{\ti\al_1}(y),\bfS^{\ti\be_1}(y')\}=
\frac{\imath N}2\left(\de_{y+\frac{\ti\be_1}{2N},y'-\frac{\ti\al_1}{2N}}
\bfS^{\widetilde{\ti\al_1+\ti\be_1}}(y+\frac{\ti\be_1}{2N})-
\de_{y-\frac{\ti\be_1}{2N},y'+\frac{\ti\al_1}{2N}}
\bfS^{\widetilde{\ti\al_1+\ti\be_1}}(y-\frac{\ti\be_1}{2N})\right)\,.
 \eq
%
%
%In particular, for $\slt$ $(\al_1,\be_1)\in(\mZ_2\oplus\mZ_2)$ and we have
% $$
%\{\bfS^0(y),\bfS^0(y')\}=0\,,~~\{\bfS^0(y),\bfS^1(y')\}=\imath(\bfS^1(y')-\bfS^1(y')
%\,,~~
% $$
%
%Similarly, let $S^\al(y)$ is defined as $\lan S^\al(y),T^\be(y')\ran=\de^{\al\be}\de(y,y')$,
%where $y=(y_1,y_2)$ for the two loop algebra, or $y=y_1$ for $L(\sln)$.
%Then
% \beq{ptr}
%\{S^\ti\al(y),S^\be(y')\}=\bfC_\f1{N}(\al,\be)S^{\al+\be}\de(y,y')\,.
% \eq

%%%%%%%%%%%%%%%%%%%%%%%%%%%%%%%%%%%%%%%%%%%%%%%%%%%%%%%%%%%%%%%%%%%%%%%%%%%%%%%%%%%

\underline{$LL(\gln)$}\vskip2mm

The two-loop algebra $LL(\gln)$ has the basis (\ref{tlg}) and the commutator (\ref{crtl}).
%The generators form
% To see it consider the basis
%$T_N^{\ti\al}$ (\ref{bat}) in $\gln$, where $\al=\ti\al+m_\al N$. We pass to the generators
% \beq{llg1}
%T^\al\to T_N^{\ti\al_1,\ti\al_2}\bfe(\al_1y_1+\al_2y_2)\,,~~\al_j=\ti\al_j+m_jN
% \eq
%with the commutator
% $$
%[T^{\ti\al}\bfe(\al\cdot y),T^{\ti\be}\bfe(\be\cdot y)]=
%\bfC_{1/N}(\ti\al,\ti\be)T^{\widetilde{\ti\al+\ti\be}}\bfe(\al+\be)\cdot y)\,.
% $$
%The generators $\{T_N^{\ti\al}\bfe(\al\cdot y)\}$ form a bases in the two-loop algebra in the principle
%gradation.
%Since $\bfC_{1/N}(\al,\be)=\bfC_{1/N}(\ti\al,\ti\be)$ the algebra $LL(\gln)$ is isomorphic to the
%algebra generated by $\{T^\al\}$.
%
For the dual variables we have the Poisson brackets
 \beq{pbr}
\{S^{\al_1,\al_2},S^{\be_1,\be_2}\}=
\bfC^{rat}_{M/N}(\al,\be)S^{\al_1+\be_1,\al_2+\be_2},\quad
\al=(\al_1,\al_2)\in\mZ^{(2)}=(\mZ\oplus\mZ)\setminus(0,0)\,.
 \eq
In terms of the currents
 \beq{par}
\bfS(y_1,y_2)=\sum_{\al\in\mZ^{(2)}}S^{\al_1,\al_2}\bfe(\al_1y_1+\al_2y_2)
 \eq
the Poison brackets are the Moyal brackets
 $$
\{\bfS(y_1,y_2),\bfF(y_1,y_2)\}^{rat}=:[\bfS(y_1,y_2),\bfF(y_1,y_2)]_{1/N}
=N\left(\bfS(y_1,y_2)\star\bfF(y_1,y_2)-\bfF(y_1,y_2)\star\bfS(y_1,y_2)\right)=
 $$
 \beq{tlb}
\bfS(y_1+\f1{2N}\p_{y_2},y_2-\f1{2N}\p_{y_1})\bfF(y_1,y_2)
-\bfS(y_1-\f1{2N}\p_{y_2},y_2+\f1{2N}\p_{y_1})\bfF(y_1,y_2)\,.
 \eq
In this way we come to the nonlocal Poisson brackets (as well as in the trigonometric case (\ref{pbtr})).  Nonlocal Poisson
brackets in integrable hierarchies were considered recently \cite{dSK,Sokolov}.

%%%%%%%%%%%%%%%%%%%%%%%%%%%%%%%%%%%%%%%%%%%%%%%%%%%%%%%%%%%%%%%%%%%%%%%%%%%%%%55

\subsection{Equations of motion}

\subsubsection{Trigonometric systems}

We consider the Poisson algebra (\ref{pgtr}) and the corresponding currents $S^{\ti\al}(y)$ (\ref{cur}).
 As in (\ref{ixv}) introduce the $x$-dependence
 \beq{cur1}
\bfS^{\ti\al_1}(x,y)=\sum_{\al_2\in\mZ}S_n^{\ti\al_1,\al_2}\bfe(nx+\al_2y)\,.
 \eq
In this way we come to the two-loop algebra in the variables $(x,y)$
with the central extension in the $x$ direction
 $$
\mC^*\times\mC^*\to\sln=\{\bfS^{\ti\al_1}(x,y)\}\,.
 $$
The Poisson brackets for the Fourier modes assume the form
 \beq{pbtm}
\{S_n^{\ti\al_1,\al_2},S_m^{\ti\be_1,\be_2}\}=S_{n+m}^{\ti\al_1+\ti\be_1,\al_2+\be_2}
\bfC^{tr}_{M/N}(\al\times\be)\,.
 \eq
The conjugate inertia tensor
 $$
(J^{tr})_{\ti\al_1,\al_2}(n)=E_2^{tr}\bigl(-\frac{\ti\al_1
+\tau\al_2}{ N}+\bar{k}n\bigr)=\pi^2\sin^{-2}\pi\bigl(-\frac{\ti\al_1
+\tau\al_2}{ N}+\bar{k}n\bigr)\,,~~ \al_2\in\mZ\,.
 $$
defines the trigonometric Euler-Arnold Hamiltonian
 \beq{htc2}
H^{tr}_\tau=k\la+
\sum_{\ti\al_1\in\mZ_N,\al_2,n\in\mZ} S_n^{\ti\al_1,\al_2}
(J^{tr})_{-\ti\al_1,-\al_2}(-n)
(S_{-n}^{-\ti\al_1,-\al_2})\,,
 \eq
or in term of velocities
$F_n^{\ti\al_1,\al_2}=(J^{tr})^{*}_{\ti\al_1,\al_2}(n)S_n^{\ti\al_1,\al_2}$
 $$
H^{tr}_\tau=k\la+
\sum_{\ti\al_1\in\mZ_N,\al_2,n\in\mZ} S_n^{\ti\al_1,\al_2}F_{-n}^{-\ti\al_1,-\al_2}\,.
 $$
Then we come to the equations of motion
 \beq{emtrc}
\ka\p_\tau S_n^{\ti\al_1\al_2}=\sum_{\ti\ga_1\in\mZ_m,\ga_2,n'\in\mZ}
F_{n-n'}^{-\ti\ga_1,-\ga_2}(x)
S_{n'}^{\widetilde{\ti\al_1+\ti\ga_1},\al_2+\ga_2}\bfC^{tr}_{M/N}(\ga,\al)\,.
 \eq
Rewrite the equations in terms of fields. Then the conjugate inertia tensor
 \beq{iotr1}
\bfJ^{tr}_{\ti\al_1,\al_2}(\p_x)=\pi^2\sin^{-2}\pi\bigl(-\frac{\ti\al_1
+\tau\al_2}{ N}+\frac{\bar{k}}{2\pi\imath}\p_x\bigr)\,,~~ \al_2\in\mZ
%=\pi^2\sin^{-2}\pi\bigl(-\frac{\al_1
%+\tau\ti\al_2}{ N}+m\tau+\frac{\bar{k}}{2\pi\imath}\p_x\bigr)\,.
 \eq
acts on the $y$ Fourier components
 $$
S^{\ti\al_1,\al_2}(x)\to(\bfJ^{tr}_{\ti\al_1,\al_2}(\p_x)S^{\ti\al_1,\al_2})(x)\,.
 $$
The Hamiltonian is the integral
 \beq{htc1}
H^{tr}_\tau=k\la+
\sum_{\ti\al_1\in\mZ_N,\al_2\in\mZ}\oint_x \bfS^{\ti\al_1,\al_2}(x)
\bfJ^{tr}_{-\ti\al_1,-\al_2}(\p_x)
(\bfS^{-\ti\al_1,-\al_2})(x)\,,
 \eq
or in terms of velocities
$\bfF_n^{\ti\al_1,\al_2}(x)=\bfJ^{tr}_{\ti\al_1,\al_2}(\p_x)\bfS^{\ti\al_1,\al_2}(x)$
 $$
H^{tr}_\tau=k\la+
\sum_{\ti\al_1\in\mZ_N,\al_2\in\mZ}\oint_x \bfS^{\ti\al_1,\al_2}(x)\bfF^{-\ti\al_1,-\al_2}(x)\,.
 $$
The equations of motion for the fields
 \beq{emtrc1}
\fbox{$
\ka\p_\tau \bfS^{\ti\al_1\al_2}(x)=\sum_{\ti\ga_1\in\mZ_m,\ga_2\in\mZ}
\bfF^{-\ti\ga_1,-\ga_2}(x)\bfS^{\widetilde{\al_1+\ga_1},\al_2+\ga_2}(x)
\bfC^{tr}_{M/N}(\ga,\al)\,.
$}
 \eq
Taking into account the equation of motion $\bar k=\frac{\tau}\ka k$ rewrite the
conjugate inertia operator in terms of the currents (\ref{cur})
 $$
(\bfJ^{tr})_{\al_1}(\p_x,\p_y)=\pi^2\sin^{-2}\pi\left(-\frac{\al_1}N
+\frac{\tau}{2\pi\imath}(\frac{1}{ N}\p_y+\frac{1}{\ka}k\p_x)\right)\,.
 $$
Thus the  inertia operator becomes the difference operator of the second order
 \beq{iop}
((\bfJ^{tr})^*_{\al_1}f)(x,y)=\frac{2f(x,y)-\bfe(-2\al_1/N)f(x\!+\!2\tau k/\ka,y\!+\!2\tau/N)- \bfe(2\al_1/N)f(x\!-\!2\tau
k/\ka,y\!-\!2\tau/N)}{\pi^{2}/4}\,.
 \eq
Then we come to the  Hamiltonian
 \beq{htat}
H^{tr}_\tau=k\la+
\sum_{\al_1\in\mZ_N}\int_{T^2} \bfS^{\al_1}(x,y)(\bfJ^{tr})^{*}_{\al_1}
(\bfS^{-\al_1})(x,y)\,,~~(\int_{T^2} =\oint_x\oint_y)\,.
 \eq
Define the dual fields
 $$
\bfF^{-\al_1}(x,y)=(\bfJ^{tr})^{*}_{\al_1}(\p_x,\p_y)(\bfS^{-\al_1})(x,y).
 $$
Then
 $$
H^{tr}_\tau=k\la+
\sum_{\al_1\in\mZ_N}\int_{T^2} \bfS^{\al_1}(x,y)\bfF^{-\al_1}(x,y).
 $$
The equations of motion assume the form
 $$
\ka\p_\tau \bfS^{\al_1}(x,y)=\frac{\imath N}2\sum_{\ga_1\in\mZ_N}
\left(\bfF^{-\ga_1}(x,y+\frac{\al_1+\ga_1}2)\bfS^{\al_1+\ga_1}(x,y-\frac{\ga_1}2)\right.-
 $$
 $$
\left.
-\bfF^{-\ga_1}(x,y-\frac{\al_1+\ga_1}2)\bfS^{\al_1+\ga_1}(x,y+\frac{\ga_1}2)\right)\,,
 $$
or in term of velocities
 $$
\ka\p_\tau \bfJ_{\al_1}^{tr}\bfF^{\al_1}(x,y)=\frac{\imath N}2\sum_{\ga_1\in\mZ_N}
\left(\bfF^{-\ga_1}(x,y+\frac{\al_1+\ga_1}2)\bfJ^{tr}_{\al_1+\ga_1} \bfF^{\al_1+\ga_1}(x,y-\frac{\ga_1}2)\right.-
 $$
 $$
\left.
-\bfF^{-\ga_1}(x,y-\frac{\al_1+\ga_1}2)\bfJ^{tr}_{\al_1+\ga_1}
\bfF^{\al_1+\ga_1}(x,y+\frac{\ga_1}2)\right)\,,
 $$

%%%%%%%%%%%%%%%%%%%%%%%%%%%%%%%%%%%%%%%%%%%%%%%%%%%%%%%%%%%%%%%%%%%%%%%%%%%%%%%%%%%%%%%

\subsubsection*{Lax representation}

Recall that (see(\ref{ixv}))
 $ T_{N,n}^{\al}=\bfe(nx+\al_2y)T_N^{\ti\al_1,\ti\al_2}$,  where
  $$
 \al\in\mZ_N^{tr(2)} ~(\ref{rtr})\,,~\al_2=\ti\al_2+mN\in\mZ\,,~0\leq \ti\al_2<N\,
 ~\al_\tau=\frac{\ti\al_1+\al_2\tau}N.
 $$
  The Lax pair assumes the following form:
  $$
 L^{tr}=\sum_{n\in\mZ}\sum_{\al\in\mZ_N^{tr(2)}} L^{tr,\al}_{n}T_{N,n}^{\al}\,,~~
 M^{tr}=\sum_{m,n\in\mZ}\sum_{\al\in\mZ_N^{tr(2)}} M^{tr,\al}_{n}T_{N,n}^{\al}\,,
  $$
  $$
 \begin{array}{ll}
L^{tr,\al}_{n}=S_{n}^{\al} X_{n}^\al &X_{n}^\al=\bfe
\Bigl(-w\frac{\al_2}N\Bigr)\phi^{tr}_{\al_\tau}(\bar{k}n,w)  \,,\\
 M^{tr,\al}_{n}=\f1{2\pi\imath}S_{n}^{\al} Y_{n}^\al &
  Y_{n}^\al=\f1{2\pi\imath}\bfe\Bigl( -\frac{w\al_2}N\Bigr)
f_{\al_\tau}^{tr}(\bar{k}n)\,,
                 \end{array}
  $$
 %where $S_{n}^{\al}=S_n^{\ti\al_1,\al_2}$,
  $$
 \phi^{tr}_{\al_\tau}(\bar{k}n,w)=
 \pi\cot\pi(-\al_\tau+\bar{k}n)+\pi\cot(\pi w)\,,~
 f_{\al_\tau}^{tr}(\bar{k}n)=-\pi^2\sin^{-2}\pi(-\al_\tau+\bar{k}n)\,,
  $$
  $$
 \bfS^{\al}(x,y)=\sum_{\al_2,n\in\mZ}S^{\ti\al_1,\al_2}_{n}\bfe(nx+\al_2y)\,,~~
 S_{n}^{\al}=S_n^{\ti\al_1,\al_2}.
%S_{m,n}^{\ti\al}=S^{\ti\al_1,\ti\al_2+mN}_{n}
  $$
 The both functions $X_{n}^\al$, $Y_{n}^\al$ are well defined on lattice $\cL^{tr}$ (\ref{etr}).
%
 %Let
%  $$
% \bfS^{\al_1,\ti\al_2}(x,y)=\sum_{n,\al_2}S_{n}^{\al_1,\ti\al_2}\bfe(nx+\al_2y)\,,~~
% (\ti \al_2=\al_2~mod(N))\,.
%  $$
 In terms of fields we have
 $$
 \begin{array}{ll}
L_n^{tr,\al}=S_n^{\al}\pi\bfe\Bigl( -\frac{w\al_2}{N})\Bigr)\phi^{tr}_{\al_\tau}(\bar{k}n,w)
%\Bigl(\cot\pi(-\al_\tau+\bar{k}n)+\cot(\pi w)\Bigr)
T_{N,n}^{\al}\,,&
M_n^{tr,\al}=-S_n^{\al}\pi^2\bfe\Bigl( -\frac{w\al_2}{N})\Bigr)f_{\al_\tau}^{tr}(\bar{k}n)
%\sin^{-2}\pi(-\al_\tau+\bar{k}n)
T_{N,n}^{\al}\,.
 \end{array}
 $$
 \begin{predl}\label{leqtr}
The Lax equation
 \beq{lm2}
\Bigl[\ka\p_\tau +M^{tr}\,,\,
\ka\p_w+k\p_x+L^{tr}\Bigr]=0
 \eq
is equivalent to the equations of motion (\ref{emtrc1})
 \end{predl}
\emph{Proof}.\vskip2mm
 \noindent The proof is the same as in the elliptic case (Proposition \ref{leq}). It is based on the addition
formulas for $\phi^{tr}_{\al_\tau}$ and $f^{tr}_{\al_\tau}$.$\blacksquare$

\subsubsection{Rational systems}

Consider the Poisson algebra $\{\bfS(y_1,y_2)\}$ corresponding to the two-loop algebra $LL(\gln)$
(\ref{tlga}), (\ref{tlg}), (\ref{crtl}).
As in (\ref{ixv})  introduce the $x$-dependence
(\ref{par}),
 \beq{strl}
\bfS(x,y_1,y_2)=\sum_{n\in\mZ,\al\in\mZ^{(2)}}S_n^{\al_1,\al_2}\bfe(nx+\al_1y_1+\al_2y_2)
 \eq
 \beq{pbr2}
\{S_n^{\al_1,\al_2},S_m^{\be_1,\be_2}\}=
\bfC^{rat}_{M/N}(\al,\be)S_{n+m}^{\al_1+\be_1,\al_2+\be_2}\,.
 \eq
The fields $\bfS(x,y_1,y_2)$ represent three-loop coalgebra (with the cocentral extension in the $x$-direction).
Define the conjugate inertia tensor
 $$
J^{rat}_{\al_1,\al_2}(n)=E_2^{rat}\bigl(-\frac{\al_1
+\tau\al_2}{ N}+\bar{k}n\bigr)=\bigl(-\frac{\al_1
+\tau\al_2}{ N}+\bar{k}n\bigr)^{-2}\,,
 $$
and the Hamiltonian
 \beq{htc}
H^{rat}_\tau=k\la+
\sum_{\al\in\mZ^{(2)},n\in\mZ} S_{n}^{\al_1,\al_2}
J^{rat}_{-\al_1,-\al_2}(-n)
(S_{-n}^{-\al_1,-\al_2})\,,
 \eq
or in terms of velocities
$F_n^{\al_1,\al_2}=J^{rat}_{\al_1,\al_2}(n)S_n^{\al_1,\al_2}$
 $$
H^{rat}_\tau=k\la+
\sum_{\al\in\mZ^{(2)},n\in\mZ} S_n^{\al_1,\al_2}F_{-n}^{-\al_1,-\al_2}\,.
 $$
Then we come to the equations of motion
 \beq{emtrc22}
   \begin{array}{|c|}
  \hline\\
\ka\p_\tau S_n^{\al_1\al_2}=\sum_{\ga\in\mZ^{(2)},n'\in\mZ}
F_{n-n'}^{-\ga_1,-\ga_2}S_{n'}^{\al_1+\ga_1,\al_2+\ga_2}\bfC^{rat}_{M/N}(\ga,\al)\\ \ \\
  \hline
  \end{array}
   \eq

\emph{The Lax representation} for these equations is based on the addition theorems
(\ref{d2}), (\ref{d3}) and similar to the trigonometric case.

%%%%%%%%%%%%%%%%%%%%%%%%%%%%%%%%%%%%%%%%%%%%%%%%%%%%%%%%%%%%%%%%%

%%%%%%%%%%%%%%%%%%%%%%%%%%%%%%%%%%%%%%%%%%%%%%%%

%%%%%%%%%%%%%%%%%%%%%%%%%%%%%%%%%%%%%%%%%%%%%%%%%%%%%%%% %%
%%%%%%%%%%%%%%%%%%%%%%%%%%%%%%%%%%%%%%%%%%%%%%%%%%%%%%%%%%%%

\section{Noncommutative Torus and Isomonodromic Deformations}
\setcounter{equation}{0}

\subsection{Non-autonomous top on NCT}
\label{sec:defNCT}

Let $\gg=sin_\theta$ be the sine-algebra and $\gg^*=sin^*_\theta$ is its Lie co-algebra.
Consider corresponding to the basis $\{T^\be\}$ (\ref{3.10}), (\ref{3.11}) in $sin_\theta$
the basis $S^\al$ in the Poisson algebra $sin^*_\theta$.
Then the Poisson-Lie structure on $sin^*_\theta$ assumes the form
 \beq{lpb}
 \{S_\al,S_\be\}=\bfC_\te(\al,\be) S_{\al+\be}\,,~~\al=(\al_1,\al_2)\in\mZ^{(2)}=(\mZ\oplus\mZ)\setminus(0,0)\,.
  \eq
% In terms of fields $\bfS(x)$ (\ref{ds}) on the NCT ${\cal T}^2_\te$
%
The function
 \beq{ca1}
C_2=\sum_{\al\in\mZ^{(2)}} S_\al S_{-\al}
 \eq
is the Casimir function with respect to (\ref{lpb}).
Consider the representation (\ref{3.2})
 \beq{bfs}
\bfS(x)=\sum_{\ga\in\mZ^{(2)}}S_\ga T^\ga(x)
 \eq
as a function on the noncommutative torus. Then $C_2$ has the integral form (\ref{3.6})
 \beq{ca2}
C_2=-\f1{4\pi^2}\int_{\cT^2_\hbar}\bfS^2(x)\,.
 \eq
Let $\ep_1$, $\ep_2$  be real numbers such that $0<\te\ep_j<1$
and $\te\ep_j$ are irrational.
 Define the Weierstrass function on $\Si_\tau$ (\ref{C0})
 \footnote{In this section we work with the Weierstrass function instead of $E_2$
 (see (\ref{a101})).}
 \beq{4.8}
\wp_\hbar(\al)=\wp(\hbar(\ep_1\al_1+\ep_2\al_2\tau)|\tau)\,.
 \eq
It determines the conjugate inertia operator
 \beq{iio}
\bfJ\,:\,sin^*_\hbar\to sin_\hbar\,,~~S_\al\to J_\al S_\al\,,~~
J_\al=\wp_\hbar(\al)\,.
 \eq
We rewrite this map   in terms of basis (\ref{ds}) of the NCT ${\cal T}^2_\te$.
For this purpose introduce a complex structure on  ${\cal T}^2_\te$ and define the operator
$\bp=\p_{\bar{Z}}$. In the basis (\ref{ds})
 \beq{4.9}
\p_{\bar{Z}}(U_1^{\al_1}U_2^{\al_2})=\frac{1}{2\pi\imath}
\bigl(\ep_1\al_1+\ep_2\ep_2\al_2\tau\bigr)(U_1^{\al_1}U_2^{\al_2})\,.
 \eq
%Then
% \beq{bpz}
%\p_{\bar{Z}}=\frac{\te}{2\pi \imath}((\ep_1\p_1+\ep_2\tau\p_2)\,,
% \eq
In these terms the conjugate inertia operator $J$ becomes the pseudo-differential operator
 $$
J(\bfS(x))=\wp(\hbar\p_{\bar{Z}})\bfS(x)\,.
 $$
Define the Hamiltonian
 \beq{h1}
H=-\oh\int_{{\cal T}^2_\te} \bfS(x)\wp(\hbar\p_{\bar Z})\bfS(x)=
-\oh\sum_{\ga\in \mZ^{(2)}} S_\ga \wp_\hbar(\ga) S_{-\ga}\,.
 \eq
The phase space of the system is defined by the conditions
 \beq{pha}
\cR^*=\{\bfS~|~ C_2(\bfS)<\infty\,,~H(\bfS)<\infty\}\,.
  \eq
  The equations of motion for the non-autonomous Euler-Arnold top
$\ka\p_\tau \bfS=ad^*_{J(\bfS)}\bfS$ for the group
$SIN_\te$ with inertia operator (\ref{iio}) take the form
 \beq{4.5}
   \begin{array}{|c|}
  \hline\\
\ka\p_\tau S_\al=\sum_{\ga\in \mZ^{(2)}} S_{\al-\ga}S_\ga \wp_\hbar(\ga)\bfC_\te(\ga,\al)\\ \ \\
  \hline
  \end{array}
   \eq
or in terms of the noncommuting fields
 \beq{4.6a}
   \begin{array}{|c|}
  \hline\\
\ka\p_\tau\bfS(x)=[\bfS(x),\wp(\hbar\p_{\bar Z})\bfS(x)]_\hbar\\ \ \\
  \hline
  \end{array}
   \eq

%%%%%%%%%%%%%%%%%%%%%%%%%%%%%%%%%%%%%%%%%%%%%%%%%%%%%%%%%

\subsection{Holomorphic bundles and NCT}

Here we give two descriptions of flat bundles relevant to our construction.
Consider first an infinite rank principal bundle
 $\cP$ with the structure group $SIN_\hbar$ (\ref{A.10}) over the deformed elliptic curve $\Si_\tau$ (\ref{dec}).
As in subsection \ref{inf} define  the adjoint bundle $E_\hbar=\cP\otimes_{SIN_\hbar} V$, where $V$ is a vector representation of
$SIN_\hbar$.
%We don't need exact form of $V$, because we will work with connections on $E_\hbar$.
Sections $s\in \Gamma(E_\hbar)$ are transformed by the transition operators
 \beq{qpc1}
\left\{
 \begin{array}{l}
  s(w+1)=U_1^{-\ep_2}s(w)  \\
   s(w+\tau)=-\bfe_\hbar\bigl(-w-\frac{\tau}2\bigr)U_2^{-\ep_1}s(w)
 \end{array}
\right.\,,~~  U_j\in SIN_\hbar \,,~(\ref{3.5})\,,\,(\ref{3.2})\,.
 \eq
The operators $U_1^{-\ep_2}$, $U_2^{-\ep_1}$ are well defined in the representation
(\ref{3.5a}), or in the Moyal representation (\ref{3.2}).
In the former case $s(w)=s(x,w)$, $x\in\mR$.

Equivalently, we can consider the projective module $E_{\hbar,\tau}$ over $\Si_\tau\times\cT^2_\hbar$.
As a module over $\Si_\tau$ it is defined in (\ref{qpc1}).
 The module over $\cT^2_\hbar$ is defined by the left action
\beq{rim}
\cT^2_\hbar \to U_1\cT^2_\hbar\,,~U_2\cT^2_\hbar\,.
\eq
 Let $\p_{\bar Z}+w$ ($w\in\mC$) be the  connection on the module,
where $\p_{\bar Z}$ is defined as (\ref{4.9}). The connection defines the
 holomorphic structure on $E_{\hbar,\tau}$. The right action on $\cT^2_\hbar$
commutes with the left action and in this way preserves the module.
By acting from the right
by $U_1^nU_2^m$ we find, that $w$ and $w+n+m\tau$ define the connection on the
equivalent modules. In other words, the moduli space of
 holomorphic structures on  $E_{\hbar,\tau}$ is isomorphic to the elliptic curve $\Si_\tau$.

%Consider  connections on $E_\hbar$
% $$
%\left\{
% \begin{array}{l}
%\ka\p_w+L(w,\ti{w},\tau)\,,\\
%\p_{\ti w}+\ti L(w,\ti{w},\tau)\,,
% \end{array}
%\right.
% $$
%where $L$ will play the role of the Lax operator.
%
%
%
%
%
%%But first we explain how the time $\tau$ is arisen \cite{LOZ2}.
%%
%
%Consider a projective module $E_{\te,\tau}$ over $\Si_\tau\times\cT^2_\hbar$ (\ref{dec}).
%Its restriction over $\cT^2_\hbar$ is $E_\hbar$ (\ref{rim}) and
%%Consider an infinite rank principal bundle
%% $\cP$ with the structure group $SIN_\hbar$ over the deformed elliptic curve $\Si_\tau$ (\ref{dec}).
%%As in subsection \ref{inf} define  the adjoint bundle $E_\hbar=\cP\otimes_{SIN_\hbar} V$, where $V$ is a vector representation of
%%$SIN_\hbar$. We don't need exact form of $V$, because we will work with connections on $E_\hbar$. Sections $s\in \Gamma(E_\hbar)$
%%are transformed by the transition operators
% \beq{qpc1}
%\left\{
% \begin{array}{l}
%  s(w+1)=U_1^{-\ep_2}s(w)   \\
%   s(w+\tau)=-\bfe_\hbar\bigl(-w-\frac{\tau}2\bigr)U_2^{-\ep_1}s(w)
% \end{array}
%\right.\,,~~  U_j\in SIN_\hbar \,,~(\ref{3.5})\,,\,(\ref{3.2})\,.
% \eq
%The operators $U_1^{-\ep_2}$, $U_2^{-\ep_1}$ are well defined in the representation
%(\ref{3.5a}), or in the Moyal representation (\ref{3.2}).
%%In the former case $s(w)=s(x,w)$, $x\in\mR$.

%%%%%%%%%%%%%%%%%%%%%%%%%%%%%%%%%%%%%%%%%%%%%%%%%%%%%%%%%

\subsection{Lax representation}

The goal of this subsection is the Lax representation of the equation
of motion (\ref{4.6a}).

Consider connections on $E_{\hbar,\tau}$ in the direction $(w,\ti w)$
 $$
\left\{
 \begin{array}{l}
\ka\p_w+L(w,\ti{w},\tau)\,,\\
\p_{\ti w}+\ti L(w,\ti{w},\tau)\,,
 \end{array}
\right.
 $$
where $L$ will play the role of the Lax operator.
%where $L(w,\ti{w},\tau)$, $\bL(w,\ti{w},\tau)$ are meromorphic maps of
%$\Si_\tau$ in $sin_\hbar$.
For  the almost all bundles (\ref{qpc1}) $\ti L$ can be gauged away:
$f^{-1}\p_{\ti w} f+f^{-1}\ti L f=0$. As a result of the symplectic reduction we
assume that the bundle is flat in the direction $(w,\ti{w})$.
As in the case of affine algebras (\ref{fl}), (\ref{guy}), (\ref{res})  the Lax operator
 can be  fixed by the following
conditions:\vskip2mm
 \noindent $\bf 1$.\emph{ The flatness}
 \beq{fl2}
\p_{\ti{w}}L=0\,.
 \eq
$\bf 2.$ \emph{The quasi-periodicity
conditions:}
 \beq{qp}
\left\{
 \begin{array}{l}
L(w+1)=U_1^{-\ep_2}L^{(1)}(w)U_1^{\ep_2}\,,  \\
L(w+\tau)=\tilde\La L(w)\tilde\La^{-1}+2\pi i\hbar\,,
 \end{array}
\right.
~~\tilde\Lambda(w,\tau)=-\bfe_\hbar\bigl(-w-\frac{\tau}2\bigr)U_2^{-\ep_1}\,.
 \eq
%It means that there are no moduli parameters for  $V_\hbar$.\\
$\bf 3.$ \emph{The quasi-parabolic structure}:
$L$ has a simple pole at $w=0$ and
 \beq{res2}
Res|_{w=0}\,L(w)=\bfS-\ka\hbar\cdot Id\,,
 \eq
 where $\bfS$ is defined by (\ref{pha}). All degrees of freedom will come from
the residue.
 \begin{lem}
The Lax operator assumes the form
 \beq{l1}
L(w)=-\ka\hbar E_1(w|\tau)Id+
\sum_{\al\in\mZ^{(2)}} S_\al\vf_{\hbar\ep\cdot\al}(w)T^\al\,,
 \eq
where  $E_1(w|\tau) $ (\ref{A.1})  and
$\vf_{\hbar\ep\cdot\al}(z)$ (\ref{CC3})).
 \end{lem}
\emph{Proof}\vskip2mm
 \noindent It follows from (\ref{mnc}) that
 $$
U_1^{-1}T^\al U_1=\bfe_\hbar(\al_2) T^\al\,,~~U_2^{-1}T^\al U_2 =\bfe_\hbar(-\al_1) T^\al\,,
~~\bfe_\hbar(1)=\exp 2\pi\imath\hbar\,.
 $$
Then (\ref{qp})  follows from (\ref{opph}), while (\ref{res2}) from (\ref{A.1}) and (\ref{A.3a}).
 $\blacksquare$

 \bigskip

%The linear system
% \beq{li}
%\left\{ \begin{array}{ll}
% i. & (\ka\p_w+L(w))\psi=0\,, \\
% ii &  \p_{\ti{w}}\psi=0
% \end{array}
%\right.
% \eq
%is compatible due to the flatness of the bundle (\ref{fl2}).
%Here $\p_w$ and  $\p_{\ti{w}}$
%were defined in (\ref{dmu}).
%The independence of the monodromy of (\ref{li}) means that the
%Baker-Akhiezer vector $\Psi$ satisfies the additional linear equation
% \beq{mono}
%iii.~(\ka\p_\tau+M)\psi=0\,.
% \eq
%The compatibility condition of $i.$ and $iii.$ is the Lax equation.
%
 \begin{predl}\label{pr3.1}
The equations of motion of the non-autonomous top (\ref{4.5})
are the monodromy preserving equations for the linear system
 $$
\left\{
 \begin{array}{l}
 ( \ka\p_w+L(w,\ti{w},\tau))\psi=0\,, \\
  \p_{\ti w}\psi=0\,.
 \end{array} \right.
 $$
They have
 the Lax representation
 \beq{ltop}
\ka\p_\tau L-\ka\p_w M+[M,L]=0\,,
 \eq
where $L$ is defined by (\ref{l1}),
 \beq{mp2}
M=-\frac{\ka}N\p_\tau\ln\vth(w;\tau) Id+\f1{2\pi i}
\sum_{\ga\in\mZ^{(2)}} S_\ga f_{\hbar\ep\ga}(w) T^\ga \,,
 \eq
and $f_{\hbar\ep\ga}(w)$ is defined by (\ref{CC3a}).
 \end{predl}
\emph{Proof}.\vskip2mm
 \noindent The proof of the equivalence between (\ref{4.5}) and (\ref{ltop}) is based on the
 addition formula (\ref{d2}) and the heat equation (\ref{A.4b}).
 Let us substitute $L$ (\ref{l1}) and $M$ in  (\ref{ltop}). Then in
 the  basis $T^\al$ we come to the equation
  \beq{ple}
\ka \p_{\tau}(S_\al\vf_{\hbar\ep\cdot\al}(w))-\ka\p_w(S_\al f_{\hbar\ep\al}(w))=
 \sum_\ga S_{\al-\ga}S_\ga\bfC_\hbar(\ga,\al)\vf_{\hbar\ep\cdot(\al-\ga)}(w)f_{\hbar\ep\ga}(w)\,.
  \eq
The l.h.s.  equals
 $$
\ka\p_{\tau}(S_\al)\vf_{\hbar\ep\al}(w)+
S_\al\Bigl(\ka \p_{\tau}\vf_{\hbar\ep\al}(w)-\f1{2\pi \imath}\ka\p_w f_{\hbar\ep\al}(w)\Bigr)\,.
 $$
The expression in the brackets $\ka \p_{\tau}\vf_{\hbar\ep\al}(w)-\f1{2\pi \imath}\ka\p_w f_{\teal}(w)$
vanishes due to the heat equation (\ref{A.4b}).
 The r.h.s. of  (\ref{ple}) gives
 $$
\sum_\ga S_{\al-\ga}S_\ga\bfC_\hbar(\ga,\al)f_{\hbar\ep(\al-\ga)}(w)\vf_{\hbar\ep\ga}(w)
\stackrel{{\rm(\ref{d2})}}{=}\sum_\ga S_{\al-\ga}S_\ga\bfC_\hbar(\ga,\al)
\wp_\hbar(\ga)-\wp_\hbar(\al-\ga))\vf_{\hbar\ep\al}(w)\,
 $$
Comparing the l.h.s. and the r.h.s. we come to the equations of motion. $\blacksquare$

\subsection{Trigonometric limit}

The trigonometric limit is associated with $Im(\tau)\rightarrow+\infty$, where $\tau$ is the modulus of torus
$\tau=\omega_2/\omega_1$. In this limit the elliptic curve degenerates into pinched torus. For our purposes it is natural to
consider it as an infinite complex cylinder with one additional point at infinity.
 In our systems $\tau$ also plays the role of time and we can no longer
consider it as a free parameter of the limit. To avoid this difficulty
we introduce the scaling of $\tau$ in the following way
 $$
\tau=\tau^{tr}\tau_2\,,
 $$
where $\tau^{tr}$ plays the role of time and $\tau_2$ is a free parameter.
With this substitution the equations of motion (\ref{4.5}) are modified in a simple way.
% that $\kappa$ is substituted by $\widetilde\kappa=\kappa/\tau_2$.
%
The parameter of the elliptic curve
$\tau$ is originally constrained to $Im(\tau)>0$. In what follows we consider $\tau^{tr}$ as
real and $\tau^{tr}>c>0$ for some fixed $c$, while $Im(\tau_2)>0$. Then we take $Im(\tau_2)\rightarrow+\infty$ and
simultaneously put $\ep_2\to 0$, $\kappa\rightarrow\infty$ in such a way that combinations
 \begin{equation*}
\hbar\epsilon_2\tau_2=\widetilde\epsilon_2,\qquad
\dfrac{\kappa}{\tau_2}=\widetilde\kappa,\qquad
\dfrac w{\tau_2}=\widetilde w
 \end{equation*}
remain fixed. One can note that with this substitution the corresponding Lax equation
 preserves the form (\ref{ltop}) with
$\widetilde\kappa$ and $\widetilde w$ instead of $\kappa$ and $w$.
Using expansion (\ref{eq:E2Series}) we get the limiting Hamiltonian
 \begin{equation}
\label{eq:NCTTrigHam}
H^{tr}=\frac{\pi^2}2\sum_{\al\in\mZ^{(2)}} S_\al S_{-\al}
\sin^{-2} \bigl(\pi(\widetilde\ep_1\al_1+\widetilde\ep_2\al_2\tau^{tr})\bigr),
 \end{equation}
where $\widetilde\epsilon_1=\hbar\epsilon_1$.
The Hamilton equations of motion are
 \beq{trnt}
   \begin{array}{|c|}
  \hline\\
\widetilde\kappa\partial_{\tau^{tr}}S_{\alpha}=\pi^2\sum_{\gamma\in\mathbb Z^{(2)}}
\dfrac{\textbf{C}_{\hbar}(\gamma,\alpha)S_{\alpha+\gamma}S_{-\gamma}}
{\sin^2\brc{\pi(\widetilde\ep_1\gamma_1+\widetilde\ep_2\gamma_2\tau^{tr})}}\,.
\\ \ \\
  \hline
  \end{array}
   \eq
These equations of motion have the form (\ref{eat}) with $\widetilde\kappa$ instead of $\kappa$ and
 $$
J=\{\pi^2\sin^{-2} \bigl(\widetilde\ep_1\al_1+\widetilde\ep_2\al_2\tau^{tr}\bigr)\}.
 $$
If we restrict ourselves to $0<Re(\widetilde w)<1$ the Hamilton equations of
motion can be presented as an equation of the zero curvature
 \begin{equation}
\label{eq:NCTTrigZeroCurvature}
\widetilde\ka\p_{\tau^{tr}} L^{tr}-\widetilde\kappa\p_{\widetilde w} M^{tr}+[M^{tr},L^{tr}]=0,
 \end{equation}
with the Lax matrices $L^{tr},M^{tr}$ constructed as the limit of the elliptic ones
 \begin{equation*}
L^{tr}=\lim_{q_2\rightarrow0}L,\qquad
M^{tr}=\lim_{q_2\rightarrow0}M,
 \end{equation*}
where $q_2 \equiv\textbf{e}(\tau_2)$. Using (\ref{eq:PhiDec}) and (\ref{eq:E1Dec}) we get the following limit of the Lax
pair:
 \begin{equation}
\label{eq:NCTTrigL}
L^{tr}(w)=
\pi\sum_{\al\in\mZ^{(2)}}
S_\al\textbf{e}\brc{\widetilde\epsilon_2\alpha_2\widetilde w}
\brc{\cot\brc{\pi(\tilde\ep_1\al_1+\tilde\ep_2\al_2\tau^{tr})}+\imath}
T^\al\,,
 \end{equation}
 \begin{equation}
\label{eq:NCTTrigM}
M^{tr}(w)=
\frac{\pi}{2\imath}\sum_{\al\in\mZ^{(2)}}
S_\al\dfrac{\textbf{e}\brc{\widetilde\epsilon_2\alpha_2\widetilde w}}
{\sin^2\brc{\pi(\tilde\ep_1\al_1+\tilde\ep_2\al_2\tau^{tr})}}
T^\al\,,
 \end{equation}

%%%%%%%%%%%%%%%%%%%%%%%%%%%%%%%%%%%%%%%%%%%%%%%%%%%%%%%%%%%%%%%%%%%%%%%%%%%%%%%%%%%%%%%%%%

\subsection{Rational limit}
\label{sec:NCTRat}

Here we reduce the system constructed in previous subsection further with a limit which we call rational.
 This limit is motivated by the analogy with rational limit of the finite-dimensional monodromy-preserving systems.
 First, we make the following substitution:
 \begin{equation*}
\tau^{tr}=\tau^ra^2,\qquad
\widetilde w=w^ra,\qquad
\widetilde\epsilon_1=\epsilon_1^ra,\qquad
\widetilde\epsilon_2=\dfrac{\epsilon_2^r}a,
 \end{equation*}
where $a$ in the finite-dimensional case had the meaning of the inverse first period
of the original curve $a\propto1/\omega_1$.
Using this substitution we adjust the scaling of the Hamiltonian and the Lax pair in such a way that equations of motion and
the equation of  zero curvature preserve the form after this substitution with new $\tau^r$, $w^r$. Then we take the limit
$a\rightarrow0$ using the above scalings
 \begin{equation*}
H^r=\lim_{a\rightarrow0}a^2H^{tr},\qquad
L^r=\lim_{a\rightarrow0}aL^{tr},\qquad
M^r=\lim_{a\rightarrow0}a^2M^{tr}.
 \end{equation*}
Thus, for the Hamiltonian we have:
 \begin{equation}
\label{eq:NCTRatHam}
H^{r}=\dfrac12\sum_{\al\in\mZ^{(2)}}
\dfrac{S_\al S_{-\al}}{\bigl(\ep_1^r\al_1+\ep_2^r\al_2\tau^r\bigr)^2}.
 \end{equation}
The corresponding Hamilton equations have the form (\ref{4.6a})
 \beq{req}
   \begin{array}{|c|}
  \hline\\
\kappa^r\p_{\tau^r}\bfS(x)=[\bfS(x),(\p_{\bar Z})^{-2}\bfS(x)]_\hbar\,.\\ \ \\
  \hline
  \end{array}
   \eq
Equation (\ref{req}) can be presented as an equation of the zero curvature with limiting Lax matrices $L^r$ and $M^r$ which
have the following form:
 \begin{equation}
\label{eq:NCTRatL}
L^{r}(w)=
\sum_{\al\in\mZ^{(2)}}
S_\al\dfrac{\textbf{e}\brc{\epsilon_2^r\alpha_2w^r}}
{\ep_1^r\al_1+\ep_2^r\al_2\tau^{r}}
T^\al\,,
 \end{equation}
 \begin{equation}
\label{eq:NCTRatM}
M^{r}(w)=\dfrac1{2\pi\imath}
\sum_{\al\in\mZ^{(2)}}
S_\al\dfrac{\textbf{e}\brc{\epsilon_2^r\alpha_2w^r}}{\brc{\ep_1^r\al_1+\ep_2^r\al_2\tau^{r}}^2}
T^\al\,,
 \end{equation}

%%%%%%%%%%%%%%%%%%%%%%%%%%%%%%%%%%%%%%%%%%%%%%%%%%%%%%%%%%%%%%%%%%%%%%%%%%%%%%%%

\subsection{Scaling limit}
\label{sec:ILtheta}
To define the limit we decompose the parameter $\tau$ of the elliptic curve as
 \beq{dectau}
\tau=\tau_1+\tau_2\,,
 \eq
where $\tau_1$ plays the role of time of the limiting system and $\tau_2$ gives
 the trigonometric limit $\Im m\tau_2\rightarrow+\infty$.
The limiting procedure under consideration consists of the shift of the spectral parameter
 \beq{wshif}
w=\wt w +\tau/2\,,
 \eq
the scalings of the coordinates
 \beq{InScale}
S_\alpha=\wt S_\alpha q_2^{-g(\alpha_2)}\,,\quad\mbox{where~} q_2 \equiv\bfe(\tau_2)\,,\quad
g(\alpha_2)=\hbar\epsilon_2\frac{1-\delta_{\alpha_2,0}}{2}\,,
\eq
and the trigonometric limit $\Im
m\tau_2\rightarrow+\infty$.
After scalings (\ref{InScale}), in the limit $\Im m\tau_2\rightarrow+\infty$
we obtain the contraction of the Poisson algebra (\ref{lpb})
 \beqnl
\bfi{\wt S_\al,\wt S_\be}=\bfC_\hbar(\al,\be)\wt S_{\al+\be}\,q_2^{g(\al_2)+g(\be_2)-g(\al_2+\be_2)},
 \eq
where $\wt S_\alpha \equiv\wt S_{\alpha_1,\alpha_2}$, $\alpha\in\mZ^{(2)}$. Scaled coordinates $\wt S_\alpha$ with the
Poisson brackets form an algebra  provided that
 \beq{LiCon}
\forall\alpha_2,\beta_2\in\mZ:\quad g(\al_2)+g(\be_2)-g(\al_2+\be_2)\geqslant0.
 \eq
For $g(\alpha_2)=\theta\epsilon_2\brc{1-\delta_{\alpha_20}}/2$ condition (\ref{LiCon}) is
trivial and we can write down all nonzero brackets corresponding to the equality in (\ref{LiCon})
 \beq{ContrAlg}
\bfi{\wt S_{\alpha_1,0},\wt S_{\beta_1,\beta_2}}=\frac1{\pi\hbar}\sin\brc{\pi\hbar\alpha_1\beta_2}\wt S_{\alpha_1+\beta_1,\beta_2}.
 \eq
An important component of the procedure is that we consider
the subsystem with arbitrary large but finite number of nonzero coordinates $S_\alpha$.
%Though this subsystem is open in the initial system,
%it turns out to be closed in the limiting one.
%\mo{Nichego ne ponial.}
To compute the limits of the Hamiltonian (\ref{h1}) and the Lax operators (\ref{l1}), (\ref{mp2}) we consider those values of
constants $\hbar, \epsilon_2$ that for all nonzero coordinates $S_\alpha$ the following condition is true:
 \beqnl
\left|\hbar\epsilon_2\alpha_2\right|<1.
 \eq
Then for the Hamiltonian of the limiting system we have
 \begin{equation}
\label{InHam}
H=-\frac{\pi^2}{2}\sum_{\alpha_1\in\mZ\backslash\bfi{0}}
\frac{\wt S_{\alpha_1,0}\wt S_{-\alpha_1,0}}{\sin^2\brc{\pi\hbar
\epsilon_1\alpha_1}} +4\pi^2q_1^{\hbar\epsilon_2} \sum_{\alpha_1\in\mathbb Z}\bfe
\brc{\hbar\epsilon_1\alpha_1}\wt S_{\alpha_1,1}\wt S_{-\alpha_1,-1},
 \end{equation}
where $q_1 \equiv\bfe(\tau_1)$. Note that coordinates included in the Hamiltonian form a subalgebra of the limiting Poisson
algebra (\ref{ContrAlg}).
 Thus, the Hamilton equations of motion for these coordinates do not depend on the coordinates
 which are not included in the Hamiltonian. %\mo{!!!}%
Equations of motion with respect to the brackets (\ref{ContrAlg})
 \beqnl
\p_{\tau_1}\wt S_{\alpha}=\bfi{H,\wt S_{\alpha}}
 \eq
can be also obtained as a limit of (\ref{4.5}).
For coordinates included in the Hamiltonian the equations of motion are of the form
 \beq{inct}
\fbox{$
 \begin{array}{l}
\p_{\tau_1}\wt S_{\gamma_1,0}=
-\frac{4\pi}\te q_1^{\te\epsilon_2}\sum_{\alpha_1\in\mZ}
\bfe\brc{\te\epsilon_1\alpha_1}\sin(\pi\te\gamma_1)
\brc{\wt S_{\gamma_1+\alpha_1,1}\wt S_{-\alpha_1,-1}-\wt
S_{\gamma_1-\alpha_1,-1}\wt S_{\alpha_11}}\,, \\
  \p_{\tau_1}\wt S_{\gamma_1,1}=-\frac\pi\te\sum_{\alpha_1\in\mZ\backslash\bfi{0}}
  \wt S_{\alpha_1+\gamma_1,1}
  \frac{\sin(\pi\te\alpha_1)}{\sin^2\brc{\pi\te\epsilon_1\alpha_1}}\wt S_{-\alpha_1,0}\,, \\
\p_{\tau_1}\wt S_{\gamma_1,-1}=\frac\pi\te\sum_{\alpha_1\in\mZ\backslash\bfi{0}}
\wt S_{\alpha_1+\gamma_1,-1}
\frac{\sin(\pi\te\alpha_1)}{\sin^2\brc{\pi\te\epsilon_1\alpha_1}}\wt S_{-\alpha_1,0}\,.
 \end{array}
$}
 \eq
%\mo{$\al_1\neq\pm\ga_1$?}
%
To construct the Lax representation for the equations (\ref{inct}) we consider
the limit of the Lax operators (\ref{l1}), (\ref{mp2}) and the equation (\ref{ltop}).
Since the shift of the spectral parameter (\ref{wshif}) in the limiting procedure
under consideration is time-dependent, equation (\ref{ltop}) turns into
 \beqnl
\p_{\tau}L-\frac{1}{2\pi \imath}\p_{\wt w}\brc{M +\pi\imath L}=\bsq{L,M},
 \eq
where $L=L\brc{\textbf{S},\wt w+\tau/2,\tau}$, $M=M\brc{\textbf{S},\wt w+\tau/2,\tau}.$ Thus, the Lax pair of the limiting system is defined via
 \beqnl
\wt L=\lim_{\Im m\tau_2\rightarrow+\infty}L,\quad \wt M=\lim_{\Im m\tau_2\rightarrow+\infty} \brc{M+\pi\imath L}
 \eq
and the Lax equation assumes the form
 \beq{InLaxEq}
\p_{\tau_1}\wt L-\frac{1}{2\pi \imath}\p_{\wt w}\wt M=\bsq{\wt L,\wt M},
 \eq
where
 \begin{eqnarray}
&\displaystyle\wt L=\pi\sum_{\alpha_1\in\mZ\backslash\bfi{0}}
\frac{\bfe\brc{-\hbar\epsilon_1\alpha_1/2}}{\sin\brc{\pi\hbar\epsilon_1\alpha_1}}\wt S_{\alpha_1,0}T^{\alpha_1,0}
-2\pi \imath q_1^{\hbar\epsilon_2/2}\sum_{\alpha_1\in\mZ}\bfe(\hbar\epsilon_2\wt w)
\wt S_{\alpha_1,1}T^{\alpha_1,1}+\nonumber\svs
&\displaystyle+2\pi \imath q_1^{\hbar\epsilon_2/2}\sum_{\alpha_1\in\mZ}
\bfe(-\hbar\epsilon_1\alpha_1-\hbar\epsilon_2\wt w)\wt S_{\alpha_1,-1}T^{\alpha_1,-1},\label{InL}
 \end{eqnarray}
 \begin{eqnarray}
&\displaystyle\wt M=-\frac{\pi^2}2\sum_{\alpha_1\in\mZ\backslash\bfi{0}} \frac{1+\bfe\brc{-\hbar\epsilon_1\alpha_1}}
{\sin^2\brc{\pi\hbar\epsilon_1\alpha_1}} \wt
S_{\alpha_1,0}T^{\alpha_1,0}
+2\pi^2 q_1^{\hbar\epsilon_2/2}\sum_{\alpha_1\in\mZ}\bfe(\hbar\epsilon_2\wt w)\wt S_{\alpha_1,1}T^{\alpha_1,1}+\nonumber\svs
&\displaystyle+2\pi^2 q_1^{\hbar\epsilon_2/2}\sum_{\alpha_1\in\mZ}
\bfe(-\hbar\epsilon_1\alpha_1-\hbar\epsilon_2\wt w)\wt S_{\alpha_1,-1}T^{\alpha_1,-1}.\label{InM}
 \end{eqnarray}
As one can see, coordinates $\wt S_{\alpha}$, $|\al_2|>1$ are not present in the Lax operators.
 Therefore, we have the Lax representation (\ref{InLaxEq}) for the equations of motion (\ref{inct}).
Since the Hamiltonian (\ref{InHam}) and Lax operators (\ref{InL}), (\ref{InM}) depend only
 on coordinates of the form $\wt S_\al$, $|\al_2|\leqslant1$,
 we can pass to the following three field variables:
 \beqnl
h=h(x_1)=\sum_{\alpha_1\in\mZ\backslash\{0\}}
\frac1{\sin^2\brc{\pi\hbar\epsilon_1\alpha_1}}\wt S_{\alpha_1,0}\bfe\brc{\alpha_1x_1},
 \eq
 \beqnl
f=f(x_1)=\sum_{\alpha_1\in\mZ}\wt S_{\alpha_1,1}\bfe\brc{\alpha_1x_1},\quad
g=g(x_1)=\sum_{\alpha_1\in\mZ}\wt S_{\alpha_1,-1}\bfe\brc{\alpha_1x_1}.
 \eq
Then the Hamiltonian (\ref{InHam}) can be rewritten as follows
 \beqnl
H=\pi^2\int_{S^1}\brc{\frac12h(x_1)\sinh^2\brc{\frac{\hbar\epsilon_1\partial_{x_1}}2}h(x_1)+
 4q_1^{\hbar\epsilon_2}g(x_1)\rme^{-\hbar\epsilon_1\partial_{x_1}}f(x_1)}\rmd x_1.
 \eq
The equations of motion %(\ref{eqS10})--(\ref{eqS1-1})% \mo{?}
in terms of field variables
 $f$, $g$, and $h$ acquire the following form:
 \begin{eqnarray}
\sinh^2\brc{\frac{\hbar\epsilon_1\partial_{x_1}}2}\p_{\tau_1} h&=&-\frac{4\pi\imath}{\hbar}
 q_1^{\hbar\epsilon_2}\sinh\brc{\frac{\hbar\partial_{x_1}}2}
 \brc{f\rme^{-\epsilon_1\partial_{x_1}}g-g\rme^{\epsilon_1\partial_{x_1}}f},\\
\p_{\tau_1}f&=&-\frac{\pi\imath}\hbar f\sinh\brc{\frac{\hbar\partial_{x_1}}2}h, \label{InEqf11}\\
\p_{\tau_1}g&=&\frac{\pi\imath}\hbar g\sinh\brc{\frac{\hbar\partial_{x_1}}2}h. \label{InEqg11}
 \end{eqnarray}
Lax operators in terms of field variables can be defined via the shift operators $\rme^{\pm\hbar\p_{x_1}}$:
 \begin{eqnarray}
\wt L= \frac1{4\hbar}\brc{h(x_1)-h(x_1-\hbar\epsilon_1)}+\frac\imath\hbar q_1^{\hbar\epsilon_2/2}
\bfe\brc{\hbar\epsilon_2\wt w}f(x_1+\frac\hbar2)\rme^{\hbar\p_{x_1}}-\nonumber\\
-\frac\imath\hbar q_1^{\hbar\epsilon_2/2}\bfe\brc{-\hbar\epsilon_2\wt w} g(x_1-\hbar\epsilon_1-\frac\hbar2)\rme^{-\hbar\p_{x_1}},
 \end{eqnarray}
 \begin{eqnarray}
\wt M=-\frac{\pi\imath}{4\hbar}\brc{h(x_1)+h(x_1-\hbar\epsilon_1)}-\frac\pi\hbar q_1^{\hbar\epsilon_2/2}
 \bfe\brc{\hbar\epsilon_2\wt w}f(x_1+\frac\hbar2)\rme^{\hbar\p_{x_1}}-\nonumber\\
-\frac\pi\hbar q_1^{\hbar\epsilon_2/2}\bfe\brc{-\hbar\epsilon_2\wt w}
g(x_1-\hbar\epsilon_1-\frac\hbar2)\rme^{-\hbar\p_{x_1}}.
 \end{eqnarray}

%%%%%%%%%%%%%%%%%%%%%%%%%%%%%%%%%%%%%%%%%%%%%%%%%%%%%%%%%%%
%%%%%%%%%%%%%%%%%%%%%%%%%%%%%%%%%%%%%%%%%%%%%%%%%%%%%%%%%%%%

\subsection{Dispersionless limit}

\subsubsection{General case}
In the "dispersionless" limit $(\hbar\to 0)$ the Lie algebra $sin_\hbar$ becomes the Poisson-Lie algebra $Ham(T^2)$ of
Hamiltonians on the two-dimensional torus, see (\ref{Ham}). It can be represented by the Lie algebra of the corresponding
divergence-free vector fields $SVect(T^2)$. More precisely, to pass from $Ham(T^2)$ to $SVect(T^2)$ one has to discard the
constant Hamiltonians, but add the ``flux'' vector fields $\partial/\partial x_1$ and $\partial/\partial x_2$ corresponding
to multivalued Hamiltonian functions $x_1$ and $ x_2$ on the torus.
We define the non-autonomous top related to
 the Lie group $SDiff(T^2)$ by
considering the limit $\hbar\to 0$ of the described above systems.
Let
 \beq{rs}
\hbar\to 0,~\ep_{1,2}\to\infty,~ ~{\rm such~that~} \lim_{\hbar\to 0}(\hbar\ep_{1,2})=\ep'_{1,2}<1\,,~
~\ep'_{1,2}~{\rm are~irrational}\,.
 \eq
In what follows we drop the superscript $'$.
Let $\bfS=\sum_\al S^{\al}\bfe(\al\cdot x)\in Ham^*(T^2)$,
where $\bfe(\al\cdot x)$ is the Fourier basis (\ref{fb}) of $Ham^*(T^2)$.
In  the Fourier basis, the linear Poisson bracket assumes the form (\ref{3.7b})
 \beq{3.7a}
\{S^\al,S^\be\}_1=(\al\times\be)S^{\al+\be}\,,~\al\times\be=\al_1\be_2-\al_2\be_1\,.
 \eq
The conjugate inertia operator $\bfJ\,:\,Ham^*(T^2)\to Ham(T^2)$  becomes
 $$
\bfJ\,:~S^\al \to \wp(\eal) S^\al\,,~~
\wp(\eal)=\wp(\ep_1\al_1+\ep_2\al_2\tau;\tau)\,,
 $$
where $\al\in\mZ^{(2)}$.
%(\ref{lat}).
The operator is well defined since $\ep_j$ are irrational.
%In other words, the operator $\bfJ$ is the pseudodifferential operator
% \beq{defJD}
%\bfJ\,:\bfS(x) \to
%\left\{
% \begin{array}{l}
%  0~ {\rm for~almost~all}~\al\in \mZ^{(2)} \\
% \wp(\eal)S^\al~   {\rm for~a~finite~subset} ~\mZ^{(2)}_f
% \subset \mZ^{(2)}\,.
% \end{array}
%\right.
% \eq
In terms of fields
 \beq{jit}
\bfJ\,:\bfS(x) \to\wp(\bp_Z)\bfS(x)
 \eq
where
 $$
\bp_{Z}=\frac{1}{2\pi \imath(\bar\tau-\tau)}(\ep_1\p_1+\ep_2\tau\p_2)
 $$
 is the operator of the complex structure on the commutative torus $T^2$.
In fact, the complex structure depends on the ratio
$\tau{\ep_2}/{\ep_1}$.
The quadratic Hamiltonian of the system is
 \beq{5.7_0}
H=-\oh\sum_\ga S^\ga\wp(\ega)S^{-\ga}=2\pi^2
\int_{T^2}\bfS(\wp(\bp_{Z})\bfS)\,,
 \eq
(see  (\ref{int})),
and the corresponding equations of motion are (see  (\ref{pal}))
 \beq{5.8}
   \begin{array}{|c|}
  \hline\\
\p_\tau\bfS=\{\bfS,\wp(\bp_{Z})\bfS\}\,.
\\ \ \\
  \hline
  \end{array}
   \eq
For the velocities the equation takes the form
 \beq{5.9}
\p_\tau(\wp^{-1}(\bp_{Z})\bfF)=\{\wp^{-1}(\bp_Z)\bfF,\bfF\}\,.
 \eq
%This form is useless, since the Weierstrass  function has zeroes in the fundamental
%domain. We consider below its the trigonometric and the rational limits.
%
Define the Lax operator
 \beq{3.7}
L(x;w)=\sum_{\al\in\mZ^{(2)}} S^{\al}\vf(\eal,w)\bfe(\al\cdot
x)\,.
 \eq
%The conditions on the phase space formulated below, see (\ref{ps0}),
%ensure that the
%operator $L^{rot}(x,z)$ is well defined
%for $z\in E_\tau,~z\neq 0$.
Notice that
 $$
\int_{T^2}
L(x,w)=0\,.
 $$
In this way $L$ defines divergence-free vector field $\p_1L\p_2-\p_2L\p_1$.
 \begin{predl}\label{pr5.1}
The equations of motion (\ref{5.8})
 have the dispersionless  Lax
representation
 $$
\p_\tau L-\p_wM=\{L,M\}
 $$
with $L$ given by (\ref{3.7}) and
 \beq{4.7}
M(x;w)=\sum_{\al\in\mZ^{(2)}} S^{\al}f(\eal,w)\bfe(\al\cdot
x)\,,
 \eq
where
 $$
f(\eal,w)=
\bfe(\ep_2\al_2w)\p_u\phi(u,w)|_{u=\ep_1\al_1+\tau\ep_2\al_2}\,.
 $$
 \end{predl}
The proof of Proposition \ref{pr5.1} is the same as for Proposition \ref{pr3.1}.
Notice that operators $L$ and $M$ satisfy the quasi-periodicity properties
 $$
 \begin{array}{cc}
L(x_1,x_2;w)=L (x_1+\ep_2,x_2;w+1)\,, &L (x_1,x_2;w)=L (x_1,x_2-\ep_1;w+\tau)\,, \\
M (x_1,x_2;w)=M (x_1+\ep_2,x_2;w+1)\,, & M (x_1,x_2;w)-M (x_1,x_2-\ep_1;w+\tau)=
2\pi i L (x_1,x_2,w)\,.
 \end{array}
 $$

%%%%%%%%%%%%%%%%%%%%%%%%%%%%%%%%%%%%%%%%%%%%%%%%%%%%%5

\subsubsection{Trigonometric  limit}

In the trigonometric case the equations of motion (\ref{4.6a})
assume the form
 \beq{4.6ad}
   \begin{array}{|c|}
  \hline\\
\p_\tau\bfS(x)=\{\bfS(x),\brc{\pi^2/\widetilde\kappa^2}\sin^{-2}(\pi\bp_{ Z})\bfS(x)\}\\ \ \\
  \hline
  \end{array}
    ~~~\bp_{
Z}=\f1{2\pi\imath}(\widetilde\ep_1\p_1+\widetilde\ep_2\tau^{tr}\p_2)\,.
 \eq
%\mo{What is $\widetilde\ep$?}
In terms of the velocities $\bfF(x)$ (\ref{3.1a})
 $$
\bfF(x)=(\bfJ^{tr})^{-1}\bfS(x)\,,~~\bfJ^{tr}=\brc{\widetilde\kappa^2/\pi^2}\sin^{2}(\pi\bp_{ Z})
 $$
 $$
((\bfJ^{tr})^{-1}\bfF)(x)=
\bfF(x_1+2\pi\imath\widetilde\ep_1,x_2+2\pi\imath\widetilde\ep_2\tau^{tr})+
\bfF(x_1-2\pi\imath\widetilde\ep_1,x_2-2\pi\imath\widetilde\ep_2\tau^{tr})-2\bfF(x)
 $$
 it can be rewritten as
 \beq{3.1ab}
   \begin{array}{|c|}
  \hline\\
\sin^{2}(\pi\bp_{ Z})\p_\tau\bfF(x) =\{\sin^{2}(\pi\bp_{ Z})\bfF(x),\bfF(x)\}-\pi\ep_2\sin(\pi\bp_{ Z})\cos(\pi\bp_{
Z})\p_2\bfF(x)\\ \ \\
  \hline
  \end{array}
   \eq
or
 \beq{32}
\p_\tau\Bigl(\bfF(x_1+2\pi\imath\widetilde\ep_1,x_2+2\pi\imath\widetilde\ep_2\tau^r)+
\bfF(x_1-2\pi\imath\widetilde\ep_1,x_2-2\pi\imath\widetilde\ep_2\tau^r)-2\bfF(x)\Bigr)=
 \eq
 $$
=\Bigl\{\bigl(\bfF(x_1+2\pi\imath\widetilde\ep_1,x_2+2\pi\imath\widetilde\ep_2\tau^{tr})+
\bfF(x_1-2\pi\imath\widetilde\ep_1,x_2-2\pi\imath\widetilde\ep_2\tau^{tr})\bigr),\bfF(x)\Bigr\}-
 $$
 $$
-\pi\widetilde\ep_2\p_2\Bigl(\bfF(x_1+2\pi\imath\widetilde\ep_1,x_2+2\pi\imath\widetilde\ep_2\tau^{tr})-
\bfF(x_1-2\pi\imath\widetilde\ep_1,x_2-2\pi\imath\widetilde\ep_2\tau^{tr})\Bigr)\,.
 $$
This equation can be presented as an equation of the zero curvature with Lax matrices
from (\ref{eq:NCTTrigL}) and (\ref{eq:NCTTrigM}). In dispersionless limit
the decomposition with respect to generators $T^{\alpha}$ turns into Fourier decomposition:
 \begin{equation}
\label{eq:NCTDispersionlessTrigL}
L^{tr}(x_1,x_2,\widetilde w)=
\pi\sum_{\al\in\mZ_f^{(2)}}
S^\al\textbf{e}\brc{\widetilde\epsilon_2\alpha_2\widetilde w}
\brc{\cot\brc{\pi(\tilde\ep_1\al_1+\tilde\ep_2\al_2\tau^{tr})}+\imath}
\textbf{e}\brc{\alpha_1x_1+\alpha_2x_2}\,,
 \end{equation}
 \begin{equation}
\label{eq:NCTDispersionlessTrigM}
M^{tr}(x_1,x_2,\widetilde w)=
\frac{\pi}{2\imath}\sum_{\al\in\mZ_f^{(2)}}
S^\al\dfrac{\textbf{e}\brc{\widetilde\epsilon_2\alpha_2\widetilde w}}
{\sin^2\brc{\pi(\tilde\ep_1\al_1+\tilde\ep_2\al_2\tau^{tr})}}
\textbf{e}\brc{\alpha_1x_1+\alpha_2x_2}\,,
 \end{equation}
Moreover, (\ref{3.1ab}) (or (\ref{32})) is the monodromy preserving condition
for the linear partial differential equation
 \beq{isot}
\p_{\widetilde w}\Psi(x_1,x_2,\widetilde w)+\{L(x_1,x_2,\widetilde w),
\Psi(x_1,x_2,\widetilde w)\}=0\,.
 \eq

%%%%%%%%%%%%%%%%%%%%%%%%%%%%%%%%%%%%%%%%%%%%%%%%%%%%%%%%
\subsubsection{Rational case}

Here we apply dispersionless limit to the rational system constructed in subsection \ref{sec:NCTRat}. In this way we come to
the equations of motion for the Fourier modes
 \begin{equation}
\label{eq:NCTDispersionlessRatEq}
\p_{\tau^r}S^\al =\sum _{\ga\in\mZ^{(2)}} S^\ga \f1{(\epsilon^r\cdot\gamma)^2}S^{\al-\ga}\,.~~~
(\epsilon^r\cdot\gamma=\ep_1^r\ga_1+\ep_2^r\ga_2\tau^{tr})\,,
 \end{equation}
where in comparison with (\ref{req}) we omitted $\kappa^r$. This can be done since in rational case it is nothing else than a scale of free parameters $\epsilon^r_1$ and $\epsilon^r_2$. In terms of fields $\bfS(x)\in Ham(T^2)$ the equations of motion (\ref{eq:NCTDispersionlessRatEq}) turn into:
 \beq{cem}
\p_{\tau^r}\bfS=\{\bfS,\bp_Z^{-2}\bfS\}\,.
 \eq
The equations of motion are defined by the corresponding quadratic Hamiltonian
 \beq{qh}
H=\frac{1}{2}\int_{T^2}\bfS \bp_Z^{-2}\bfS=
\frac{1}{2}\sum_{\al\in\mZ^{(2)}} \f1{(\epsilon^r\cdot\alpha)^2} S^{\al} S^{-\al}\,.
 \eq
with respect to the linear Poisson bracket on $Ham(T^2)$.
It is worth noting that the equation (\ref{cem}) is equivalent to some partial differential
 equation. To show that, we rewrite it in terms of the angular velocities $\bfF(x)$
 \beq{meq}
   \begin{array}{|c|}
  \hline\\
\p_{\tau^r}\bp^2_Z\bfF(x,\tau^r)-\{\bp^2_Z\bfF(x,\tau^r),\bfF(x,\tau^r)\}+
\ep_2^r\p_2\bp_Z\bfF(x,\tau^r)\\ \ \\
  \hline
  \end{array}
   \eq
where $\bp_Z=\dfrac{\ep_1^r\p_1-\ep_2^r\tau^r\p_2}{2\pi\imath}.$
%\mo{$\ep_1^r\p_1+\ep_2^r\tau^r\p_2$?}
%
This equation can be presented as an equation of the zero curvature.
Using (\ref{eq:NCTRatL}) and (\ref{eq:NCTRatM}) we get
 \begin{equation}
\label{eq:NCTDispersionlessRatL}
L^{r}(x_1,x_2,w^r)=
\sum_{\al\in\mZ^{(2)}}
S^\al\dfrac{\textbf{e}\brc{\epsilon_2^r\alpha_2w^r}}
{\ep_1^r\al_1+\ep_2^r\al_2\tau^{r}}
\textbf{e}\brc{\alpha_1x_1+\alpha_2x_2}=\brc{\bar\partial_Z}^{-1}\bfS(x_1,x_2+\epsilon_2^r w^r)\,,
 \end{equation}
%\mo{Why $\textbf{e}\brc{\epsilon_2^r\alpha_2w^r}$ acts on $\bfS$?}
 \begin{equation}
\label{eq:NCTDispersionlessRatM}
M^{r}(x_1,x_2,w^r)=\dfrac1{2\pi\imath}
\sum_{\al\in\mZ^{(2)}}
S^\al\dfrac{\textbf{e}\brc{\epsilon_2^r\alpha_2w^r}}{\brc{\ep_1^r\al_1+\ep_2^r\al_2\tau^{r}}^2}
\textbf{e}\brc{\alpha_1x_1+\alpha_2x_2}
 \end{equation}
 $$
=\f1{2\pi\imath}\brc{\bar\partial_Z}^{-2}\bfS(x_1,x_2+\epsilon_2^r w^r)\,.
 $$
%\mo{I did not succeed to check $\p_\tau L-\p_w M=\{L,M\}$ in terms of the fields $\bfS$.}
The equation (\ref{meq})  is the monodromy preserving condition
for the linear partial differential equation
 \beq{isot2}
\p_{w^r}\Psi(x_1,x_2,\tau^r,w^r)+\{L^r(x_1,x_2,\tau^r,w^r),\Psi(x_1,x_2,\tau^r,w^r)\}=0\,,
 \eq
because the Baker-Akhiezer function satisfies the equation
 $$
\p_{\tau^r}\Psi(x_1,x_2,\tau^r,w^r)+\{M^r(x_1,x_2,\tau^r,w^r),\Psi(x_1,x_2,\tau^r,w^r)\}=0\,.
 $$
%%%%%%%%%%%%%%%%%%%%%%%%%%%%%%%%%%%%%%%%%%%%%%%%%%%%%%%

\subsubsection{Scaling limit}
In this subsection we construct the limiting procedure analogous to the one proposed in section \ref{sec:ILtheta}. We use the same decomposition (\ref{dectau}) of the parameter $\tau$ of the elliptic curve
 \beqnl
\tau=\tau_1+\tau_2,
 \eq
where $\tau_1$ plays the role of time of the limiting system and $\tau_2$ gives the trigonometric limit $\Im m\tau_2\rightarrow+\infty$.
The limiting procedure consists of the shift of the spectral parameter
 \beq{wshifd}
w=\wt w +\tau/2,
 \eq
the scalings of the coordinates
 \beq{InScaleD}
S^\alpha=\wt S^\alpha q_2^{-g(\alpha_2)},\quad q_2 \equiv\bfe(\tau_2),\quad
g(\alpha_2)=\epsilon_2\frac{1-\delta_{\alpha_2,0}}{2},  \eq and the trigonometric limit $\Im m\tau_2\rightarrow+\infty$.
After scalings (\ref{InScale}), we obtain the contraction of Poisson algebra (\ref{3.7a}) in the limit $\Im m\tau_2\rightarrow+\infty$. All nonzero brackets for coordinates $\wt S^{\al}$ of the limiting system have the form
 \beq{ContrAlgD}
\bfi{\wt S^{\alpha_1,0},\wt S^{\beta_1,\beta_2}}=\alpha_1\beta_2\wt S^{\alpha_1+\beta_1,\beta_2},
 \eq
where $\wt S^{\alpha_1,\alpha_2} \equiv\wt S^\alpha$, $\alpha\in\mZ^{(2)}$.
Again, it is important that we consider  the subsystem with arbitrary large but finite number of nonzero coordinates
$S^\alpha$. Though this subsystem is open in the initial system, it turns out to be closed in the limiting one. To compute
the limits of the Hamiltonian (\ref{5.7_0}) and the Lax operators (\ref{3.7}), (\ref{4.7}) we consider those values of
constant $\epsilon_2$ such that for all nonzero coordinates $S^\alpha$ the following condition is true:
 \beqnl
\left|\epsilon_2\alpha_2\right|<1.
 \eq
Then for the Hamiltonian of the limiting system we have
 \begin{equation}
\label{InHamD}
H=-\frac{\pi^2}{2}\sum_{\alpha_1\in\mZ\backslash\bfi{0}}
\frac{\wt S^{\alpha_1,0}\wt S^{-\alpha_1,0}}{\sin^2\brc{\pi\epsilon_1\alpha_1}}
+4\pi^2q_1^{\epsilon_2} \sum_{\alpha_1\in\mathbb Z}\bfe\brc{\epsilon_1\alpha_1}\wt S^{\alpha_1,1}\wt S^{-\alpha_1,-1},
 \end{equation}
where $q_1 \equiv\bfe(\tau_1)$. Note that  coordinates included in the Hamiltonian form a subalgebra of the limiting Poisson
algebra (\ref{ContrAlgD}). Thus, the Hamilton equations of motion for these coordinates
 \beqnl
\p_{\tau_1}\wt S^{\alpha}=\bfi{H,\wt S^{\alpha}}
 \eq
do not depend on the coordinates which are not included in the Hamiltonian.
For the coordinates included in the Hamiltonian the equations of motion are of the form
 \begin{equation}
\label{eqDS10}
\p_{\tau_1}\wt S^{\gamma_1,0}=-4\pi^2q_1^{\epsilon_2}\sum_{\alpha_1\in\mZ}\gamma_1
\brc{\wt S^{\alpha_1+\gamma_1,1}\bfe\brc{\epsilon_1\alpha_1}\wt S^{-\alpha_1,-1}-\wt S^{\alpha_1+\gamma_1,-1}
\bfe\brc{-\epsilon_1\alpha_1}\wt S^{-\alpha_1,1}},
 \end{equation}
 \begin{equation}
\label{eqDS11}
\p_{\tau_1}\wt S^{\gamma_1,1}=-\pi^2\sum_{\alpha_1\in\mZ\backslash\bfi{0}}\wt S^{\alpha_1+\gamma_1,1}
\frac{\alpha_1}{\sin^2\brc{\pi\epsilon_1\alpha_1}}\wt S^{-\alpha_1,0},
 \end{equation}
 \begin{equation}
\label{eqDS1-1}
\p_{\tau_1}\wt S^{\gamma_1,-1}=\pi^2\sum_{\alpha_1\in\mZ\backslash\bfi{0}}\wt S^{\alpha_1+\gamma_1,-1}
\frac{\alpha_1}{\sin^2\brc{\pi\epsilon_1\alpha_1}}\wt S^{-\alpha_1,0}.
 \end{equation}
These equations of motion also can be obtained by applying the dispersionless limit (\ref{rs}).
%to the equations
%(\ref{eqS10})--(\ref{eqS1-1}) from section \ref{sec:ILtheta}.
%
We define Lax pair of the limiting system as limits of operators (\ref{3.7}) and (\ref{4.7}):
 \begin{equation}
\label{InLaxEqD4} \wt L=\lim_{q_2\rightarrow0}L,\quad \wt M=\lim_{q_2\rightarrow0}\brc{M+\pi iL}.
 \end{equation}
This pair admits Lax equation
 \begin{equation}
\label{InLaxEqD}
\p_{\tau_1} \wt L -\frac1{2\pi i}\p_{\wt w}\wt M=\bsq{\wt L,\wt M},
 \end{equation}
where
 \begin{eqnarray}
&\displaystyle\wt L=\pi\sum_{\alpha_1\in\mZ\backslash\bfi{0}} \frac{\bfe\brc{-\epsilon_1 \alpha_1/2}}{\sin\brc{\pi\epsilon_1\alpha_1}}\wt S^{\alpha_1,0}\bfe(\alpha_1x_1)
-2\pi iq_1^{\epsilon_2/2}\sum_{\alpha_1\in\mZ}\bfe(\epsilon_2\wt w)\wt S^{\alpha_1,1}\bfe(\alpha_1x_1+x_2)+\nonumber\svs
&\displaystyle+2\pi iq_1^{\epsilon_2/2}\sum_{\alpha_1\in\mZ}\bfe(-\epsilon_1\alpha_1-\epsilon_2\wt w)\wt S^{\alpha_1,-1}\bfe(\alpha_1x_1-x_2),
 \end{eqnarray}
 \begin{eqnarray}
&\displaystyle\wt M=-\frac{\pi^2}2\sum_{\alpha_1\in\mZ\backslash\bfi{0}} \frac{1+\bfe\brc{-\epsilon_1\alpha_1}}{\sin^2\brc{\pi\epsilon_1\alpha_1}} \wt
S^{\alpha_1,0}\bfe(\alpha_1x_1)
+2\pi^2 q_1^{\epsilon_2/2}\sum_{\alpha_1\in\mZ}\bfe(\epsilon_2\wt w)\wt S^{\alpha_1,1}\bfe(\alpha_1x_1+x_2)+\nonumber\svs
&\displaystyle+2\pi^2 q_1^{\epsilon_2/2}\sum_{\alpha_1\in\mZ}\bfe(-\epsilon_1\alpha_1-\epsilon_2\wt w)\wt S^{\alpha_1,-1}\bfe(\alpha_1x_1-x_2).
 \end{eqnarray}
Equation (\ref{InLaxEqD}) is equivalent to equations of motion (\ref{eqDS10})--(\ref{eqDS1-1}). Thus,  in the limiting system
Lax representation describes the equations of motion only for the coordinates included in the Hamiltonian (\ref{InHamD}).
Since the Hamiltonian (\ref{InHamD}) depends only on coordinates of the form $\wt S^\al$, $|\al_2|\leqslant1$, we  can pass
to the following three field variables:
 \beqnl
h=h(x_1)=\sum_{\alpha_1\in\mZ\backslash\{0\}}\wt S^{\alpha_1,0}\bfe\brc{\alpha_1x_1},
 \eq
 \beqnl
f=f(x_1)=\sum_{\alpha_1\in\mZ}\wt S^{\alpha_1,1}\bfe\brc{\alpha_1x_1},\quad
g=g(x_1)=\sum_{\alpha_1\in\mZ}\wt S^{\alpha_1,-1}\bfe\brc{\alpha_1x_1}.
 \eq
Then the Hamiltonian (\ref{InHamD}) can be rewritten as follows
 \beqnl
H=\pi^2\int_{S^1}\brc{2h(x_1)\sinh^{-2}\brc{\frac{\epsilon_1\partial_{x_1}}2}h(x_1)+
 4q_1^{\epsilon_2}g(x_1)\rme^{-\epsilon_1\partial_{x_1}}f(x_1)}\rmd x_1.
 \eq
Equations of motion (\ref{eqDS10})--(\ref{eqDS1-1}) in terms of field  variables $f$, $g$, and $h$ acquire the following
form:
 \begin{eqnarray}
\p_{\tau_1} h&=&2\pi\imath q_1^{\epsilon_2}
\partial_{x_1}\brc{f\rme^{-\epsilon_1\partial_{x_1}}g-g\rme^{\epsilon_1\partial_{x_1}}f},\\
\p_{\tau_1}f&=&f\frac{\pi\imath}{2\sinh^2\brc{\epsilon_1\partial_{x_1}/2}}\partial_{x_1} h, \label{InEqf}\\
\p_{\tau_1}g&=&-g\frac{\pi\imath}{2\sinh^2\brc{\epsilon_1\partial_{x_1}/2}}\partial_{x_1} h. \label{InEqg}
 \end{eqnarray}
These equations are again Hamiltonian
 \beqnl
\p_{\tau_1}h=\bfi{H,h},\quad\p_{\tau_1}f=\bfi{H,f},\quad\p_{\tau_1}g=\bfi{H,g}
 \eq
with respect to the following "Poisson" brackets:
 \begin{eqnarray}
&\displaystyle\bfi{h(x),f(y)}=\frac{f(y)}{2\pi\imath}\delta'(x-y),\quad
\bfi{h(x),g(y)}=-\frac{g(y)}{2\pi\imath}\delta'(x-y),\nonumber\\
&\bfi{f(x),g(y)}=0,
 \end{eqnarray}
where $\delta'(x-y)$ is the first derivative of the Dirac delta function.
Notice that equations (\ref{InEqf}) and (\ref{InEqg}) imply
 \beqnl
\partial_{\tau_1}\brc{f(x_1,\tau_1)g(x_1,\tau_1)}=0.
 \eq

%%%%%%%%%%%%%%%%%%%%%%%%%%%%%%%%%%%%%%%%%%%%%%%%%%%%%
%%%%%%%%%%%%%%%%%%%%%%%%%%%%%%%%%%%%%%%%%%%%%%%%%%%%%

\section{Appendix}

%\vspace{1.0cm}
%\bigskip
% {\bf \ \ \ List of abbreviations:}
% \vskip5mm

\subsection{List of abbreviations}

{\small
\begin{itemize}
\item PI,..,PVI -- Painlev\'e I,..,VI equations  \hskip26.6mm $\bullet\ $ $Fl^{Aff}$ -- affine flag variety

\item PVI$^{FT}$ -- Painlev\'e VI field theory   \hskip32mm $\bullet\ $ $\om^{KK}$ -- Kirillov-Kostant symplectic form

\item NCT -- noncommutative torus                 \hskip37.2mm $\bullet\ $ $Conn$ -- space of smooth connections

\item NAVZG -- non-autonomous Zhukovsky-Volterra  \hskip10.3mm $\bullet\ $ $FConn$ -- space of flat connections

gyrostat                                         \hskip70.8mm $\bullet\ $ $FBun$ -- moduli space of flat connections

\item $sin_\hbar$ -- noncommutative analogue of sin-algebra \hskip10.8mm $\bullet\ $ $SDiff(M)$ -- group of volume-preserving

\item $SIN_\hbar$ -- group corresponding to $sin_\hbar$            \hskip32mm  diffeomorphisms of $M$

%\item $SDiff(M)$ - group of the volume-preserving

%diffeomorphisms of $M$

%\item $Fl^{Aff}$ - affine flag variety

%\item $\om^{KK}$ - Kirillov-Kostant symplectic form

%\item $Conn$ - space of smooth connections

%\item $FConn$ - space of flat connections

%\item $FBun$ - moduli space of flat connections

\end{itemize}
}

\subsection{Appendix A: Elliptic functions}
\setcounter{equation}{0}
\def\theequation{A.\arabic{equation}}

\emph{Notations.}
 $$
\bfe(x)=\exp\left(2\pi\imath \,x\right)\,,~~q=\bfe(\tau)\,,
 $$
 $$
\om_1\,,~\om_2-{\rm~fundamental~half-periods}\,,~\tau=\om_2/\om_1\,.
 $$
\bigskip
\noindent
{\it The theta  function}:
 \beq{A.1a}
\vth(z|\tau)=\sum_{n\in {\bf Z}}\bfe\left(\oh\left(n+\oh\right)^2\tau+\left(n+\oh\right)\left(z+\oh\right)\right)=
 \eq
 $$
=-2q^{\f1{8}} \sum_{n=0}^\infty(-1)^{n}q^{\oh n(n+1)}\sin\left(2n+1\right)\pi z.
 $$
\bigskip
\noindent
{\it The  Eisenstein functions}
 \beq{A.1}
E_1(z|\tau)=\p_z\log\vth(z|\tau), ~~E_1(z|\tau)|_{z\to0}\sim\f1{z}-2\eta_1(\tau)z\,,
 \eq
 \beq{A.2}
E_2(z|\tau)=-\p_zE_1(z|\tau)=- \p_z^2\log\vth(z|\tau)\,,
~~E_2(z|\tau)|_{z\to0}\sim\f1{z^2}+2\eta_1(\tau)\,.
 \eq
Here
 \beq{A.6}
\eta_1(\tau)=-\frac{1}{6}\frac{\vth'''(0|\tau)}{\vth'(0|\tau)}=-\frac{2\pi
i}{3}\p_\tau \log\vth'(0|\tau)\,.
 \eq

\bigskip
\noindent
{\it Relation to the Weierstrass functions}
 \beq{a100}
\zeta(z,\tau)=E_1(z,\tau)+2\eta_1(\tau)z\,,
 \eq
 \beq{a101}
\wp(z,\tau)=E_2(z,\tau)-2\eta_1(\tau)\,.
 \eq

\bigskip
\noindent
\emph{Important functions:}
 \beq{A.3}
\phi(u,z)=
\frac
{\vth(u+z)\vth'(0)}
{\vth(u)\vth(z)}\,.
 \eq
It has a pole at $z=0$ and
 \beq{A.3a}
\phi(u,z)=\frac{1}{z}+E_1(u)+\frac{z}{2}(E_1^2(u)-\wp(u))+\ldots\,.
 \eq
Let
 \beq{A3c}
f(u,z)=\p_u\phi(u,z)
=\phi(u,z) (E_1(u+z)-E_1(u))\,.
 \eq

\bigskip
\noindent
{\it Addition theorems.}
 \beq{d2}
\phi(u,z)f(v,z)-\phi(v,z)f(u,z)=(E_2(v)-E_2(z))\phi(u+v,z)\,.
 \eq
 \beq{d3}
\phi(u,z)\phi(-u,z)=(E_2(z)-E_2(u))\,.
 \eq
 \beq{ft5}
\phi(u,z)\phi(-u,w)=\phi(u,z-w)[E_1(u+z-w)-E_1(u)+E_1(w)-E_1(z)]\,.
 \eq

\bigskip
\noindent
{\it Heat equation}
 \beq{A.4b}
\p_\tau\phi(u,w)-\f1{2\pi i}\p_u\p_w\phi(u,w)=0\,.
 \eq

\bigskip
\noindent
{\it Parity}
 \beq{A.300}
\phi(u,z)=\phi(z,u)\,,~~\phi(-u,-z)=-\phi(u,z)\,.
 \eq
 \beq{pei}
E_1(-z)=-E_1(z)\,,~~E_2(-z)=E_2(z)\,.
 \eq
 \beq{pfu}
f(-u,-z)=f(u,z)\,.
 \eq

\bigskip
\noindent
{\it Quasi-periodicity}
 \beq{A.11}
\vth(z+1)=-\vth(z)\,,~~~\vth(z+\tau)=-q^{-\oh}e^{-2\pi iz}\vth(z)\,,
 \eq
 \beq{A.12}
E_1(z+1)=E_1(z)\,,~~~E_1(z+\tau)=E_1(z)-2\pi i\,,
 \eq
 \beq{A.13}
E_2(z+1)=E_2(z)\,,~~~E_2(z+\tau)=E_2(z)\,,
 \eq
 \beq{A.14}
\phi(u,z+1)=\phi(u,z)\,,~~~\phi(u,z+\tau)=e^{-2\pi \imath u}\phi(u,z)\,.
 \eq
 \beq{A.15}
f(u,z+1)=f(u,z)\,,~~~f(u,z+\tau)=e^{-2\pi \imath u}f(u,z)-2\pi\imath\phi(u,z)\,.
 \eq

\bigskip
\noindent \emph{Particular values}\vskip2mm
 \beq{pv}
E^{(2j+1)}_2(\tau/2)=E^{(2j+1)}_2(1/2+\tau/2)=E^{(2j+1)}_2(1/2)=0\,,~~(E_2^{(j)}(u)=\p^j_uE_2(u))\,.
 \eq

\bigskip
 \noindent \emph{Degenerations I}.\vskip2mm
 \noindent Let  $\Im m\tau\to+\infty$.
 Then
 \beq{c3}
\vartheta(z|\tau)\sim \sin(\pi z)\,,
 \eq
 \beq{c3a}
E^r_1(z)= \pi\cot(\pi z)\,,
 \eq
 \beq{c4}
\phi^{tr}(u,z)= \pi(\cot \pi u+\cot \pi z)\,,
 \eq
 \beq{c4a}
f(u,z)^{tr}=-\pi^2\sin^{-2} \pi u\,,
 \eq
 \beq{c5}
E^{tr}_2(z)=\frac{\pi^2}{\sin^2(\pi z)}\,.
 \eq
The rational limit of the trigonometric functions
assumes the form
 \beq{c6b}
E^r_1(z)= \f1{ z}\,,
 \eq
 \beq{c6}
\phi^r(u,z)= \f1{u}+\f1{z}\,,
 \eq
 \beq{c6a}
f^r(u,z)= -\f1{u^2}\,,
 \eq
 \beq{c7}
E^r_2(z)=\frac{1}{z^2}\,.
 \eq
The both types of functions satisfy the addition formulae (\ref{d2}), (\ref{d3}).

\bigskip
\noindent \emph{Degenerations II}.

 \noindent To evaluate limits of the Hamiltonian we  need the decomposition of the second Eisenstein
function with shifted argument in trigonometric limit $Im(\tau)\rightarrow+\infty$, or equivalently
$q=\textrm{e}^{2\pi\imath\tau}\rightarrow0$. Directly from definition we get:
 \begin{equation}
\label{eq:E2Series}
E_2\brc{u-g\tau}=\left\{\matr{ll}{
-4\pi^2\dfrac{\textbf{e}\brc{u}}{\brc{1-\textbf{e}\brc{u}}^2},&\bfi{g}=0,\\
-4\pi^2\textbf{e}\brc{-u}q^{g},&0<\bfi{g}<\dfrac12,\\
-4\pi^2q^{1/2}\brc{\textbf{e}\brc{-u}+\textbf{e}\brc{u}},&\bfi{g}=\dfrac12,\\
-4\pi^2\textbf{e}\brc{u}q^{1-\bfi{g}},&\dfrac12<\bfi{g}<1.\\
}\right.
 \end{equation}
To evaluate various limits of Lax pair we need the decomposition  of $\phi$ function with shifted arguments in trigonometric
limit. Using definition (\ref{A.3}) we reduce the expansion of $\phi(u-\sigma\tau,z-\varsigma\tau)$ to the expansion of theta
functions:
 \begin{equation}
\label{eq:ThetaDec}
\phi(u-\sigma\tau,z-\varsigma\tau)=
\dfrac{\vartheta(u+z-(\sigma+\varsigma)\tau)\vartheta'(0)} {\vartheta(u-\sigma\tau)\vartheta(z-\varsigma\tau)},
 \end{equation}
and for the expansion of theta function we have:
 \begin{equation}
\vartheta\brc{z+\sigma\tau}=
\bsq{1+\textbf{o}\brc{1}}q^{\brc{-\flr{\sigma}^2/2+\frac18-\flr{\sigma}\bfi{\sigma}-\bfi{\sigma}/2}}\;
\rme\brc{-\flr{\sigma}z-\frac{\flr{\sigma}}2}\times
 \end{equation}
 \begin{equation}
\times\left\{\matr{ll}{
-2\sin\brc{\pi z},&\bfi{\sigma}=0,\svs
-\imath\;\rme\brc{-\dfrac z2},&\bfi{\sigma}>0,
}\right.
 \end{equation}
where $\flr{\sigma}$ is the integer part of $\sigma$ and $\bfi{\sigma}$ is  the fractional part of $\sigma$. This gives the
following answer:
 \begin{eqnarray}
&&\phi(u+\sigma_u\tau,z+\sigma_z\tau)=\brc{1+\textbf{o}\brc{1}}\times\nonumber\svs
&&\left\{\matr{llll}{
-2\pi\imath q^{-\sigma_u\sigma_z+\bfi{\sigma_u}\bfi{\sigma_z}}\rme
\brc{-\flr{\sigma_z}u-\flr{\sigma_u}z},& \bfi{\sigma_u}>0,&\bfi{\sigma_z}>0,&\bfi{\sigma_u}+\bfi{\sigma_z}<1,\svs
\matr{r}{
4\pi q^{-\sigma_u\sigma_z+\bfi{\sigma_u}\bfi{\sigma_z}}\sin\brc{\pi\brc{u+z}}
\times\cr \times\rme\brc{-\brc{\frac12+\flr{\sigma_z}}u-\brc{\frac12+\flr{\sigma_u}}z}
},& \bfi{\sigma_u}>0,&\bfi{\sigma_z}>0,&\bfi{\sigma_u}+\bfi{\sigma_z}=1,\\[15pt]
\matr{r}{
2\pi\imath q^{-\sigma_u\sigma_z+\bfi{\sigma_u}\bfi{\sigma_z}-\bfi{\sigma_u+\sigma_z}}\times\cr
\times\rme\brc{-(1+\flr{\sigma_z})u-(1+\flr{\sigma_u})z}
},& \bfi{\sigma_u}>0,&\bfi{\sigma_z}>0,&\bfi{\sigma_u}+\bfi{\sigma_z}>1,\\[15pt]
\pi q^{-\sigma_u\sigma_z}\rme\brc{-\flr{\sigma_z}u-\sigma_uz}
\dfrac{\rme\brc{-\frac u2}}{\sin\brc{\pi u}},& \bfi{\sigma_u}=0,&\bfi{\sigma_z}>0,\\[15pt]
\pi q^{-\sigma_u\sigma_z}\rme\brc{-\sigma_zu-\flr{\sigma_u}z}
\dfrac{\rme\brc{-\frac z2}}{\sin\brc{\pi z}},& \bfi{\sigma_u}>0,&\bfi{\sigma_z}=0,\\[15pt]
\pi q^{-\sigma_u\sigma_z}\rme\brc{-\sigma_zu-\sigma_uz}
\dfrac{\sin\brc{\pi(u+z)}}{\sin\brc{\pi z}\sin\brc{\pi u}},& \bfi{\sigma_u}=0,&\bfi{\sigma_z}=0.
}\right.
\label{eq:PhiDec}
 \end{eqnarray}
To evaluate the limits of $f_{\alpha}\brc{u+\omega_{\beta},z}$
we use the identity (\ref{A3c}) and the expansion of $E_1(u-\sigma\tau)$:
 \begin{equation}
\label{eq:E1Dec}
E_1(u-\sigma\tau)=2\pi\imath\flr{\sigma}+\left\{\matr{ll}{
\pi\cot(\pi u)+\textbf{o}\brc{1},&
\{\sigma\}=0,\\
\pi\imath+2\pi\imath q^{\{\sigma\}}\textbf{\rme}(-u)+\textbf{o}\brc{q^{\{\sigma\}}},&
0<\{\sigma\}<\dfrac12,\\
\pi\imath+2\pi\imath q^{\frac12}\brc{\textbf{\rme}(-u)-\textbf{\rme}(u)}+\textbf{o}\brc{q^{\frac12}},&
\{\sigma\}=\dfrac12,\\
\pi\imath-2\pi\imath q^{1-\{\sigma\}}\textbf{\rme}\brc{u}+\textbf{o}\brc{q^{1-\{\sigma\}}},&
\dfrac12<\{\sigma\}<1.
}\right.
 \end{equation}

%%%%%%%%%%%%%%%%%%%%%%%%%%%%%%%%%%%%%%%%%%%%%%%%%%%%%%%%%%%
%%%%%%%%%%%%%%%%%%%%%%%%%%%%%%%%%%%%%%%%%%%%%%%%%%%%%%%%%%%%

\subsection{Appendix B: Noncommutative torus}
\setcounter{equation}{0}
\def\theequation{B.\arabic{equation}}

In this Appendix we use \cite{FFZ,Ho,Ri}.

\noindent {\sl 1. Definition and representation}.
 \noindent The noncommutative torus ${\cal T}^2_\hbar$ is a unital algebra with the
 two generators $(U_1,U_2)$ that satisfy the relation
 \beq{3.1}
U_1U_2=\bfe_\hbar^{-1} U_2U_1,~\bfe_\hbar=e^{ 2\pi i \hbar},~ \hbar\in[0,1)\,.
 \eq
Elements of ${\cal T}^2_\hbar$ are the double sums
 \beq{ds}
{\cal T}^2_\hbar=
\left\{X=\sum_{a_1,a_2\in{\mathbb Z}}c_{a_1,a_2}U_1^{a_1}U_2^{a_2}~|~~
c_{a_1,a_2}\in\mathbb C\right\}\,.
 \eq
It is convenient to introduce the following basis in ${\cal T}^2_\hbar$
 \beq{3.10}
T^a=\frac{i}{2\pi\hbar}\bfe
\left(-\frac{a_1a_2}{2}\hbar
\right)U_1^{a_1}U_2^{a_2}\,~~a\in{\mZ}^{2}=\mZ\oplus\mZ\,.
 \eq
It follows from (\ref{3.1}) that
 \beq{mnc}
T^aT^b=-2\pi\imath\hbar\bfe_\hbar\Bigl(\frac{a\times b}2\Bigr)T^{a+b}\,,~~(a\times b=
a_2b_1 -a_1b_2)\,.
 \eq
Therefore,
 \beq{mnc1}
T^aT^b=\bfe_\hbar(a\times b)T^{b}T^a\,.
 \eq
We can identify $U_1,U_2$ with matrices from GL$(\infty)$.
Define GL$(\infty)$ as the associative algebra of infinite matrices
$c_{jk}E_{jk}$, where $E_{jk}=||\delta_{jk}||$, such
that
 $$
{\rm sup}_{j,k\in\mZ}|c_{jk}|^2|j-k|^n<\infty\, ~{\rm for}~n\in\mN\,.
 $$
Consider the following two matrices from GL$(\infty)$:
 \beq{th}
Q=\di (\bfe(j\hbar))~~{\rm and }~~\La=||\delta_{j,j+1}||\,,~j\in\mZ\,.
 \eq
We have the following identification
 \beq{3.5}
U_1\to Q,~U_2\to \La\,.
 \eq
Another useful realization of $\cT^2_\hbar$ by the operators acting on the Schwartz space
$\cS(\mR)$
 \beq{3.5a}
U_1f(x)=f(x-\hbar)\,,~~U_2f(x)=\exp(2\pi ix)f(x)\,.
 \eq
The trace functional  on ${\cal T}^2_\hbar$ is defined as
 \beq{tr0}
\langle X\rangle= \tr(X)=c_{00}\,.
 \eq
It  satisfies the evident identities
 \beq{tr}
\lan1\ran=1\,,~~~\lan XY\ran=\lan YX\ran\,.
 \eq
The relation of $\cT^2_\hbar$  with the commutative algebra of smooth functions
on the two-dimensional torus $T^2=\{\mR^2/\mZ\oplus\mZ\}\,\sim \,\{0<x_1\leq 1,\,0<x_2\leq 1\}$
comes from the identification
 $$
U_1\to\bfe(x_1)\,,~U_2\to\bfe(x_2)\,,~~
\bfe(x_1)*\bfe(x_2)=e^{-2\pi\imath\hbar}\bfe(x_2)*\bfe(x_1)\,,
 $$
 \beq{3.2}
f(x)=\sum_{a\in\mZ^{2}}f_aT_a(x)\,, ~~T^a=T^a(x)=\frac{i}{2\pi\hbar}\bfe
\left(\frac{a_1a_2}{2}\hbar
\right)\bfe(a_1x_1)\bfe(a_2x_2)\,.
 \eq
The multiplication on $T^2$ becomes the Moyal multiplication:
 \beq{3.3}
(f*g)(x):=fg+
\sum_{n=1}^\infty\frac{(i\pi\hbar)^n}{n!}
\ve_{r_1S^1}\ldots\ve_{r_nS^n}(\ti\p^n_{r_1\ldots r_n}f)
(\ti\p^n_{S^1\ldots S^n}g)\,,~~~\ti\p_j=\f1{2\pi i}\p_j\,.
 \eq
The trace functional (\ref{tr0}) in the Moyal identification
is the integral
 \beq{3.6}
\tr f=-\f1{4\pi^2}\int_{{\cal T}^2_\hbar}fdx_1dx_2=f_{00}\,.
 \eq

%Define a projective left module $E_\hbar$ over $\cT^2_\hbar$ \cite{CR}. Its elements are functions
%from the Schwartz space $\cS(\mR)$ and the action is (see (\ref{3.5a}))
%\beq{rim}
%E_\hbar=\{f(x)\in\cS(\mR)\,|\,~ U_1 f(x)=f(x-\hbar)\,,~U_2 f(x)=\bfe(x)f(x)\}\,.
%\eq
%The analogues of degree and the rank are
%\beq{rd}
%deg(E_\hbar)=1\,,~~rk(E_\hbar)=\hbar^{-1}\,.
%\eq

%%%%%%%%%%%%%%%%%%%%%%%%%%%%%%%%%%%%%%%%%%%%%%%%%%%%%%%%%%%%%

\bigskip
\noindent {\sl 2. $sin$-algebra.}%\vskip2mm

 \noindent We denote by $\gg=sin_\hbar$ the Lie algebra with the generators $T^{\al}$ $(\al\in\mZ^{2})$
over the ring $\gS$
% \beq{lat}
%\mZ^{(2)}=\{\al=(\al_1,\al_2)\,,\al_j\in\mZ,\,\al\neq(0,0)\}\,,
% \eq
 \beq{3.12}
sin_\hbar=\{\psi=\sum_{\al\in\mZ^{2}}\psi_{\al}T^{\al},~~\psi_{\al}\in\gS\}\,,~~
\gS=\{\psi_\al=0~~
{\rm almost~ for~all }\,,~\al\in\mZ^{2}\}\,.
 \eq
In other words, $sin_\hbar$ is a noncommutative analog of the algebra of  the trigonometric
 polynomials.
 It follows from (\ref{mnc}) that the commutator has the form
 \beq{3.11}
[T^{\al},T^{\be}]=\bfC_\hbar(\al,\be)T^{\al+\be}\,,
 \eq
where
 \beq{3.11a}
\bfC_\hbar(\al,\be)=\f1{\pi\hbar}\sin\pi\hbar(\al\times \be)\,,~~~\al\times \be=\al_2\be_1-\al_1\be_2\,.
 \eq
In the Moyal representation (\ref{3.3}) the commutator of $sin_\hbar$
has the form
 \beq{3.14}
[f(x_1,x_2),g(x_1,x_2)]_\hbar:=\f1{\hbar}(f*g-g*f)\,
 \eq
The trace functional (\ref{tr0}) allows one to define the coalgebra $\gg^*=sin^*_\hbar$.
It is defined as
 the linear space of distributions
 \beq{A.20}
sin^*_\hbar=\left\{
\bfS\,\,|\, \int_{\cT^2_\hbar}\bfS\cdot\psi<\infty\,,~{\rm for}~\psi\in sin_\hbar\right\}
 \eq
The group $SIN_\hbar$ is the group of invertible elements
from ${\cal T}^2_\hbar$.
 \beq{A.10}
SIN_\hbar=\{\Psi=\sum_{a\in\mZ^{2}}\Psi_{a}T^{a}\}\,.
 \eq
In this group $\frac{\imath}{2\pi\hbar}T^0$ plays the role of the identity element and
$(T^a)^{-1}=T^{-a}$.

%%%%%%%%%%%%%%%%%%%%%%%%%%%%%%%%%%%%%%%%%%%%%%%%%%%%%%%%%%%%%%%%%%%%%%%%%%%%%%%%%%%%%%%%%%%%
\bigskip
\noindent \emph{3. sine basis in $\gln$}.
 \noindent This  basis is a finite-dimensional version of (\ref{3.10}). Let
  \beq{z2n}
 \mZ^{(2)}_N=(\mZ_N\oplus\mZ_N)\setminus(0,0)\,.
  \eq
 Then
 \beq{ta2}
T^\al=e^{\frac{2\pi i}Na_1a_2}Q^{\al_1}\La^{\al_2}\,,~\al\in \mZ^{(2)}_N\,,
 \eq
 where $Q$ and $\La$ are the t'Hooft matrices
 $$
Q=\di(\bfe_N(1),\bfe_N(2),\ldots,1)\,, ~~~\bfe_N(x)=\exp\frac{2\pi i}N(x)\,,
 $$
 \beq{la}
\La= \left( \begin{array}{ccccc}
0&1&0&\cdots&0\\
0&0&1&\cdots&0\\
\vdots&\vdots&\ddots&\ddots&\vdots\\
0&0&0&\cdots&1\\
1&0&0&\cdots&0
 \end{array}\right)\,.
 \eq
 \beq{ta}
[T^{\al},T^{\be}]=\bfC(\al,\be)T^{\al+\be}\,,~~
\bfC_N(\al,\be)=\frac{N}{\pi}\sin\frac{\pi}N(\al\times \be)
 \eq
The action of the transition matrices is diagonalized in this basis
 \beq{atm}
QT_\al Q^{-1}=\bfe(-\al_2/N)T^\al \,,~~\La T^\al \La^{-1}=\bfe(\al_1/N)T^\al \,.
 \eq
If $a\in\mZ^{(2)}_N\cup(0,0)$ the $\{T^a\}$ forms a basis $\gln$.

%%%%%%%%%%%%%%%%%%%%%%%%%%%%%%%%%%%%%%%%%%%%%%%%%%%%%%%%%%%%%%%%%%%%%%%%%%%%%%%%%%

\bigskip
\noindent \emph{4.NCT and two-loop and one-loop algebras}.

 \noindent Define the two-loop algebra as
 \beq{tlga}
LL(\gln)\,:\, \mC^*\times\mC^*\to\gln\,.
 \eq
 This algebra has the basis
 \beq{tlg}
T_N^\al=T^{ \al_1,\al_2}\bfe(\al_1y_1+\al_2y_2)\,,~~\al_s=\ti \al_s+j_sN\,,~
0\leq\ti \al_j<N\,,~j_s\in\mZ\,,~s=1,2\,,
 \eq
with commutation relations
 \beq{crtl}
[T^\al_N,T^\be_N]=\bfC^{rat}_N(\ti \al,\ti \be)T^{\al+\be}_N\,,
 \eq
 $$
\bfC^{rat}_N=(-1)^\ep\bfC_N\,,~~\ep=\ti\al_1k_2+\ti\be_2j_1
-\ti\al_2k_1-\ti\be_1j_2+N(j_1k_2-j_2k_1)\,,
 $$
where $\bfC_N$ is defined by (\ref{ta}). This definition of $LL(\gln)$ differs
from the standard one due to the factor $(-1)^\ep$. The Jacobi identity for
the brackets follows from the isomorphism of $LL(\gln)$ and $sin_\hbar$ for
 $\hbar=M/N$, $M<N$ and $(M,N)$ are coprime numbers.
 The isomorphism is provided by the
map of the basis $T^\al\to T_N^\al$.
 For $\hbar=M/N$ the commutator (\ref{crtl}) coincides
with (\ref{3.11}).

The reduction to the one-loop algebra
\beq{tlga1}
L(\gln)\,:\, \mC^*\to\gln
 \eq
leads to the basis
 \beq{tlg1}
T_N^\al=T^{ \al_1,\al_2}\bfe(\al_2y)\,,~~\al_2=\ti \al_2+jN\,,~
0\leq\al_1\,, \ti \al_2<N\,,~j\in\mZ\,.
 \eq
% with the commutation relations
Then in this basis the commutator assumes the form
 \beq{ctr}
[T_N^{\ti\al_1,\ti\al_2}\bfe(\al_2y),T_N^{\ti\be_1,\ti\be_2}\bfe(\be_2y)]=
\bfC^{tr}_{1/N}(\ti\al,\ti\be)
T_N^{\widetilde{\ti\al_1+\ti\be_1},
\widetilde{\ti\al_2+\ti\be_2}}\bfe((\al_2+\be_2)y)\,,
 \eq
 $$
 \bfC^{tr}_{1/N}(\ti\al,\ti\be)=(-1)^\ep\bfC_{1/N}(\ti\al,\ti\be)\,,~~
 \ep=\al_2n-\be_2m\,.
 $$
The notion $\widetilde{\ti\al_1+\ti\be_1}$ means that the sum is taken $mod\,N$.
The algebra has the representation by the basis in $\gln$
 \beq{bat}
T_N^{\ti\al_1,\ti\al_2}=\bfe_N(\ti\al_1\al_2)Q^{\ti\al_1}\La_{tr}^{\ti\al_2}\,,
 \eq
 $$
Q=\di(1,\bfe_N(1),\ldots,\bfe_N(N-1))\,,~~\La_{tr}^{\ti\al_2}=\{E_{j,j+\ti\al_2}\}\,.
 $$
%This parametrization means  that we use the principle gradation of $L(\sln)$
%\cite{Ka} (chapter 14) defined on the generators as
% $$
%yE_{j,j+1}~(j=1,\ldots,N-1)\,,~~y^{-1}E_{j+1, j}~(j=2,\ldots,N)\,.
% $$
%To pass to the standard gradation
%$L(\gln)=\{T_N^{\ti\al_1,\ti\al_2}\bfe(m\xi)\}$ $(\al_2=\ti\al_2+mN)$
%one should conjugate the basis (\ref{trba}) by the diagonal matrix $h=\di(y,y^2,\ldots,y^N)$
% $$
%T_N^{\ti\al_1,\ti\al_2}\bfe(m\xi)=h T_N^{\ti\al_1,\ti\al_2}\bfe(\al_2y)h^{-1}\,,~\xi=y^N\,.
% $$

%%%%%%%%%%%%%%%%%%%%%%%%%%%%%%%%%%%%%%%%%%%%%%%%%%%%%%%

\bigskip
\noindent {\sl 5. SL$(2,\mC)$ case}.

 \noindent For   SL$(2,\mC)$  the basis $T_\al$ is the basis of sigma-matrices
 $$
\si_0=Id\,,~~\si_1=i\pi T^{0,1}\,,~~\si_2=i\pi T^{1,1}\,,~~\si_3=-i\pi
T^{1,0}\,,
 $$
 \beq{100}
\{\si_a\}=\{\si_0,\si_\al\}\,,(a=0,\al)\,,(\al=1,2,3)
\,,~~[\si_\al,\si_\be]=\epsilon_{\al,\be,\ga}\si_\ga\,,
 \eq
 $$
\si_1=\left(
         \begin{array}{cc}
          0 & 1 \\
          1 & 0 \\
         \end{array}
      \right)\,,~
      \si_2=\left(
         \begin{array}{cc}
          0 & -\imath \\
          \imath & 0 \\
         \end{array}
      \right)\,,~
      \si_3=\left(
         \begin{array}{cc}
          1 & 0 \\
          0 & -1 \\
         \end{array}
      \right)\,.
 $$
The standard theta-functions with the characteristics are
 \beq{101}
\te_{0,0}=\te_3\,,~~\te_{1,0}=
\te_2\,,~~\te_{0,1}=\te_4\,,~~\te_{1,1}=\te_1\,.
 \eq
Half-periods:
 \beq{ome}
\om_a=(\om_0,\om_\al)\,,~~\om_\al=\frac{\al_1+\al_2\tau}2\,,~~\al=(\al_1,\al_2)\,,~\al_j=0,1\,.
 \eq
 \begin{center}
\vspace{3mm}
 \begin{tabular}{|c||c|c|c|}
\hline
$\al$ & (1,0) & (0,1) & (1,1) \\
\hline
$\si_\al$ &
$\si_3$   &
$\si_1$ &
$\si_2$ \\
\hline
 half-periods & $\om_1=\oh$ & $\om_2=\frac{\tau}{2}$ &
$\om_3=\frac{1+\tau}{2}$ \\
\hline
$\phi_\al (z)$    &
$\frac{\te_2(z)\te'_1(0)}{\te_2(0)\te_1(z)}$ &
$\frac{\te_4(z)\te'_1(0)}{\te_4(0)\te_1(z)}$ &
$\frac{\te_3(z)\te'_1(0)}{\te_3(0)\te_1(z)} $\\
\hline
 \end{tabular}
 \end{center}
%\caption
%{$\si_\al$ and half-periods}
\vspace{3mm}
Let
 \beq{var}
\varphi_\al(u,z)=\bfe(-\al_2z/2)\phi(-\om_\al+u,z)\,,
 \eq
 \begin{equation}
\label{eq:fadef}
f_{\alpha}(u,z)=\bfe(-\alpha_2 z/2)f(-\omega_{\alpha}+u,z)\,.
 \end{equation}
% \beq{efi02}
%\vf_\al(z)\vf_\al(z-\om_\al)=-
%\bfe_1(-\om_\al\p_\tau\om_\al)\left(\frac{\vth'(0)}{\vth(\om_\al)}\right)^2
% \eq
Then from (\ref{A.300}),  (\ref{A.14})
 \beq{sc}
\vf_\al(-u,-z-\om_b)=
\left\{
 \begin{array}{lc}
 - \vf_\al(u,z) & b=0\,,\\
  -\bfe(-b_2u)\vf_\al(u,z-\om_\al) & b=\al\,, \\
     \bfe(-b_2u)\vf_\al(u,z-\om_b) & b\neq\al\,,
 \end{array}
\right.
 \eq
 \beq{mal}
\vf_{-\al}(u,z-\om_b)= \vf_\al(u,z-\om_b)\,.
 \eq

%%%%%%%%%%%%%%%%%%%%%%%%%%%%%%%%%%%%%%%%%%%%%%%%%%%%%%%%

\bigskip
\noindent {\sl 6. Elliptic constants related to NCT $\cT_\hbar$}.

 \noindent  Introduce two  numbers
 $\ep=(\ep_1,\ep_2)$   such that  $\ep_a\hbar<1$ and $\ep_a\hbar$
are irrational. Consider the  dense
set  $\mZ_{\hbar,\ep}(\tau)$ in $E_\tau$:
 \beq{Z_eq}
\mZ_{\hbar,\ep}(\tau)=\{(\ep_1\ga_1+\tau\ep_2\ga_2)\hbar=\hbar\ep\cdot\ga\in E_\tau~~|~
~(\ga_1,\ga_2)\in\mZ^{(2)}\}\,.
 \eq
The corresponding
elliptic functions with the arguments from $\mZ_{\hbar,\ep}(\tau)$
 are as follows:
 \beq{C0}
\vth(\hbar\ep\cdot\ga)\,,~~\zeta(\hbar\ep\cdot\ga)\,,~~\wp(\hbar\ep\cdot\ga)\,,~~
E_1(\hbar\ep\cdot\ga)\,,~~E_2(\hbar\ep\cdot\ga)\,,
 \eq
 \beq{CC3}
\vf_{\hbar\ep\cdot\ga}(z)=\bfe_\hbar(\ep_2\ga_2z)\phi(\hbar\ep\cdot\ga,z)\,.
 \eq
 \beq{CC3a}
f_{\hbar\ep\cdot\ga}(z)=\bfe_\hbar(\ep_2\ga_2z)\p_u\phi(u,z)|_{u=\hbar\ep\cdot\ga}\,.
 \eq
It follows from (\ref{A.14}) that
 \beq{opph}
\vf_{\hbar\ep\cdot\ga}(z+1)=\bfe_\hbar(\ep_2\ga_2)\vf_{\hbar\ep\cdot\ga}(z)\,,~~~
\vf_{\hbar\ep\cdot\ga}(z+\tau)=\bfe_\hbar(-\ep_1\ga_1)\vf_{\hbar\ep\cdot\ga}(z)\,.
 \eq

\bigskip
\noindent {\sl 7. Dispersionless limit.}

 \noindent In the limit $\hbar\to 0$ the Lie group $SIN_\hbar$ becomes the group of the volume preserving diffeomorphisms
$SDiff(T^2)$ of the two-dimensional torus $T^2$ and the algebra $sin_\hbar$ becomes the Lie algebra of Hamiltonian functions
 \beq{Ham}
Ham(T^2)\sim C^\infty(T^2)/\mC
 \eq
 equipped with the canonical Poisson brackets.
In $Ham(T^2)$ we have the Fourier
basis
 \beq{fb}
\bfe(\al\cdot x)=\exp (2\pi i(\al_1x_1+\al_2x_2))
 \eq
instead of the basis (\ref{3.10}).
 The commutator
(\ref{3.11}) becomes
 \beq{3.7b}
[\bfe(\al x),\bfe(\be x)]=(\al\times\be)\bfe((\al+\be)\cdot x)\,,
 \eq
or in terms of functions $f,g\in Ham(T^2)$
 \beq{pal}
[f,g]:=\{f,g\}=\p_1f\p_2g-\p_2f\p_1g\,,~~\p_j=\p_{x_j}\,.
 \eq
The algebra $Ham(T^2)$ (without constant Hamiltonians) is isomorphic
to Lie algebra $SVect_0(T^2)$
of the divergence-free {\it zero-flux} vector fields on $T^2$
equipped with the area
form $-4\pi^2dx_1dx_2$. Let $h(x_1,x_2)\in Ham(T^2)$. Then
the Hamiltonian field $V_h$ corresponding to the Hamiltonian function $h$ is
 \beq{3.20}
V_h=-\f1{4\pi^2}((\p_2h)\p_1-(\p_1h)\p_2)\,,
 \eq
while
 \beq{3.21}
[V_{h},V_{h'}]=V_{\{h,h'\}}\,.
 \eq
For $f(x)=\sum_\al f_\al\bfe(\al\cdot x)$
 \beq{int}
\int_{T^2}f=-\f1{4\pi^2}f_0\,.
 \eq

\end{document}